\documentclass[preprint,12pt]{elsarticle}
\usepackage{amssymb}
\usepackage{mathrsfs}
\usepackage{epsfig}
\usepackage{graphicx}
\usepackage{dcolumn}
\usepackage{bm}
\usepackage{times}
\usepackage{xcolor}
\usepackage[percent]{overpic}
\usepackage{cancel}
\usepackage{amsmath}
\usepackage{multirow}
\usepackage{booktabs} 
\usepackage{hyperref}
\usepackage[hyphenbreaks]{breakurl}
\usepackage{mathrsfs}
\usepackage{multirow}
\usepackage{longtable}
\usepackage{tabu}
\usepackage{lscape} 
\usepackage{setspace}
\usepackage{dsfont}
\usepackage{extarrows}
\usepackage{bbm}
\usepackage{fancybox}
\usepackage{subfigure}

\begin{document}
\begin{sloppypar}

\begin{frontmatter}

\title{Network Alignment}

\author{Rui Tang$^{1,2,3}$, Ziyun Yong$^{1}$, Shuyu Jiang$^{1}$, Xingshu Chen$^{1,2,3}$, Yaofang Liu$^{*4}$, Yi-Cheng Zhang$^{5}$, Gui-Quan Sun$^{*6,7}$, Wei Wang$^{*8}$} 
\cortext[cor1]{Corresponding Authors: Yaofang Liu, Gui-Quan Sun, and Wei Wang are co-corresponding authors. \\
E-mail addresses: sclzlyf001@swmu.edu.cn (YaoFang Liu), gquansun@126.com (Gui-Quan Sun), wwzqbx@hotmail.com (Wei Wang).}

\address{1. School of Cyber Science and Engineering, Sichuan University, Chengdu, 610065, China}
\address{2. Cyber Science Research Institute, Sichuan University, Chengdu, 610065, China}
\address{3. Key Laboratory of Data Protection and Intelligent Management, Ministry of Education, Sichuan University, Chengdu, 610065, China}
\address{4. Department of Reproductive Technology, The Affiliated Hospital of Southwest Medical University, Luzhou, 646000, China}
\address{5. Physics Department, University of Fribourg, Chemin du Musée 3, 1700 Fribourg, Switzerland}
\address{6. Sino-Europe Complex Science Center, School of Mathematics, North University of China, Shanxi, Taiyuan 030051, China}
\address{7. Complex Systems Research Center, Shanxi University, Shanxi, Taiyuan 030006, China}
\address{8. School of Public Health, Chongqing Medical University, Chongqing, 400016, China}

\begin{abstract}
Complex networks are frequently employed to model physical or virtual complex systems. When certain entities exist across multiple systems simultaneously, unveiling their corresponding relationships across the networks becomes crucial. This problem, known as network alignment, holds significant importance. It enhances our understanding of complex system structures and behaviours, facilitates the validation and extension of theoretical physics research about studying complex systems, and fosters diverse practical applications across various fields. However, due to variations in the structure, characteristics, and properties of complex networks across different fields, the study of network alignment is often isolated within each domain, with even the terminologies and concepts lacking uniformity. This review comprehensively summarizes the latest advancements in network alignment research, focusing on analyzing network alignment characteristics and progress in various domains such as social network analysis, bioinformatics, computational linguistics and privacy protection. It provides a detailed analysis of various methods' implementation principles, processes, and performance differences, including structure consistency-based methods, network embedding-based methods, and graph neural network-based (GNN-based) methods. Additionally, the methods for network alignment under different conditions, such as in attributed networks, heterogeneous networks, directed networks, and dynamic networks, are presented. Furthermore, the challenges and the open issues for future studies are also discussed.
\end{abstract}


\end{frontmatter}

\tableofcontents
\newpage
\section{Introduction} \label{intro}

Complex networks are frequently employed to model a wide range of physical and virtual systems, from biological networks, power grids, and transportation systems to software systems, the Internet, and the World Wide Web~\cite{huberman1999growth,albert2000error,strogatz2001exploring,albert2002statistical}. The components within complex systems are represented as nodes, and the relationships or interactions between these components are represented as links. Scientists use this paradigm to accurately and comprehensively describe these systems, enabling them to uncover and explore structural patterns, such as clustering, motifs, and community structures, as well as the dynamics, including spreading processes~\cite{de2018fundamentals,wang2017unification,pastor2015epidemic,sun2024dynamics,liu2023global,chang2022sparse}, synchronization~\cite{arenas2008synchronization,lu2014synchronization,dorfler2014synchronization,wu2024synchronization}, and games~\cite{szabo2007evolutionary,perc2017statistical,wang2015evolutionary,wang2015coupled}. For example, in the aspect of network structures, Shen-Orr et al.~\cite{shen2002network} analyzed the network motifs within gene regulation networks and identified three significant motifs in the transcriptional interaction network of Escherichia coli, which were proven to play specific roles in determining gene expression. This method offers a framework for defining the basic computational elements of other biological networks, driving a series of related studies. Building on these efforts, Shoval et al.\cite{shoval2010snapshot} identified several motifs shared by diverse organisms, such as plants and humans, and demonstrated that each motif carries out specific dynamic functions in cellular computation. Zu et al.~\cite{zu2023single} compared these common network motifs to help uncover the regulatory mechanisms within various cell types. Regarding dynamics, specifically the spreading process, scientists have extensively studied how information, diseases, or behaviours propagate through networks. Pastor-Satorras and Vespignani~\cite{pastor2001epidemic} studied the spread of computer viruses on scare-free networks. They found that, within a certain range of the degree exponent, an epidemic threshold and its associated critical behaviour do not exist. This insight has profound implications for understanding and controlling the spread phenomena in real-world networks, such as improving our ability to better predict infectious diseases disease by accounting for more topological features of these networks~\cite{hufnagel2004forecast,belik2011natural} or finding structural diversity in social contagion~\cite{ugander2012structural,su2018optimal}.

Unveiling the corresponding relationships of components to build a bridge of connections from one network to another becomes crucial. This challenge, known as network alignment, is essential for advancing the understanding of complex systems, validating and extending theoretical research, and enabling practical applications across various fields. For instance, in social network analysis, network alignment can be used to integrating the different online social networks (OSN), enabling a more comprehensive study of social network dynamics~\cite{tejedor2018diffusion,cozzo2013contact}. In bioinformatics, protein-protein interaction (PPI) network alignment can be utilized to establish node mappings between networks of different species, thereby predicting protein function from a well-studied species to a poorly studied one~\cite{clark2014comparison}. In computational linguistics, each language, whether English, Chinese, German, etc., has a network structure where words are considered nodes and co-occurrence in sentences as links~\cite{cancho2001small}. In each language, words, phrases, people, places, organizations, events, and activities, among other entities or concepts, can be represented as nodes. A knowledge graph (KG) for the language is constructed by representing the relationships between these nodes as edges or links, forming a structured network of interconnected knowledge~\cite{trisedya2019entity}. Network alignment can be used to align the KGs of two languages, aiding cross-language translation, cross-language information integration, etc.

Due to variations in the structure, characteristics, and properties of complex networks across different fields, the study of network alignment is often isolated within each domain, with even the terminologies and concepts lacking uniformity. 
For example, in social network analysis scenarios, this problem is typically referred to as user identity linkage, while in the privacy protection domain, it is known as de-anonymization. For researchers from other fields, these terms may not clearly indicate their relevance to network alignment. This leads to several disadvantages. First, the latest insights, technologies, and methods related to network alignment in various fields are difficult to quickly disseminate and be understood by other domains, hindering the exchange and collision of ideas. Second, inconsistencies in terminology and concepts may cause confusion and misunderstanding among researchers, making it challenging to compare results, replicate studies, or advance cross-domain research. Third, the lack of a unified framework or consensus may result in researchers unknowingly duplicating efforts to develop similar network alignment solutions, wasting resources and time. Finally, the lack of a holistic perspective impedes this issue's comprehensive understanding and widespread application.

To address these issues and provide a holistic perspective that promotes the comprehensive understanding and widespread application of network alignment, we present a comprehensive review of the latest advancements in this field.  Specifically, there are three main reasons for us to write this review on network alignment. First, despite the publication of numerous papers on network alignment in recent years, there remains a lack of a thorough review that systematically organizes recent developments, identifies key challenges, and outlines open questions for future research. Second, although network alignment has shown promising applications in fields such as social network analysis, bioinformatics, and computational linguistics,  it remains unclear which specific problems in these areas are directly related to network alignment or can be addressed through network alignment techniques, as well as how such transformations can be made. This gap highlights the need for a comprehensive overview of applications across these domains. Third, the existing literature often employs different terminologies and frameworks to address similar alignment problems, especially across different fields, creating an urgent need to unify descriptions and standardize notations. By offering this review, we aim to clarify these issues and foster further development and integration of network alignment research. 

This review systematically explores the progress of network alignment research on complex networks. In section 2, we introduce the fundamental concepts of complex networks related to network alignment, the definition of the alignment problem and various evaluation metrics. In section 3, we examine four key fields, analyzing which specific problems in these areas can be considered network alignment problems or how they can be transformed into alignment tasks, presenting a statistical summary of the current research situation in each area. we provide a detailed overview of the latest developments in network alignment, focusing on two major approaches: structure consistency-based methods and machine learning-based methods. The structure consistency-based methods are divided into local and global structure consistency-based methods. The machine learning-based methods are categorized into network embedding-based methods, GNN-based methods, and feature extraction-based methods. For each category, we thoroughly explain the underlying principles and discuss recent advancements in the field. In section 5, we extend the discussion to specific network conditions, including attributed networks, heterogeneous networks, directed networks, dynamic networks, and alignment without seeds, analyzing how to effectively perform alignment under these conditions. Finally, in section 6, we summarize the key findings, highlight ongoing challenges, and discuss future research directions. In summary, this review primarily discusses three questions: What is network alignment? Where is network alignment needed? And how can network alignment be effectively achieved?

\section{Basic concepts and metrics} \label{Theoretical studies}
This section will introduce the fundamental concepts of network alignment, including relative concepts, problem definition and evaluation metrics.

\subsection{Relative concepts}
Complex networks are widely used to describe complex systems. In these networks, nodes represent elements, and edges represent the relationships among nodes. We briefly introduce some critical concepts. 

A complex network $G$ can be described by an adjacency matrix $\bm{\mathrm{A}}$, which $\bm{\mathrm{A}}(i,j)=1$ if there is a link between nodes $v_i$ and $v_j$,  and 0 otherwise. If node $v_i$ has a link with node $v_j$, $v_i$ can be referred to as the neighbour of node $v_j$, and the neighbour set for node $v_j$ can be denoted as $\Gamma_1(v_j)$. Network $G$ is directed or undirected depending on whether the network links have a direction.  
For an undirected network, its adjacency matrix is symmetric, meaning that if $\bm{\mathrm{A}}(i,j)=1$, then $\bm{\mathrm{A}}(j,i)=1$. An undirected edge between nodes $v_i$ and $v_j$ can be denoted as $(v_i,v_j)$ or $(v_j,v_i)$. In contrast, its adjacency matrix is not symmetric for a directed network. $\bm{\mathrm{A}}(i,j)=1$ only indicates that there is a link (sometimes referred to as an arc) from $v_i$ to $v_j$ in the network. A directed edge between nodes $v_i$ and $v_j$ can be denoted as $<v_i,v_j>$, where the direction of the edge is from $v_i$ to $v_j$. For an undirected network, the degree of node $v_i$ is the number of links between $v_i$ and other nodes. It can be represents as $d(v_i)=\sum_{j=1}^{|V|} \bm{\mathrm{A}}(i,j)=\sum_{i=1}^{|V|} \bm{\mathrm{A}}(i,j)$. For a directed network, the degree of nodes includes both out-degree and in-degree. The out-degree is the number of arcs link from node $v_i$ to the other nodes, denoted as $d_{out}(v_i)=\sum_{j=1}^{|V|} \bm{\mathrm{A}}(i,j)$; the in-degree is the number of arcs link from the other nodes to node $v_i$, denoted as $d_{in}(v_i)=\sum_{j=1}^{|V|} \bm{\mathrm{A}}(j,i)$. The matrix $\bm{\mathrm{D}}$ is used to record the degrees for all nodes in the network $G$, where $\bm{\mathrm{D}}(i,i)=d(v_i)$ for any $i \in [1,|V|]$, and all other entries are 0. To better reflect the similarity and structural characteristics between nodes, the adjacency matrix $\bm{\mathrm{A}}$ can be normalized as $\mathcal{A}=\bm{\mathrm{D}}^{-\frac{1}{2}}\bm{\mathrm{A}}\bm{\mathrm{D}}^{-\frac{1}{2}}$. This normalization reduces the bias caused by high-degree nodes in the overall computation when using the adjacency matrix, thereby balancing the influence of different nodes in network analysis and more accurately reflecting the global structural features of the network. In many studies on network alignment, directed networks are often converted into undirected networks to simplify the analysis. In this review, unless otherwise specified, the networks we refer to are undirected.
For network $G$, its Laplacian matrix is defined as $\bm{\mathrm{L}}=\bm{\mathrm{D}}-\bm{\mathrm{A}}$. To overcome the limitations of the standard Laplacian matrix when dealing with networks that have significant differences in node degrees, the Laplacian matrix can be normalized as $\mathcal{L}=\bm{\mathrm{D}}^{-1/2}\bm{\mathrm{L}}\bm{\mathrm{D}}^{1/2}$. The network spectrum is the set of eigenvalues of a matrix representing the network, such as its adjacency matrix or Laplacian matrix. The Laplacian matrix and the spectrum are useful for analyzing networks. For example, by analyzing the spectrum of the Laplacian matrix, we can identify community structures or clusters within the network. Complex systems are typically modelled as static networks. However, most real-world systems are dynamic as they evolve. Such systems can be better modelled as dynamic networks, which capture the moments when interaction begins and ends. A dynamic network $G(t) = (V, T)$ consists of a node-set $V$ and an event set $T$, where an event is a temporal edge. An event consists four elements $(v_i,v_j,t_s,t_e)$, representing an interaction between nodes $v_i$ and $v_j$ from time $t_s$ to time $t_e$. Although time is captured continuously, dynamic network data is often provided as a discrete temporal sequence of static network snapshots, such as $G{(t_1)}(V{(t_1)}, E{(t_1)}),\cdots, G{(t_k)}(V{(t_k)}, E{(t_k)})$.

\subsection{Problem definition}\label{definition}
\begin{figure*} [ht!]
    \centering
\includegraphics[width=1.0\textwidth]{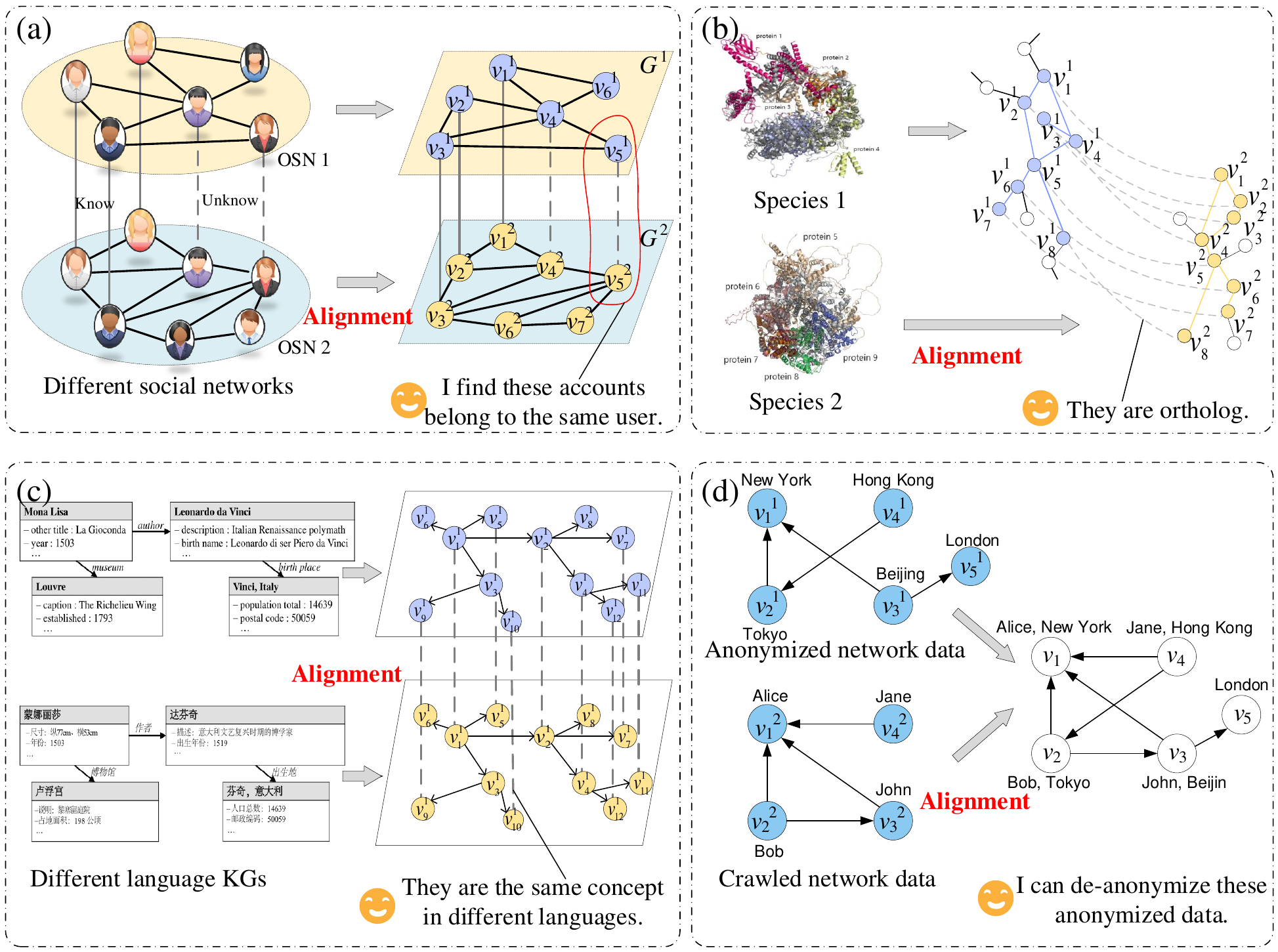}
    \caption{Illustration of network alignment. (a) Network alignment in social network analysis. Suppose there are two OSN platforms. Several tasks in this field, such as cross-platform account linkage, can be framed as network alignment problems. (b) Network alignment in bioinformatics. Suppose there are two species, and the proteins within their cells interact with one another. These protein interactions form PPI networks for the different species. Several tasks in this field, such as ortholog identification, can be transformed into network alignment problems. (c) Network alignment in computational linguistics. Suppose there are two knowledge graphs in different languages. Several tasks in this field, such as cross-language node linkage, can be translated into network alignment problems. (d) Network alignment in privacy protection. Suppose an organization shares anonymized network data of its employees for data-sharing purposes. Meanwhile, an attacker collects data on the organization's personnel through web scraping. The problem of de-anonymizing the shared data can be transformed into a network alignment problem. More problems related to network alignment across various fields can be found in Sec.~\ref{Sec: Network alignment in different fields}.
    \\Source: Reproduced from Ref.~\cite{tang2020interlayer,burke2023towards,zhang2019multi,shao2019fast}.
    }
    \label{pic:framework}
\end{figure*}

In the real world, there are scenarios where some nodes in one network also exist in other networks. For example, a molecule in the PPI network of one species may also be present in the PPI network of another species; lots of users in one OSN may also have accounts on other OSNs; and some concepts in the KG of one language may exist in the KGs of different languages. When we aim to analyze one network using another or to integrate several networks as a whole for analysis, it is crucial to correctly align these nodes that represent the same entities across different networks. However, alignment is very challenging, as the correspondence between nodes in other networks is often unknown, e.g., a user's accounts on different OSNs are not publicly known. This challenge is called the network alignment problem, as shown in Figure.~\ref{pic:framework}.

Using a graph $G(V,E,C)$ to represent a complex network and denoting $V$ as the node-set, which means all the components in a complex system, $E$ as the edges set, which represents all the relationships between any two nodes in $V$, and $C$ as the attribute set of the nodes. To differentiate various networks, we can use superscript indices for the symbols, representing the networks as $G^{l_i}(V^{l_i},E^{l_i},C^{l_i})$, where $l_i$ indices network $i$. 

Next, we use the alignment of two networks as an example to define network alignment from the perspective of the entire network and the individual nodes. Given two networks $G^{l_1}$ and $G^{l_2}$, if any node $v^{l_1}_i \in V^{l_1}$ and $v^{l_2}_j \in V^{l_2}$ represent the same entity, node pair $(v^{l_1}_i,v^{l_2}_j)$ can be refered to as corresponding node pair. (a) The perspective of the entire network: the goal of network alignment is to find a function $f: V^{l_1} \rightarrow V^{l_2}$ that identifies all the corresponding node pairs across $G^{l_1}$ and $G^{l_2}$. (b) Perspective of the individual nodes: given any two nodes $v^{l_1}_i$ and $v^{l_2}_j$, the goal of network alignment is to find a function $f(v^{l_1}_i, v^{l_2}_j) \rightarrow 0,1 $ such that 
\begin{equation}
f(v^{l_1}_i,v^{l_2}_j) = \left\{
\begin{array}{ll}
1, & \mbox{if $v^{l_1}_i$ and $v^{l_2}_j$ is the same entity} \\
0, & \mbox{otherwise}
\end{array}.
\right.
\label{eq_objectivefunction_na}
\end{equation}
In the alignment process, we have prior knowledge of some corresponding node pairs, referred to as observed corresponding node pairs (i.e., matched node pairs). In contrast, the remaining ones are called unobserved corresponding node pairs (i.e., unmatched node pairs). We denote the sets of the corresponding node pairs, the observed corresponding node pairs, and the unobserved corresponding node pairs as $\Psi$,$\Psi^o$, and $\Psi^u$, respectively. The nodes in $\Psi^o$ can be called matched nodes, while others can be called unmatched nodes. One node in a corresponding node pair can be called the corresponding node for the other node in the pair. We use the term corresponding node pair and matched node pair interchangeably in this review. It is worthwhile noting that since the alignment of three or more networks can be reduced to pairwise alignments between these networks, in this review, unless otherwise specified, the alignment we refer to is assumed to be between two networks. 

\subsection{Evaluation metrics}
Due to their importance in academic, societal, and economic value, hundreds of network alignment methods applicable to various fields and scenarios have been proposed by researchers in recent years (detailed statistical information can be found in Sec.~\ref{Sec: Network alignment in different fields}). Evaluating the effectiveness and superiority of these methods using appropriate evaluation metrics is crucial. Researchers have developed over ten different evaluation metrics. Based on the types of data used for evaluation, we categorize these metrics into three main types: correspondence-based metrics, which rely on the valid corresponding node pairs to assess alignment quality; structure-based metrics, which evaluate the degree of edge overlap or structural similarity between the aligned networks; and function-based metrics, which measure the functional coherence or similarity of the aligned nodes based on their biological roles. A detailed introduction to these metrics will be provided in the remainder of this subsection.

\subsubsection{Correspondence-based metrics} 
To validate the effectiveness of network alignment methods, researchers typically prepare datasets in advance. For example, in a dataset composed of two networks, the dataset includes all the nodes, links, and corresponding node pairs, along with optional node attributes. During the validation process, the corresponding node pairs are divided into a training and testing set~\cite{gu2020data}. In some machine learning methods, a validation set is included alongside the training and testing sets~\cite{feng2019dplink}. Using the nodes, links, and other relevant information from the two networks, along with the training set as input, a network alignment method will output the alignment results for the unmatched nodes. Therefore, the effectiveness of the network alignment method can be evaluated by comparing the output results with the corresponding node pairs in the testing set. Obviously, for methods that do not require observed corresponding node pairs, all the corresponding node pairs in the dataset are used as the testing set.

Metrics based on the true corresponding node pairs can be further divided into accuracy and ranking metrics. The former primarily reflects the proportion of correct results compared to the true results in the predictions made by a network alignment method. In contrast, the latter reflects the ability of a method to rank correct results near the top of a list of candidates, ordered by the predicted probability of correctness. The accuracy metrics include Recall~\cite{ZhangYutao2015}, Precision~\cite{ZhangYutao2015},  F1~\cite{ZhangYutao2015}, and Accuracy~\cite{ZhangJiawei2015-DM,vijayan2017multiple}. The ranking metrics include Precision@N~\cite{ZhouFan2018}, Hit@N~\cite{MuXin-KDD2016}, mean average precision (MAP)~\cite{ManTong2016-IJCAI}, and area under the receiver operating characteristics curve (AUC)~\cite{ShuKai2017}. The definitions of these metrics are as follows.

Recall primarily reflects the proportion of the results that needed to be predicted and were correctly predicted. It is derived from the standard evaluation of binary classification problems. Denoting true positive (TP) as the set of correctly predicted corresponding node pairs, true negative (TN) as the set of correctly predicted non-corresponding node pairs, false negative (FN) as the set of corresponding node pairs incorrectly predicted as non-corresponding node pairs, and false positive (FP) as the set of non-corresponding node pairs incorrectly predicted as corresponding node pairs, it is defined as
\begin{equation}
Recall=\frac{|TP|}{|TP|+|FN|}.
\label{eq_recall}
\end{equation}
Since network alignment methods typically do not make predictions for non-corresponding node pairs, the denominator of Eq.~(\ref{eq_recall}) can be replaced with the number of unobserved corresponding node pairs in the testing set, i.e.:
\begin{equation}
Recall=\frac{|TP|}{|\Psi^u|}.
\end{equation}

Precision primarily reflects the proportion of correct results among all the predicted results. It is defined as
\begin{equation}
Precision=\frac{|TP|}{|TP|+|FP|}.
\end{equation}

The F1 score combines Recall and Precision to provide a more balanced evaluation. Relying solely on Precision or Recall can be insufficient. For instance, predicting only a small number of easily identifiable corresponding node pairs can yield a high Precision, while making many predictions might result in a higher Recall. It is defined as
\begin{equation}
F1=\frac{2 \cdot Recall \cdot Precision}{Recall + Precision}.
\end{equation}

Accuracy considers the correct identification of both corresponding node pairs and non-corresponding node pairs. It is defined as
\begin{equation}
Accuracy=\frac{|TP|+|TN|}{|TP|+|TN|+|FP|+|FN|}.
\end{equation}

For a given node, some methods only provide a single node from another network to form a corresponding node pair. In such cases, the aforementioned four metrics can be used for evaluation. However, for some methods, a top-$N$ list of potential corresponding node pairs is provided for each node. Precision@N, Hit@N, MAP, and AUC can be used to measure the effectiveness of these methods.

Precision@N is defined as
\begin{equation}
Precision@N=\frac{\sum_{i=1}^{|\Psi^u|}{\mathds{1}_i\{success@N\}}}{{|\Psi^u|}},
\end{equation}
where $N$ is the freely adjustable size of the list, with values ranging from 1 to the number of unmatched nodes in the other network. ${\mathds{1}_i\{success@N\}}$ indicates whether the correct corresponding node pair in the top-$N$ list for node $u^{l_1}_i$. If the correct pair is in the list, ${\mathds{1}_i\{success@N\}}=1$; otherwise, ${\mathds{1}_i\{success@N\}}=0$. 

Hit@N is defined as
\begin{equation}
Hit@N=\sum_{i=1}^{|\Psi^u|}\frac{N-(position(i)-1)}{N},
\end{equation}
where $position(i)$ indicates the position of the correct corresponding node pair in the top-$N$ list for node $u^{l_1}_i$.

MAP is defined as
\begin{equation}
MAP=\frac{\sum_{i=1}^{|\Psi^u|}{1/position(i)}}{|\Psi^u|}.
\end{equation}
 
AUC is defined as
\begin{equation}
AUC=(\sum_{i=1}^{|\Psi^u|}{\frac{n_i+1-position(i)}{n_i}})/|\Psi^u|,
\label{eq_auc}
\end{equation}
where $n_i$ is the number of negative candidate pairs for unmatched node $u^{l_1}_i$. 

\subsubsection{Structure-based metrics}
This type of metric evaluates the network alignment methods by calculating the edge overlap or structural similarity between the aligned networks, containing the edge correctness (EC)~\cite{kuchaiev2010topological}, interaction correctness (IC)~\cite{milenkovic2010optimal}, induced conserved structure (ICS)~\cite{patro2012global}, symmetric substructure score ($S^3$)~\cite{saraph2014magna}, and the largest common connected subgraph (LCCS)~\cite{saraph2014magna}.

In these structure-based metrics, EC computes the proportion of overlapping edges—edges that appear in both networks after alignment—relative to the total number of edges in the smaller of the two networks being aligned. This metric assesses how well the alignment method preserves the edge structure of the smaller network within the context of the larger network. It is one of the most straightforward methods to evaluate the quality of an alignment. It is defined as
\begin{equation}
EC = \frac{|\{(v^{l_{1}}_{i},v^{l_{1}}_{j})| (v^{l_{1}}_{i},v^{l_{1}}_{j})\in E^{l_{1}} \wedge (f(v^{l_{1}}_{i}),f(v^{l_{1}}_{j}))\in E^{l_{2}}\}|}{|E^{l_{1}}|},
\end{equation}
where $f(\cdot)$ represents the operation of getting the corresponding node for the node inside the parentheses using a given network alignment method since in Eq.~(\ref{eq_objectivefunction_na}), $f(v^{l_1}_i,v^{l_2}_j)=1$ can be rewritten as $f(v^{l_1}_i)=v^{l_2}_j$.

IC is a modified version of EC that considers whether the identified corresponding nodes are the true corresponding nodes for the given nodes. It can be defined as
\begin{equation}
\begin{aligned}
    &IC = |\{(v^{l_{1}}_{i},v^{l_{1}}_{j})|(v^{l_{1}}_{i},v^{l_{1}}_{j})\in E^{l_{1}} \\ & \wedge(f(v^{l_{1}}_{i}),f(v^{l_{1}}_{j}))\in E^{l_{2}} \wedge f(v^{l_{1}}_{i})=f^{*}(v^{l_{1}}_{i}) \wedge f(v^{l_{1}}_{j})=f^{*}(v^{l_{1}}_{j})\}|/
    |E^{l_{1}}|,
\end{aligned}
\end{equation}
where $f^{*}(\cdot)$ represents the true corresponding node for the node inside the parentheses.

ICS is also a modified version of EC, which considers not only the number of overlapping edges but also the similarity between $G^{l_1}$ and the induced subgraph~\cite{kloks2000finding} in ${G^{l_2}}$ formed by the corresponding nodes of $V^{l_1}$ and all the edges between them. It is defined as
\begin{equation}
    ICS = \frac{|\{(v^{l_{1}}_{i},v^{l_{1}}_{j})|(v^{l_{1}}_{i},v^{l_{1}}_{j})\in E^{l_{1}} \wedge (f(v^{l_{1}}_{i}),f(v^{l_{1}}_{j}))\in E^{l_{2}}\}|}{|{E^{l_{2}}}'|},
\end{equation}
where ${E^{l_{2}}}'$ is the edge set of the induced subgraph~\cite{kloks2000finding} in ${G^{l_2}}$ introduced above.

$S^3$ further improves ICS by penalizing the non-overlapping edges between $G^{l_1}$ and the induced subgraph in $G^{l_2}$. It is defined as
\begin{equation}
    S^{3} = \frac{|\{(v^{l_{1}}_{i},v^{l_{1}}_{j})|(v^{l_{1}}_{i},v^{l_{1}}_{j})\in E^{l_{1}} \wedge (f(v^{l_{1}}_{i}),f(v^{l_{1}}_{j}))\in E^{l_{2}}\}|}{|E^{l_{1}}|+|{E^{l_{2}}}'|-|\{(v^{l_{1}}_{i},v^{l_{1}}_{j})|(v^{l_{1}}_{i},v^{l_{1}}_{j})\in E^{l_{1}} \wedge (f(v^{l_{1}}_{i}),f(v^{l_{1}}_{j}))\in E^{l_{2}}\}|}.
\end{equation}

Inspired by the similarity between $G^{l_1}$ and the induced subgraph in $G^{l_2}$ considered in the previous metrics, Saraph et al.~\cite{saraph2014magna} suggested that it would be more straightforward to quantify the extent of a connected subgraph shared by both $G^{l_1}$ and $G^{l_2}$. They introduced the LCCS, which measures the size of the largest connected subgraph preserved by the aligned nodes in both networks.

\subsubsection{Function-based metrics}
In bioinformatics, since the gold standard alignment has not yet been established, directly comparing alignment results is difficult~\cite{liao2009isorankn}. Several researchers assess the effectiveness of biological network alignment methods by evaluating the biological functional similarity of aligned nodes. 

A typical approach involves comparing the Gene Ontology (GO)~\cite{gene2008gene} annotations of aligned nodes to determine how well their biological functions, processes, and components are preserved across different networks~\cite{hayes2018sana}. This includes metrics such as functional coherence (FC)~\cite{singh2008global}, the entropy of the GO / Kyoto encyclopedia of genes and genomes (GO/KEGG)~\cite{liao2009isorankn}, gene ontology consistency (GOC) and its normalized version (NGOC)~\cite{aladaug2013spinal}, and Resnik's semantic similarity (RSS)~\cite{resnik1995using}. 

FC computes the ratio of the intersection size of the standardized GO terms associated with aligned protein nodes to the union set size of the GO terms associated with those nodes in two aligned PPI networks. It measures the function consistency of the aligned protein nodes by using the common GO terms assigned to proteins. The FC value of two aligned PPI networks is the average pairwise FC value of the protein node pairs. The FC value is higher, the proteins in the alignment perform more similar functions. The FC of each aligned protein node pair is defined as 
\begin{equation}
    FC(v^{l_{1}}_i,v^{l_{2}}_j) = \frac{|GO_{v^{l_{1}}_i}\cap GO_{v^{l_{2}}_j}|}{|GO_{v^{l_{1}}_i}\cup GO_{v^{l_{2}}_j}|},
\end{equation}
where $GO_{v^{l_1}_i}$ and $GO_{v^{l_2}_j}$ are standardized the GO terms of the two protein nodes in the input networks, which in this case the ancestors of GO term at a distance 5 from the root of the GO tree.

GOC uses the Jaccard index on sets of GO terms between the protein pair, which computes like FC by summing the fraction of all aligned node pairs. It is defined as
\begin{equation}
    GOC(G^{l_{1}},G^{l_{2}}) = \sum\limits_{(v^{l_{1}}_i,f(v^{l_{1}}_i))\in \Psi^u}\frac{|GO(v^{l_{1}}_i)\cap GO(f(v^{l_{1}}_i))|}{|GO(v^{l_{1}}_i)\cup GO(f(v^{l_{1}}_i))|},
\end{equation}
where $(v,f(v^{l_{1}}))$ is the aligned node pair.

NGOC is the normalized version of GOC, which is defined as
\begin{equation}
    NGOC(G^{l_{1}},G^{l_{2}}) = \frac{1}{n_{a}} \left(\sum\limits_{(v^{l_{1}}_i,f(v^{l_{1}}_i))\in \Psi^u}\frac{|GO(v^{l_{1}}_i)\cap GO(f(v^{l_{1}}_i))|}{|GO(v^{l_{1}}_i)\cup GO(f(v^{l_{1}}_i))|}\right),
\end{equation}
where $n_{a}$ is the number of the aligned node pairs.

The entropy of GO/KEGG~\cite{liao2009isorankn} computes the entropy of a given cluster. The cluster here is the functionally conserved interaction components, a set of network-aligned proteins based on the similarity scores. Also, it can use entropy to measure the within-cluster consistency and enrichment of the output clusters in the many-to-many alignment methods. It is a GO-derived measurement and the entropy of a cluster is lower, the GO/KEGG are more within-cluster consistent. The entropy of a cluster is defined as
\begin{equation}
    H(G^s) = H(p_{1},…,p_{n}) = -\sum\limits_{i=1}^{n}p_{i}\log p_{i},
\end{equation}
where $G^s$ is a given cluster and $p_{i}$ is the fraction of $G^s$ with GO/KEGG group ID.

Resnik’s semantic similarity is a semantic similarity metric based on the information content of GO. It considers the hierarchical structure of the ontology and can be used in the GO consistency measure. It is defined as
\begin{equation}
\begin{aligned}
    &S_{Res}(G^{l_{1}},G^{l_{2}}) = \frac{1}{|V^{l_{1}}|}\sum\limits_{v^{l_{1}}_i\in V^{l_{1}},f(v^{l_{1}}_i)\in V^{l_{2}}}Sim_{Res}(GO(v^{l_{1}}_i),GO(f(v^{l_{1}}_i))),\\
    &Sim_{Res}(t_{1},t_{2}) = INCO(t_{MICA}),\\
    &INCO(t) = -\log p(t),
\end{aligned}
\end{equation}
where $t_{MICA}$ is the most informative common ancestors (MICAs) between GO term sets, INCO is the information content, and $p$ is the probability of occurrence of the GO term.

\section{Network alignment in different fields}
\label{Sec: Network alignment in different fields}
In this section, we begin by detailing the methodology used to search and filter the literature about network alignment. We then identify the prominent fields that, according to our analysis of this literature, demonstrate the most incredible focus on network alignment. Lastly, we provide a comprehensive overview of each field, analyzing the underlying reasons for their interest in network alignment. 

We used keywords such as "network alignment," "graph alignment," "entity matching," and "user identity linkage" to search on Google Scholar and downloaded the resulting papers. Meanwhile, we also identified and downloaded papers that cited these documents, based on the titles and abstracts, that appeared relevant to network alignment. In total, we collected nearly 600 papers. After selecting those published in Science Citation Index Expanded (SCIE) journals or reputable international conferences and excluding low-quality or irrelevant papers, we retained over 400 papers on network alignment from the past decade. A brief analysis of these papers revealed that the most active research areas in network alignment are social network analysis, bioinformatics, computational linguistics, and privacy protection. In the following sections, we will provide an overview of these areas.

\subsection{Network alignment in social network analysis}
\label{Sec: Network alignment in social network analysis}
\begin{figure*} [!t]
    \centering
    \includegraphics[width=0.7\textwidth]{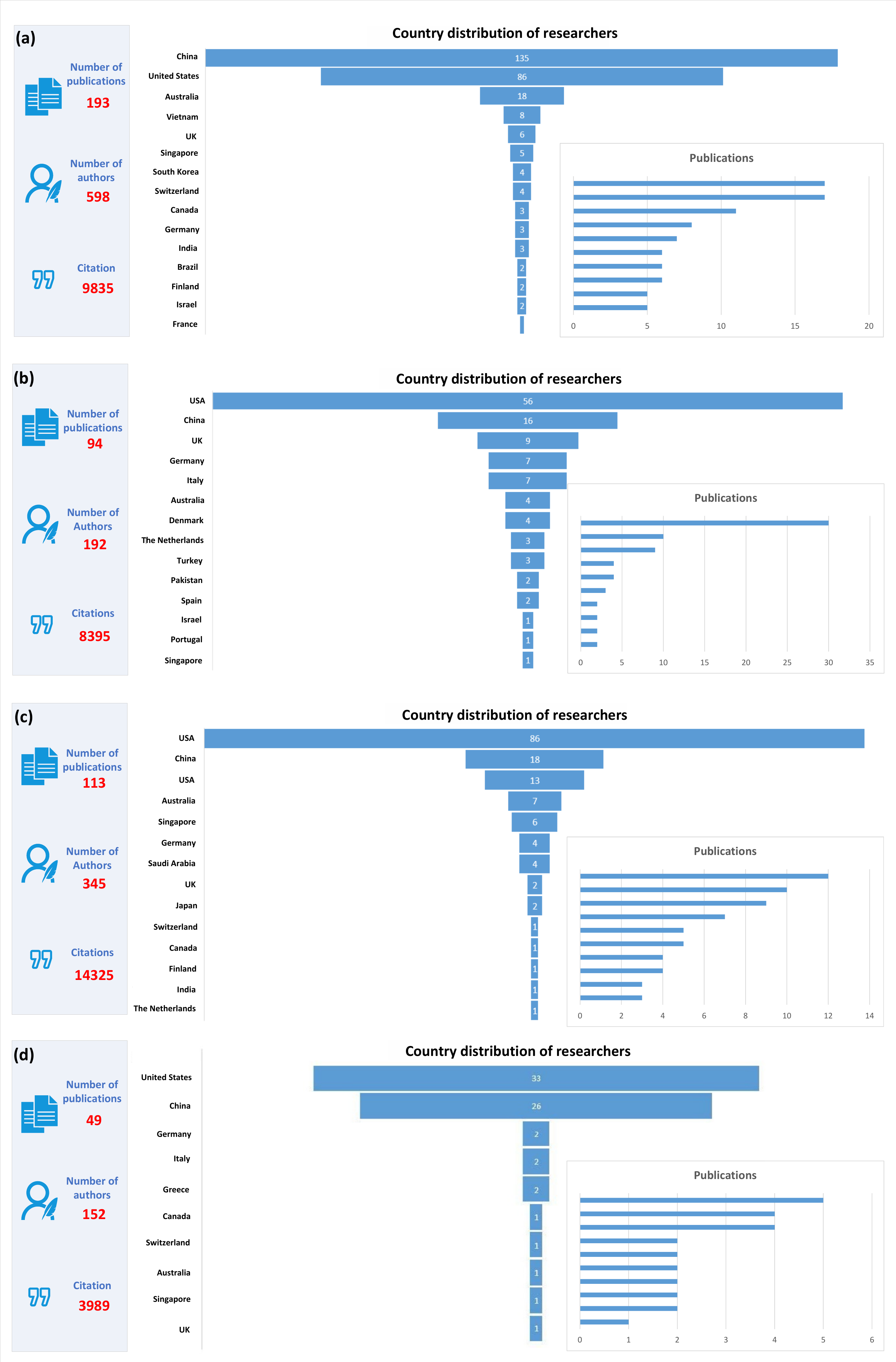}
    \caption{
    The statistics of academic papers published over the past decade related to network alignment in different fields: (a) social network analysis, (b) bioinformatics, (c) computational linguistics, and (d) privacy protection. In each subfigure, the left part displays the total number of publications, authors, and citations. The top-right part shows the number of papers published by researchers from various countries, while the bottom-right part highlights the number of papers published in different representative journals or conferences.}
    \label{pic_social_survay}
\end{figure*}
In a social network, each person can be represented as a node, and the friendship or interaction between two people can be represented as an edge. With the development of the Internet and mobile Internet, people increasingly engage with various online social network (OSN) platforms daily. For example, many people use X (formerly Twitter) or Facebook to discuss trending issues and share their daily thoughts and feelings, LinkedIn for job searching, and WhatsApp or Telegram to communicate with others. According to We Are Social's Digital Report published in 2024, as of July 2024, the number of global OSN users has reached approximately 5.17 billion, accounting for about $63.7\%$ of the world's population\footnote{\url{https://wearesocial.com/uk/blog/2024/01/digital-2024-5-billion-social-media-users/}}. People’s work, study, life, and entertainment have become inseparable from OSNs. The massive user engagement makes social network analysis indispensable for marketing, public safety, and political elections. By conducting an in-depth study of various tasks in the field of social network analysis, it is evident that tasks such as node identity recognition, the integration of multiple social networks, and cross-OSN platform account linkage are all network alignment problems.

\begin{itemize}
\item \textbf{Node identity recognition.} Given an anonymous social network, we aim to determine the real-world identity of each node. This is essentially a problem of aligning an anonymous network with a real-name network, which is particularly important in combating cybercrime. For example, cybercriminals may use multiple OSN platforms (Telegram, WhatsApp, or WeChat) to communicate covertly while engaging in illegal activities. To thoroughly investigate and identify the participants in these criminal activities, it is crucial to uncover who the anonymous accounts interacting with these cyber criminals are in the real world. We can construct a communication network based on interaction records from these social platforms and simultaneously construct a real-world relationship network, such as a phone network or a friendship network where users are willing to reveal their real identities. By aligning the communication network with the real-world relationship network, we can identify the real-world individuals behind the anonymous social accounts. This process not only aids in investigating cybercrimes but also helps uncover potential threats hidden within anonymous networks.

\item \textbf{Integration of multiple social networks.} We often study how the same hot topic or event spreads across different OSN platforms. Within each platform, users' message-forwarding activities form a dissemination network. When the same hot topic or event is being discussed across multiple OSN platforms, how can we map these different dissemination networks onto real-world individuals or organizations? This essentially involves the alignment of multiple information dissemination networks. By aligning the dissemination networks from different OSNs for the same topic or event, we can gain a more comprehensive understanding of the dynamics of information propagation and the participants and dissemination chains behind the event. This, in turn, provides valuable insights for public opinion analysis and research on the influence of significant events, thus enabling a more comprehensive study of social network dynamics~\cite{tejedor2018diffusion,cozzo2013contact}.

\item \textbf{Cross-platform account linkage.} Users often participate in multiple OSN platforms simultaneously, and the accounts they use on these platforms may have different usernames or profile information, hence unknown to the public. Identifying which accounts belong to the same user is a fundamental step for the downstream tasks such as cross-network product and content recommendations~\cite{KongXiangnan2013,ren2024efficiency}, user communities detections~\cite{luo2020local}, target users identifications~\cite{shen2020network}, and cross-network link predictions~\cite{zhan2019integrated,jiao2019collective}.
\end{itemize}

Figure~\ref{pic_social_survay} (a) shows the statistical information of academic papers published over the past decade related to network alignment in social network analysis. An impressive 193 papers have been published, with significant contributions from 598 authors. According to data from Google Scholar, these papers have been cited 9,835 times, highlighting the significant impact of network alignment research in this field. Most research originates from a few countries, such as China, the USA, Australia, Vietnam, and the UK. The numbers of articles published in each of the top ten journals and conferences that have published the most related articles are listed in the bottom-right part, with the top three being 17, 17, and 11 articles, respectively.

\subsection{Network alignment in bioinformatics}
\label{SubSec: Network alignment in bioinformatics}
With the rapid advances in molecular biology~\cite{han2008understanding} and high-throughput experimental technologies~\cite{uetz2000comprehensive, krogan2006global}, many biological datasets involving molecular interactions have emerged, such as gene co-expression network data~\cite{yang2023gcna,ficklin2011gene}, metabolic network data~\cite{cheng2009metnetaligner}, gene regulatory network data~\cite{guelsoy2012topac}, and protein-protein interaction (PPI) network data~\cite{guzzi2018survey}. Researching these network data helps to elucidate the complex relationships between genes and proteins, reveal the functional mechanisms of biological systems, and consequently promote the study of disease mechanisms, discovery of new drug targets, and more~\cite{rasti2019survey}. Issues in this field, such as identifying conserved functional components, ortholog identification~\cite{hulsen2006benchmarking,burke2023towards}, and protein function prediction, also cause network alignment problems.

\begin{itemize}
    \item \textbf{Conserved functional components identification.} Proteins serve as the main executors of biological functions for different species~\cite{jensen2003functionality}. PPI networks are critical for understanding cellular functions, signal transduction, and the formation of protein complexes~\cite{singh2007pairwise}. In cross-species, PPI network comparison, key questions include: which protein-protein interactions are conserved across species, and which functional components formed by these proteins are conserved~\cite{kalaev2008networkblast}. This essentially involves aligning PPI networks across different species. To achieve this, homologous protein nodes from different PPI networks are aligned. We can determine whether protein-protein interactions are conserved across species by comparing the edges between these aligned nodes. Similarly, by examining whether groups of interacting aligned nodes retain their interactions in the PPI networks of different species, we can assess whether the functional components formed by these proteins are conserved.

    \item \textbf{Ortholog identification.} Since orthologous proteins often participate in similar biological processes and interact with similar proteins, their identification can be framed as a PPI network alignment problem. Orthologs are genes or proteins that evolved from a common ancestral gene in different species and typically perform similar biological functions across species~\cite{gabaldon2013functional}. If two matched protein nodes exhibit highly similar local topological structures and interaction patterns with their neighbouring proteins, they are likely orthologs. The identification or prediction of orthologs is crucial for cross-species functional annotation and evolutionary biology research, as it allows researchers to infer the functions of uncharacterized proteins in one species based on known functions in another species and to understand how proteins and biological functions have evolved and been conserved across different species~\cite{altenhoff2009phylogenetic}.

    \item \textbf{Protein function prediction.} The functions of many proteins remain unannotated for the species that are understudied or unknown. This is also a network alignment problem. By aligning the PPI network of the unknown or understudied species, including its known proteins, with the PPI networks of well-studied species, the functions of unknown proteins can be predicted based on similar interaction patterns~\cite{clark2014comparison}. 
\end{itemize}

Additionally, comparative analysis of different PPI networks is also a network alignment problem. For the same species, aligning and comparing the PPI network under normal conditions with the PPI network under diseased conditions allows researchers to identify which protein-protein interactions have changed. This helps discover critical proteins related to the disease and potential drug targets.

Figure~\ref{pic_social_survay} (b) shows the statistical information of the academic papers published over the past decade related to network
alignment in bioinformatics. An impressive 94 papers have been published, with significant contributions from 192 authors. According to data from Google Scholar, these papers have been cited 8395 times. The USA, China, the UK, Germany, and Italy are among the top countries in terms of research publications. The numbers of articles published in each of the top ten journals and conferences that have published the most related articles are listed in the bottom-right part, with the top three being 30, 10, and 9 articles, respectively.

\subsection{Network alignment in computational linguistics}
\label{Sec: Network alignment in linguistics}
Computational linguistics is crucial for enabling computers to understand human language and developing technologies that facilitate natural language communication between humans and machines. Integrating linguistic theories with computational methods enables the processing, analysis, and generation of human language, powering many of today’s widely used tools and systems. One approach to making natural language computable is representing different languages as networks. For example, each language can be represented as a network by treating words as nodes and their co-occurrence in sentences as edges~\cite{cancho2001small, xuan2009node}. Additionally, knowledge can be stored and processed as a knowledge graph, where words, phrases, people, places, organizations, events, and activities in a language are represented as nodes. The relationships between these nodes—such as hierarchy or association—are represented as edges. Each node contains descriptive information, which serves as the node’s attributes. Knowledge graphs in different languages are stored in computers and support various services. After constructing these network representations, tasks in computational linguistics, such as cross-language ontology or word linkage, word sense inference and disambiguation, and cross-language text similarity evaluation, can all be framed as network alignment problems.

\begin{itemize}
    \item \textbf{Cross-language node linkage.} In different language networks or knowledge graphs, many nodes correspond to the same concept. For example, a node in an English network might represent the same concept as a node in a Chinese network. Linking these nodes, representing the same concept across languages, is a network alignment problem. This is similar to cross-platform account linkage in social network analysis or ortholog identification in PPI networks. With thousands of languages in the world today, by utilizing constructed language networks, KGs, and well-known concepts or manually labelled equivalent concepts, we can align different language networks or knowledge graphs. This enables us to derive equivalence relationships for many other concepts, completing the cross-language node linkage process. This is highly significant for cross-language translation, research on lesser-used languages, and studying language evolution and historical linguistics.
    \item \textbf{Word sense inference and disambiguation.} In tasks like text understanding and cross-language translation, word sense inference and disambiguation are crucial in natural language processing. These tasks involve correctly identifying and inferring the meaning of polysemous words in specific contexts. By constructing a context network from the sentences or paragraphs where the word appears and aligning this network with a pre-established network of known word senses, the correct meaning of the word in the current context can be determined. For instance, in the sentence ``I went to the bank to deposit money," the context network for ``bank" may include nodes like ``deposit" and ``money," which helps infer that in this context, ``bank" refers to a financial institution rather than a riverbank.
    \item \textbf{Cross-language text similarity evaluation.} By constructing context networks for paragraphs or documents in different languages and aligning the source language’s context network with the target language’s context network, it is possible to evaluate their semantic similarity. For example, consider the English sentence ``I went to the bank to deposit money" and its corresponding Spanish sentence ``Fui al banco a depositar dinero" For the English sentence, a context network can be built with nodes such as ``bank" ``deposit" and ``money". Similarly, a semantic network can be constructed for the Spanish sentence with nodes like ``banco", ``depositar" and ``dinero". By aligning the corresponding nodes between these context networks (e.g., ``bank" aligning with ``banco", "deposit" aligning with ``depositar", and ``money" aligning with ``dinero"), we can determine that the two sentences express similar meanings in different languages. Such evaluation has significant applications in cross-language plagiarism detection, cross-language information retrieval, multilingual question-answering systems, and cross-language text summarization.
\end{itemize}

Figure~\ref{pic_social_survay} (c) shows the statistical information of the academic papers published over the past decade related to network
alignment in computational linguistics. An impressive total of 113 papers have been published, with significant contributions from 345 authors. According to data from Google Scholar, these papers have been cited 14325 times. China, the USA, Australia, Singapore, and Germany are among the top countries in terms of research publications. The numbers of articles published in each of the top ten journals and conferences that have published the most related articles are listed in the bottom-right part, with the top three being 12, 10, and 9 articles, respectively.

\subsection{Network alignment in privacy protection}
\label{Sec: Network alignment in privacy protection}
Network data, such as healthcare, contact, and financial transaction data, are vital for applications in healthcare, government, and commerce. Therefore, these data are often shared with researchers, government agencies, commercial partners, and other organizations for data mining and analysis. During data sharing, each node’s attribute information is closely linked to the network structure, containing rich privacy details and inference paths~\cite{ji2014structural,ji2015secgraph}. Attackers can exploit this data to infer private information about users or systems~\cite{ji2016relative}. Data holders aim to anonymize the data before sharing it, ensuring usability while preventing excessive privacy leakage. For instance, a hospital may wish to share a large amount of patient, disease, and treatment data with a capable data analysis institution to research the causes of diseases and interaction patterns and provide more targeted and personalized treatment recommendations. However, they do not want the shared data linked to specific patients, so they aim to anonymize the network data. De-anonymization~\cite{ji2018quantifying,ji2017sag} of the anonymous network by attackers is a typical network alignment problem. Suppose the attacker possesses an auxiliary network containing node identities. By aligning the anonymous network with the auxiliary network, the attacker can re-identify nodes in the anonymous network, thereby completing the de-anonymization process~\cite{zhang2021anonymizing}. Data holders can perform this de-anonymization to evaluate whether the shared data has truly been anonymized, ensuring privacy protection.

Figure~\ref{pic_social_survay} (d) shows the statistical information of the academic papers published over the past decade related to network
alignment in the field of privacy protection. An impressive total of 49 papers have been published, with significant contributions from 152 authors. According to data from Google Scholar, these papers have been cited 3989 times. the USA, China, Germany, Italy, and Greece are among the top countries in terms of research publications. The numbers of articles published in each of the top ten journals and conferences that have published the most related articles are listed in the bottom-right part, with the top three being 5, 4, and 4 articles, respectively.

\section{Network alignment methods}
\label{Sec: Network alignment methods}
In the following, we focus on the various methods used for alignment, which can be classified into structure consistency-based methods and machine learning-based methods, depending on whether the networks and observed corresponding node pairs are used directly for computation. Structure consistency methods directly compare similarities between different networks to determine whether nodes can be matched as corresponding node pairs. In contrast, machine learning-based methods extract features from the nodes, train classification models on these features, and then use the trained models to make the determinations for the unmatched nodes. In the following sections, we will introduce each type of method in detail.

\subsection{Structure consistency-based methods}
Structure consistency-based methods assume that the same node maintains a consistent connectivity structure across different networks~\cite{ZhangSi2016-KDD} and can be divided into local and global consistency-based methods, depending on whether they rely on local or global topology information of the networks when determining whether two unmatched nodes belong to a corresponding node pair.

\subsubsection{Local structure consistency-based methods}
\begin{figure*} [th!]
    \centering
    \includegraphics[width=0.7\textwidth]{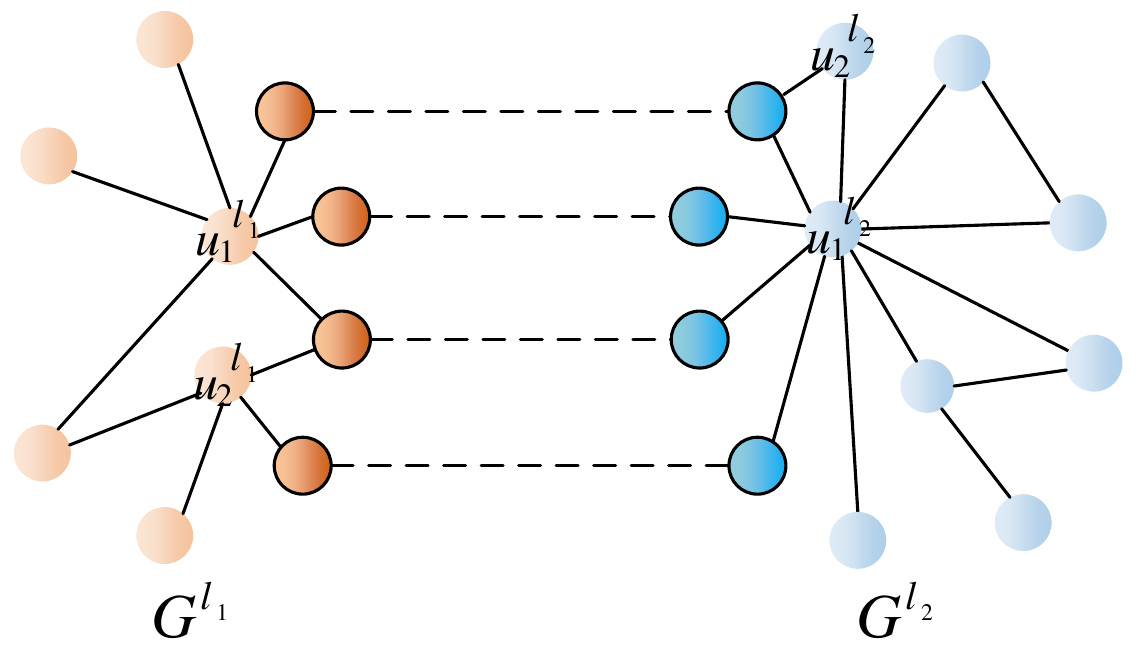}
    \caption{
    Example of the local structure consistency-based methods. There are two networks $G^{l_1}$ and $G^{l_2}$. The nodes connected by dashed lines represent observed corresponding nodes, while the others are unmatched. $u^{l_1}_1$ and $u^{l_2}_1$ share three corresponding node pairs, $u^{l_1}_1$ and $u^{l_2}_2$ share one pair, and $u^{l_1}_2$ and $u^{l_2}_1$ share two pairs. For local structure consistency-based methods, the structural similarity between $u^{l_1}_1$ and $u^{l_2}_1$ is higher due to the greater number of shared corresponding node pairs, making them more likely to be considered as corresponding nodes. 
    }
    \label{pic_example_locconsis}
\end{figure*}

The core idea behind local structure consistency-based methods is that the neighbourhood structures of the same nodes on different networks are similar. As illustrated in Figure~\ref{pic_example_locconsis}, the corresponding node for a node can be found by analyzing its neighbours. This is particularly evident in social networks, where a user's social circle is highly unique, making it nearly impossible for two people to have identical social circles~\cite{ZhouXiaoping2016}. 

The main steps consist of two: calculating the local structural similarity scores between unmatched nodes and deriving the alignment results based on these scores. 
The most straightforward approach to getting a local structural similarity score is counting the number of commonly identified neighbours (CIN)~\cite{KongXiangnan2013,korula2014efficient}, similar to the common neighbours (CN) method ~\cite{LvLinyuan2011} in link prediction. Denoting $\Gamma_1(u^{l_1}_i)$ as the set of first-hop neighbours $u^{l_1}_i$ has on network $G^{l_1}$, $\Gamma_1(u^{l_2}_j)$ as $u^{l_2}_j$ on $G^{l_2}$, commonly identified neighbours are the set of entities that coexist in both $G^{l_1}$ and $G^{l_2}$. The nodes representing these entities are neighbours of $u^{l_1}_i$ in $G^{l_1}$ and neighbours for $u^{l_2}_j$ in $G^{l_2}$. It can be formalized as
\begin{equation}
\begin{array}{l}
CIN(u^{l_1}_i,u^{l_2}_j)=\{(v^{l_1}_a,v^{l_2}_b)|v^{l_1}_a \in G^{l_1}, v^{l_2}_b\in G^{l_2},(v^{l_1}_a,v^{l_2}_b)\in \Psi^o, \\
\\v^{l_1}_a\in \Gamma_1(u^{l_1}_i), v^{l_2}_b\in \Gamma_1(u^{l_2}_j)\}.
\end{array}
\end{equation}
Usually, the identified neighbours for $u^{l_1}_i$ and $u^{l_2}_j$ can be denoted as $\Gamma_i(u^{l_1}_i)$ and $\Gamma_i(u^{l_2}_j)$ respectively. For nodes $u^{l_1}_i$ and $u^{l_2}_j$, their similarity score is
\begin{equation}
r^{CIN}_{ij}=|CIN(u^{l_1}_i,u^{l_2}_j)|=|\Gamma_1(u^{l_1}_i) \cap \Gamma_1(u^{l_2}_j)|,
\label{eq_localcons_cn}
\end{equation}
where $\cap$ denotes the intersection operation between two sets. 

Calculating similarity scores based on the number of commonly identified neighbours is often insufficient for achieving optimal results. When several nodes in $G^{l_2}$ have the same largest number of common identified neighbours with $u^{l_1}_i$, i.e. $|\Gamma_1(u^{l_1}_i) \cap \Gamma_1(u^{l_2}_j)|=|\Gamma_1(u^{l_1}_i) \cap \Gamma_1(u^{l_2}_k)|$, it is hard to determine which one of $u^{l_2}_j$ and $u^{l_2}_k$ is the corresponding node of $u^{l_1}_i$. Zhou et al.~\cite{ZhouXiaoping2016} extended the CIN counting by adding a term ranging from 0 to 1 to $r^{CIN}$, ensuring that the node pair with a higher proportion of CINs among all identified neighbours is matched as the corresponding node pair. This method, known as the Friend Relationship-Based User Identification (FRUI) algorithm, is primarily used in social network alignment. 
Its similarity score is
\begin{equation}
r^{FRUI}_{ij}=|\Gamma_1(u^{l_1}_i) \cap \Gamma_1(u^{l_2}_j)| + \frac{|\Gamma_1(u^{l_1}_i) \cap \Gamma_1(u^{l_2}_j)|}{\min(|\Gamma_1(u^{l_1}_i)|,|\Gamma_1(u^{l_2}_j)|)},
\end{equation}
where $\min(\cdot)$ represents the minimum value within the set inside the parentheses.

Li et al.~\cite{yongjun2018comment} found that if the seed-identified nodes are unsuitable, FRUI will stop early. They modified FRUI by 
\begin{equation}
r^{FRUI*}_{ij}=|\Gamma(u^{l_1}_i) \cap \Gamma(u^{l_2}_j)| + \frac{\min(CC(u^{l_1}_i),CC(u^{l_2}_j))}{\max(CC(u^{l_1}_i),CC(u^{l_2}_j))},
\end{equation}
where $CC(u^{l_1}_i)$ is closeness centrality~\cite{lu2016vital} of node $u^{l_1}_i$.

Considering that many real-world networks exhibit a scale-free degree distribution, with many nodes having a small degree and a few nodes having large degrees, the identified neighbours with different degrees should contribute differently to the alignment. To distinguish the contribution, Tang et al.~\cite{tang2020interlayer} proposed an iterative degree penalty (IDP) method which assigns higher weights to smaller-degree commonly identified neighbours, and the similarity score is
\begin{equation}
r^{IDP}_{ij}=\sum_{\substack{\forall(v^{l_1}_a,v^{l_2}_b)\in CIN(u^{l_1}_i,u^{l_2}_j)}} \log^{-1}(d({v^{l_1}_a})+1)+\log^{-1}(d({v^{l_2}_b})+1).
\label{eq_IDP}
\end{equation}

Since many real-world networks are sparse, where some unmatched nodes may have few or no commonly identified neighbours, relying solely on CINs for network alignment will result in low and indistinguishable similarity scores for many unmatched node pairs. Ding et al.~\cite{ding2021soidp} proposed a second-order CINs-based IDP (SOIDP) algorithm in which the similarity score is
\begin{equation}
r^{SOIDP}_{ij}=r^{IDP}_{ij}+\epsilon \cdot \sum_{\substack{\forall(v^{l_1}_c,v^{l_2}_d)\in \Psi^o, \\ v^{l_1}_c \in\Gamma_2(u^{l_1}_i),\\v^{l_2}_d\in\Gamma_2(u^{l_2}_j)}} \log^{-1}(d(v^{l_1}_c)+1)+\log^{-1}(d(v^{l_2}_d)+1),
\label{eq_SOIDP}
\end{equation}
where $\epsilon$ is an adjustable parameter and $\Gamma_2(u^{l_1}_i)$ and $\Gamma_2(u^{l_2}_j)$ are the second-order neighbour sets for nodes $u^{l_1}_i$ and $u^{l_2}_j$ respectively.

The similarity scores between nodes across networks calculated by the above methods may exhibit significant variations. In Ref.~\cite{bartunov2012joint}, the similarity scores are normalized to a range between 0 and 1 in the following approach.
\begin{equation}
r^{JLA}_{ij}=\frac{2 \times w(\Gamma_1(u^{l_1}_i) \cap \Gamma_1(u^{l_2}_j))}{w(\Gamma_1(u^{l_1}_i)) + w(\Gamma_1(u^{l_2}_j))},
\end{equation}
where $w(\Gamma_1(u^{l_1}_i))=\Sigma_{v\in \Gamma_1(u^{l_1}_i)}1/d(v)$. 

Xuan et al.~\cite{xuan2009node} and Kong et al.~\cite{KongXiangnan2013} extend the Jaccard's coefficient (EJC) for the normalization, which calculates the similarity score by
\begin{equation}
r^{EJC}_{ij}=\frac{|\Gamma_1(u^{l_1}_i) \cap \Gamma_1(u^{l_2}_j)|}{|\Gamma_1(u^{l_1}_i)|+|\Gamma_1(u^{l_2}_j)|-|\Gamma_1(u^{l_1}_i) \cap \Gamma_1(u^{l_2}_j)|}.
\label{eq_localcons_ejc}
\end{equation}

Kong et al.~\cite{KongXiangnan2013} also extend the Adamic/Adar (EAA) index~\cite{adamic2003friends} for the network alignment, which is
\begin{equation}
r^{EAA}_{ij}=\sum_{\substack{\forall(v^{l_1}_a,v^{l_2}_b)\in CIN(u^{l_1}_i,u^{l_2}_j)}} \log^{-1}(\frac{|\Gamma_1(v^{l_1}_a)|+|\Gamma_1(v^{l_2}_b)|}{2}).
\label{eq_EAA}
\end{equation}

Tang et al.~\cite{tang2022interlayer} calculate the similarity score based on multiple structural attributes, which contain the number of possible matched closed triad, degree of unmatched nodes, commonly identified neighbours, and the similarity of the links of all the neighbours.

The second step is determining the alignment results. The approaches contain directly comparing similarity scores to obtain all results at once or iteratively identifying the possible results in each round. 

Some researchers directly compare the similarity scores between a node in one network and all nodes in the other network, aligning the node with the one that has the highest score as the result. For example, Tang et al.~\cite{tang2022interlayer} calculated the similarity scores between all pairs of unmatched nodes in the two networks, forming a similarity score matrix $\bm{\mathrm{R}}$. The element in the $i$th row and $j$th column of the matrix $\bm{\mathrm{R}}$ is the similarity score between node $u^{l_1}_i$ and node $u^{l_2}_j$. They then sorted each row or column in descending order based on the scores to obtain the results.

Researchers identify the most possible results for the set of matched node pairs and then repeat the calculation for the remaining unmatched nodes. This iterative process continues until all unmatched nodes are aligned or all similarity scores become zero. Korula and Lattanzi~\cite{korula2014efficient} align nodes $u^{l_1}_i$ and $u^{l_2}_j$ if $u^{l_2}_j$ is the node in $\Gamma_2(u^{l_2}_j)$ with maximum similarity score to $u^{l_1}_i$ and vice versa. To avoid the misaligned of low-degree nodes, they take an iterative strategy that only allows nodes of degree larger or equal to $d/2^i$ to be aligned, where $d$ is a parameter related to the largest degree and $i$ is the count of the iteration. Zhou et al.~\cite{ZhouXiaoping2016} select the maximum similarity score element from the entire similarity matrix in each iteration for aligning. The results are then added to the set of observed corresponding node pairs, and the process of calculating the similarity scores for the remaining unobserved nodes is repeated for the next iteration. To reduce the computational time complexity, they propose two propositions. Tang et al.~\cite{tang2020interlayer} argue that relying only on selecting the maximum value from the similarity matrix in each iteration may lead to too many iterations. They leverage an additional control parameter between 0 and 1 for efficiency improvement. In each iteration, pairs with similarity scores greater than the control parameter multiplied by that iteration's maximum similarity score are matched pairs. Since alignment results can sometimes conflict, for example, when both $u^{l_1}_i$ and $u^{l_1}_k$ have the highest similarity score with $u^{l_2}_j$, the question arises as to which node's corresponding node is $u^{l_2}_j$. Xuan et al.~\cite{xuan2009node} regard $\bm{\mathrm{R}}$ as a bipartite graph and transfer the aligning problem to a maximum matching problem for the bipartite graph. 

\subsubsection{Global structure consistency-based methods}
\label{Sec: Methods based on global structure consistency}

The core idea behind global structure consistency-based methods is that the overall topological structure of the networks can be used to determine whether two unmatched nodes belong to a corresponding node pair. These methods not only consider the immediate neighbours of the unmatched nodes but also consider the connectivity patterns over the entire networks~\cite{kollias2011network}, hence helping the alignment that local structure consistency-based methods might miss due to their limited scope. 

The global structure consistency-based method has two steps. First, calculate the global structural similarity scores between unmatched nodes across networks. Second, the alignment results are derived based on these scores. While the second step is almost identical to local structure consistency-based methods, the first differs significantly. This is understandable because, for global structure consistency-based methods, the central focus is on constructing the similarity matrix $\bm{\mathrm{R}}$, where each element $r_{ij}$ denotes the similarity score between node $u^{l_1}_i$ in $G^{l_1}$ and node $u^{l_2}_j$ in $G^{l_2}$. The challenge lies in how to accurately, effectively, and efficiently capture the global topological similarities between all unmatched nodes to obtain the optimal matrix $\bm{\mathrm{R}}$.

\begin{figure*} [th!]
    \centering
    \includegraphics[width=0.8\textwidth]{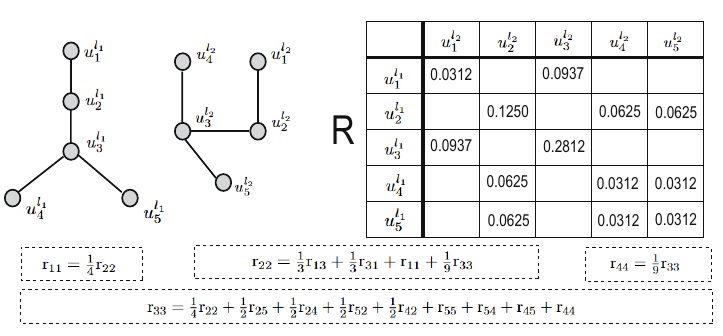}
    \caption{Illustration of calculating similarity score in IsoRank. There are two small networks. For each possible pair of unmatched nodes across two networks, its similarity score is constrained to depend on the scores from the neighbourhoods recursively.
    \\Source: Reproduced from Ref.~\cite{singh2007pairwise}
    }
    \label{pic_isorank}
\end{figure*}

Singh et al.~\cite{singh2007pairwise} proposed IsoRank to capture the global topological similarities for the unmatched nodes, which following the intuition that two unmatched nodes $u^{l_1}_i$ and $u^{l_2}_j$ across different networks are topologically similar if their respective neighbours are also topologically similar with each other, as shown in Figure~\ref{pic_isorank}. The similarity score $r_{ij}^{IsoRank}$ between $u^{l_1}_i$ and $u^{l_2}_j$ is determined by the similarity scores of their neighbours, which, in turn, depend on the similarity scores of their neighbours' neighbours, creating a recursive dependency throughout the networks. $r_{ij}^{IsoRank}$ can be calculated by
\begin{equation}
    r_{ij}^{IsoRank} = \sum\limits_{u^{l_1}_a\in \Gamma_1(u^{l_1}_i)} \sum\limits_{u^{l_2}_b\in \Gamma_1(u^{l_2}_j)} \frac{1}{|\Gamma_1(u^{l_1}_a)||\Gamma_1(u^{l_2}_b)|}r_{ab}^{IsoRank}.
\end{equation}
Since the recursive computations for all unmatched node pairs are highly complex, the solution to get the similarity score matrix $\bm{\mathrm{R}}$ is described as an eigenvalue problem. In this solution, $\bm{\mathrm{R}}$ can be calculated by 
\begin{equation}
\bm{\mathrm{R}}=\bm{\mathrm{MR}}
\label{eq_IsoRank_calcR}
\end{equation}
and updated by
\begin{equation}
\bm{\mathrm{R}}^{(k+1)}=\frac{\bm{\mathrm{MR}}^{(k)}}{|\bm{\mathrm{MR}}^{(k)}|},
\end{equation}
where $\bm{\mathrm{M}}$ is a $|V^{l_{1}}||V^{l_{2}}|\times |V^{l_{1}}||V^{l_{2}}|$ matrix, $\bm{\mathrm{R}}^{(k)}$ is the value of $\bm{\mathrm{R}}$ in the $k$-th iteration. $\bm{\mathrm{M}}[i,j][a,b]$ represents the element at the row indexed by $(i,j)$ and the column indexed by $(u,v)$, where both the row and column are doubly indexed. Its value is $1/|\Gamma_1(u^{l_1}_a)||\Gamma_1(u^{l_2}_b)|$ if link $(u_i^{l^1},u_a^{l^1})$ exists in $G^{l_1}$ and $(u_j^{l^2},u_b^{l^2})$ exists in $G^{l_2}$, and 0 otherwise. When someone wants to incorporate the other protein node information into this solution, Eq.~(\ref{eq_IsoRank_calcR}) can be modified to
\begin{equation}
\bm{\mathrm{R}}=\alpha \bm{\mathrm{MR}}+(1-\alpha)\bm{\mathrm{H}},
\label{eq_IsoRank_Mat}
\end{equation}
where $\bm{\mathrm{H}}$ is the matrix of the similarity scores of other information, such as the sequence similarity between proteins, $\alpha$ is a parameter that controls the weights between structural similarity and other information's similarity. 

To improve the efficiency of calculating the similarity scores for all unmatched nodes, Kollias et al.~\cite{kollias2011network} proposed the network similarity decomposition (NSD) method, which incorporates un-coupling and decomposing strategies for similarity computations. In the similarity calculation process, uncoupling independently preprocesses each network, allowing for more efficient handling. Decomposing, on the other hand, utilizes Singular Value Decomposition (SVD) or other matrix decomposition methods to decompose the similarity matrix \bm{\mathrm{H}} in IsoRank into a series of components. These components are then utilized incrementally in the similarity calculations for all unmatched node pairs, improving the overall efficiency. Processed by a PageRank~\cite{page1998pagerank} inspired algorithm, Eq.~(\ref{eq_IsoRank_Mat}) is iterated $k$ rounds by
\begin{equation}
\bm{\mathrm{R}}^{(k)}=(1-\alpha)\sum \limits_{h=0}^{k-1}\alpha^{h}{\mathcal{A}^{l_2}}^{h}\bm{\mathrm{H}}({\mathcal{A}^{l_1}}^T)^{h} + \alpha^{k}{\mathcal{A}^{l_2}}^{k}\bm{\mathrm{H}}({\mathcal{A}^{l_1}}^T)^{k},
\label{eq_NSD_1}
\end{equation}
where $\bm{\mathrm{R}}^{(0)}=\bm{\mathrm{H}}$. Performing SVD on $\bm{\mathrm{H}}$, it can be expressed as
\begin{equation}
\bm{\mathrm{H}}=\sum \limits_{i=0}^{s} \delta_i \bm{\mathrm{w}}_i \bm{\mathrm{z}}_i^T,
\label{eq_SVD_H}
\end{equation}
where $s \leq \min{(|V^{l_1}|,|V^{l_2}|)}$ is the rank of $\bm{\mathrm{H}}$, and $\delta_i>0$, $\bm{\mathrm{w}}_i$, $\bm{\mathrm{z}}_i$ for $i=1,2,\cdots,s$ are, respectively, the singular values, left singular vectors and right singular vectors of $\bm{\mathrm{H}}$. Inserting Eq.~(\ref{eq_SVD_H}) into Eq.~(\ref{eq_NSD_1}), and denoting $\bm{\mathrm{w}}_i^{(h)}={\mathcal{A}^{l_2}}^{h}\bm{\mathrm{w}}_i$ and $\bm{\mathrm{z}}_i^{(h)}={\mathcal{A}^{l_1}}^{h}\bm{\mathrm{z}}_i$, Eq.~(\ref{eq_NSD_1}) can be replaced by
\begin{equation}
\begin{array}{l}
\bm{\mathrm{R}}_i^{(k)}=\delta_i[(1-\alpha)\sum \limits_{h=0}^{k-1}\alpha^{h}\bm{\mathrm{w}}_i^{(h)}{\bm{\mathrm{z}}_i^{(h)}}^T+ \alpha^{k} \bm{\mathrm{w}}_i^{(k)}{\bm{\mathrm{z}}_i^{(k)}}^T],\\
\bm{\mathrm{R}}^{(k)}=\sum \limits_{i=1}^s\bm{\mathrm{R}}_i^{(k)}.
\end{array}
\end{equation}

Patro et al.~\cite{patro2012global} proposed GHOST, which uses multiscale spectral signatures to measure the topological similarity between the induced subnetworks formed by two unmatched nodes and their neighbours within $k$-hops. The spectral signature of a subnetwork is the spectral density of the Laplacian matrix for the subnetwork. It measures how eigenvalues are distributed over their potential range, hence obtaining the structural features of a node. By comparing the distance between the spectral signatures of two unmatched nodes, their local or global topological similarity (when $k$ exceeds a certain value) can be assessed. For more accurate alignment, the similarity of other information is also considered, and the spectral signatures of the subnetworks constructed by a node and its neighbours within $1$-hop, $2$-hops, ..., and $k$-hops, respectively, are incorporated. 

Neyshabur et al.~\cite{neyshabur2013netal} proposed NETAL, which improves the accuracy by adding an interaction score for each unmatched node pair in the similarity score matrix $\bm{\mathrm{R}}$. The interaction score of nodes $u^{l_1}_i$ and $u^{l_2}_j$ estimates how many links will be conserved if $u^{l_2}_j$ is the corresponding node for $u^{l_1}_i$~\cite{elmsallati2015global} and can be calculated by 
\begin{equation}
    r^{Int}_{ij} = \frac{\min (\sum\limits_{u^{l_{1}}_{a}\in \Gamma_1(u^{l_{1}}_{i})} \frac{1}{|\Gamma_1(u^{l_{1}}_{a})|},{\sum\limits_{u^{l_{2}}_{b}\in \Gamma_1(u^{l_{2}}_{j})} \frac{1}{|\Gamma_1(u^{l_{2}}_{b})|}} )}{\max\{|\Gamma_1(k)||k\in{V^{l_{1}}\cup V^{l_{2}}}\}}.
\end{equation}

Hashemifar et al.~\cite{hashemifar2014hubalign} introduced Hubalign for PPI network alignment, which is based on the insight that topologically and functionally significant nodes, such as hubs (nodes with numerous connections)~\cite{han2004evidence} and bottlenecks (nodes with high betweenness centrality)~\cite{yu2007importance}, are more likely to be conserved in different PPI networks, making them more likely to be accurately aligned. They designed a minimum-degree heuristic approach to estimate the relative importance scores of nodes, capturing their topological and functional significance. If two unmatched nodes across different networks have similar importance scores, it suggests that they play comparable important roles in their respective networks, making them more likely to be a corresponding node pair. The alignment process begins with the most critical nodes and proceeds incrementally to align less essential nodes. Such a method is suitable for biological network alignment and leads to a much faster and more accurate alignment process.

\begin{figure*} [th!]
    \centering
    \includegraphics[width=0.8\textwidth]{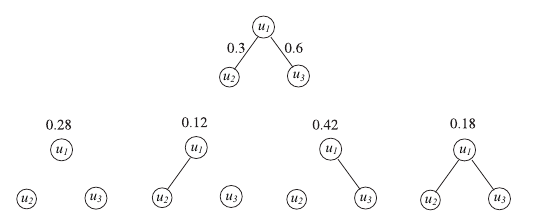}
    \caption{Example of the probabilistic and deterministic networks. The top subfigure represents a probabilistic network, while the four subfigures below are deterministic networks derived from it based on the probabilities of the edge's existence. The number above each deterministic network indicates the probability of obtaining that deterministic network from the probabilistic network.
    \\Source: Reproduced from Ref.~\cite{todor2012probabilistic}
    }
    \label{pic:proalign}
\end{figure*}

Considering that most studies model complex systems as deterministic networks, where edges between nodes either exist or do not. However, some PPI networks present a different scenario: an edge may or may not exist with some probability, influenced by factors such as the size, abundance, or proximity of the interacting molecules~\cite{bader2004gaining}. This phenomenon also exists in social networks, where two real-world friends may never interact on an OSN. To better address this situation, Todor et al.~\cite{todor2012probabilistic} proposed probAlign, which assumes that a system containing at least one probabilistic interaction is a probabilistic network $G = (V, E, p(\cdot))$, where $p(\cdot)$ is a function indicating the probability of existence for edges, as illustrated in Figure~\ref{pic:proalign}. Based on this assumption, probAlign modifies the matrix $\bm{\mathrm{M}}$ in Eq.~(\ref{eq_IsoRank_calcR}) as follows
\begin{equation}
    \\\bm{\mathrm{M}}[i,j][a,b] = \left\{
    \begin{aligned}
        &\frac{1}{d(u^{l_{1}}_{a})d^{p} (u^{l_{2}}_{b})},\quad &&if~(u^{l_{1}}_{i},u^{l_{1}}_{a})\in E^{l_{1}}~and~(u^{l_{2}}_{j},u^{l_{2}}_{b})\in E^{l_{2}}, \\
        &\frac{1}{|u^{l_{1}}||u^{l_{2}}|}, \quad &&if~ d(u^{l_{1}}_{a})d^p(u^{l_{2}}_{b})=0, \\
        &0, \quad &&otherwise.
    \end{aligned}
    \right.
\end{equation}
where $d^p(\cdot)$ represents the degree of a node in deterministic networks. Such an approach makes the alignment more biologically relevant.

Saraph et al.\cite{saraph2014magna} proposed MAGNA, which maximizes the amount of conserved edges by finding a global optimum over the search space consisting of all possible matches. Conserved edges refer to edges between matched nodes in all networks, representing structural similarity between the networks. Considering that the exhaustive search is computationally intractable due to the large size of the search space, they used genetic algorithms to obtain an approximate solution. Vijayan et al.~\cite{vijayan2015magna++} further proposed MAGNA++, which speeds up the alignment process through parallelization and more efficient edge conservation calculation. These improvements allow MAGNA++ to handle larger networks with increased computational efficiency while maintaining alignment accuracy.

\begin{figure*} [th!]
    \centering
    \includegraphics[width=0.95\textwidth]{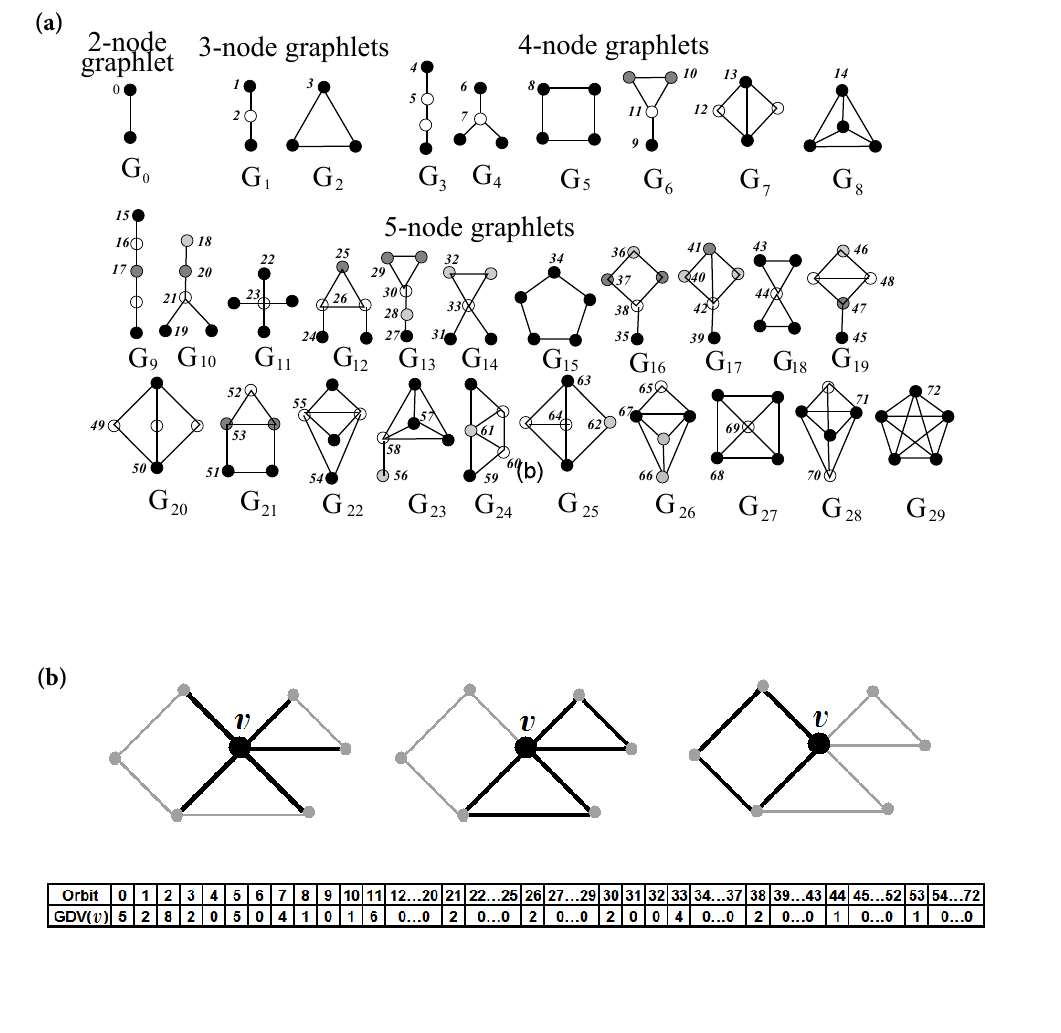}
    \caption{Examples of graphlets and graphlet degree vectors.
    (a) All possible graphlets consisting of up to five nodes. These graphlets contain 73 unique node types, referred to as orbits. (b) An illustration of the graphlet degree vector of node $v$: counts the number of graphlets the node touches.
    \\Source: Reproduced from Ref.~\cite{kuchaiev2010topological}
    }
    \label{pic:graal}
\end{figure*}

Kuchaiev et al.~\cite{kuchaiev2010topological} noted that no method for purely topological alignment had been devised. To address this gap, they proposed GRAAL, which calculates the structural similarity score based on the concepts of graphlet and orbit. The graphlet is a small, connected, induced subnetwork of a larger network. Figure~\ref{pic:graal}(a) shows the 30 graphlets with 2 to 5 nodes. The different types of nodes in these graphlets are referred to as orbits. By numbering the different orbits from 0 to 72, each node in a network can be represented as a 73-dimensional graphlet degree vector, as shown in Figure ~\ref{pic:graal}(b). The structural similarity score between two unmatched nodes is calculated by those vectors. The distance $dist_{k}(u^{l_{1}}_i,u^{l_{2}}_j)$ between the $i^{th}$ orbits is defined as
\begin{equation}
    dist_{k}(u^{l_{1}}_i,u^{l_{2}}_j) = w_{k}\times \frac{log(n^{u^{l_{1}}_i}_{k}+1)-log(n^{u^{l_{2}}_j}_{k}+1)}{\log(\max\{n^{u^{l_{1}}_i}_{k},n^{u^{l_{2}}_j}_{k}\}+2)}
\end{equation}
where $n^{u^{l_{1}}_i}_{k}$ is the number of times node $u^{l_{1}}_i$ touched by an orbit $k$ and $w_{k}$ is a weight of orbit $k$ that accounts for dependencies between orbits. The total distance $dist(u^{l_{1}}_i,u^{l_{2}}_j)$ is
\begin{equation}
    dist(u^{l_{1}}_i,u^{l_{2}}_j) = \frac{\sum^{72}_{k=0}dist_{k}(u^{l_{1}}_i,u^{l_{2}}_j)}{\sum^{72}_{k=0}w_{k}}.
\end{equation}
The final similarity score $r_{ij}^{GRAAL}$ is calculated by
\begin{equation}
r_{ij}^{GRAAL}=2-((1-\alpha)\frac{d(u^{l_{1}}_i)+d(u^{l_{2}}_j)}{\max(\bm{\mathrm{D}}^{l_1})+\max(\bm{\mathrm{D}}^{l_2})}+\alpha (1-dist(u^{l_{1}}_i,u^{l_{2}}_j))).
\end{equation}

Some studies have explored ways to improve GRAAL. For instance, Kuchaiev et al.~\cite{kuchaiev2011integrative} incorporated node similarity measures, such as sequence and functional similarity, to produce more stable alignments. Milenković et al.~\cite{milenkovic2010optimal} proposed H-GRAAL to enhance the alignment stage of GRAAL. Malod-Dognin et al.~\cite{malod2015graal} introduced L-GRAAL, which uses evolutionary relationships to calculate sequence similarity between unmatched nodes and adds this similarity to the score obtained by GRAAL. Memišević et al.~\cite{memivsevic2012c} developed C-GRAAL, which first aligns nodes with the highest combined neighbourhood density and then extends the alignment to include the common neighbours of already aligned nodes. 

\begin{figure*} [th!]
    \centering
    \includegraphics[width=0.8\textwidth]{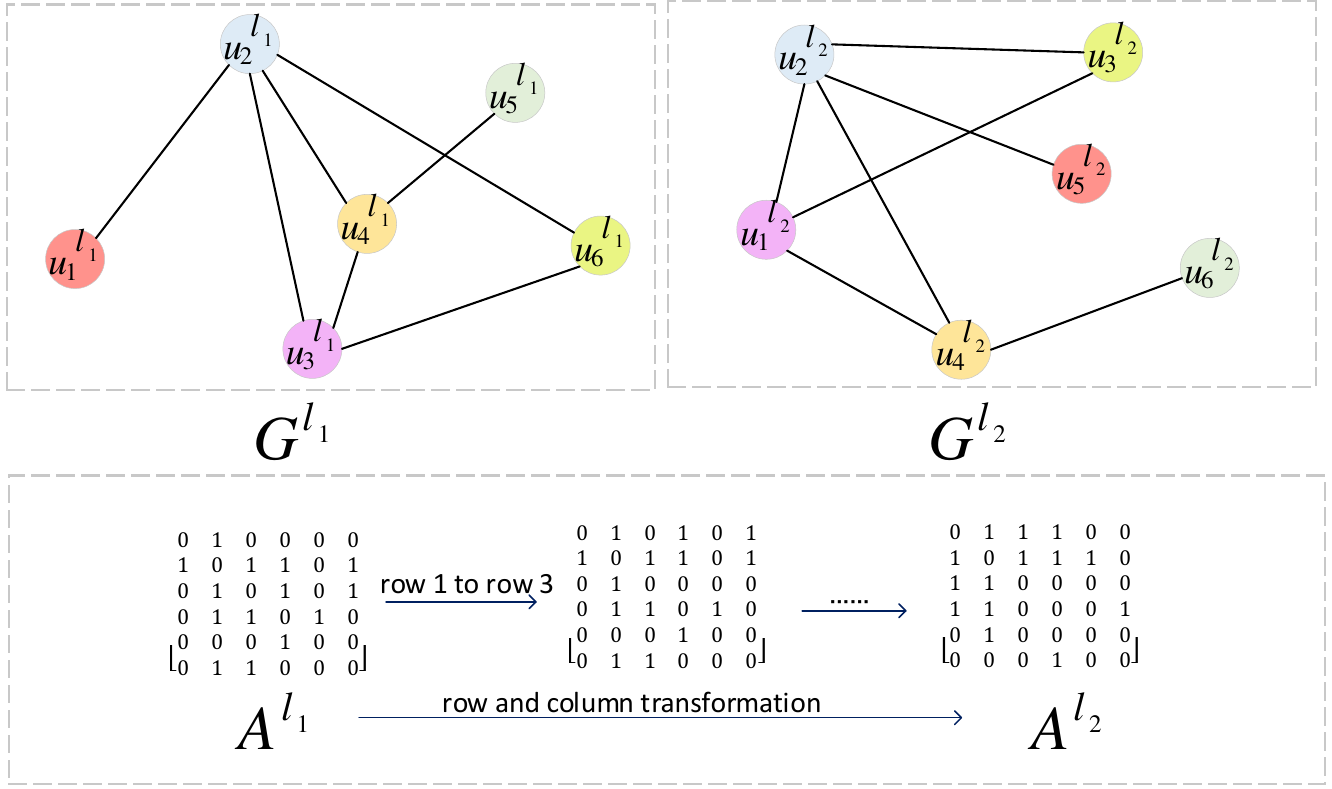}
    \caption{
    Example of considering network alignment as a graph-matching problem. Networks $G^{l_1}$ and $G^{l_2}$ represent the same network but differ in node ordering. Nodes of the same colour correspond to the same entities, meaning the corresponding node pairs are $(u^{l_1}_1,u^{l_2}_5)$, $(u^{l_1}_2,u^{l_2}_2)$, $(u^{l_1}_3,u^{l_2}_1)$, $(u^{l_1}_4,u^{l_2}_4)$, $(u^{l_1}_5,u^{l_2}_6)$, $(u^{l_1}_6,u^{l_2}_3)$. The adjacency matrix $\bm{\mathrm{A}}^{l_1}$ can be converted to $\bm{\mathrm{A}}^{l_2}$ through specific row and column transformations to $\bm{\mathrm{A}}^{l_1}$. 
    }
    \label{pic_graphmatchingproblem}
\end{figure*}

Some researchers considered network alignment as a graph-matching problem, where the adjacency matrix of one network is a noisy permutation of another network~\cite{ZhangSi2016-KDD}. Under this assumption, the network alignment task becomes applying row and column transformations to the adjacency matrix of one network to closely resemble the adjacency matrix of the other. The alignment of the two networks can be implicitly revealed through this transformation process. As illustrated in Figure~\ref{pic_graphmatchingproblem}, two identical networks with different node orders can be aligned by applying row and column transformations to their adjacency matrices. Global structure consistency-based methods adopt this idea to uncover the corresponding node pairs.

Given the adjacency matrices $\bm{\mathrm{A}}^{l_1}$ and $\bm{\mathrm{A}}^{l_2}$ for two networks to be aligned, the goal of global structure consistency network alignment methods is to find the permutation matrix $\bm{\mathrm{P}}$ that minimizes the objective function
\begin{equation}
Obj=||\bm{\mathrm{P}}\bm{\mathrm{A}}^{l_1}\bm{\mathrm{P}}^T-\bm{\mathrm{A}}^{l_2}||^2_F,
\label{eq_modeling_NA_as_permutaion}
\end{equation}
where $(\cdot)^T$ denotes the transpose of a matrix or vector in parentheses, and $||\cdot||_F$ denotes the Frobenius norm. As illustrated above, the nature of permutation matrix $\bm{\mathrm{P}}$ is to reorder the rows of $\bm{\mathrm{A}}^{l_1}$ while $\bm{\mathrm{P}}^T$ reorders the columns of $\bm{\mathrm{A}}^{l_1}$. After the optimization is complete, $\bm{\mathrm{P}}$ will indicate the best matching between the nodes of the two networks. 

Since solving Eq.~(\ref{eq_modeling_NA_as_permutaion}) is equivalent to the quadratic assignment problem (QAP), which is NP-hard, researchers have explored various approaches to solve it within a limited time frame~\cite{caetano2009learning}. There are two main strategies: one focuses on calculating similarity scores between unmatched nodes across networks based on global structural information to derive the row and column transformations, which yield the final alignment results, while the other developing effective polynomial-time approximation algorithms\cite{konar2020graph}. The first strategy is exactly what methods like IsoRank and GRAAL are striving to achieve. Many researchers have also explored the second strategy. For example, Koutra et al.~\cite{koutra2013big} considered social networks consisting of two types of nodes: users and groups. They transformed the objective function 
\begin{equation}
\begin{aligned}
& Obj=||\bm{\mathrm{P}}\bm{\mathrm{A}}^{l_1}\bm{\mathrm{Q}}-\bm{\mathrm{A}}^{l_2}||^2_F\\
& =||\bm{\mathrm{P}}\bm{\mathrm{A}}^{l_1}\bm{\mathrm{Q}}-\bm{\mathrm{A}}^{l_2}||^2_F+\lambda \sum\limits_{i,j}\bm{\mathrm{P}}_{ij}+\mu \sum\limits_{i,j}\bm{\mathrm{Q}}_{ij}\\
& =Tr(\bm{\mathrm{P}}\bm{\mathrm{A}}^{l_1}\bm{\mathrm{Q}}(\bm{\mathrm{P}}\bm{\mathrm{A}}^{l_1}\bm{\mathrm{Q}})^T-2\bm{\mathrm{P}}\bm{\mathrm{A}}^{l_1}\bm{\mathrm{Q}}{\bm{\mathrm{A}}^{l_2}}^T)+\lambda \bm{\mathrm{1}}^T\bm{\mathrm{P}}\bm{\mathrm{1}}+\mu \bm{\mathrm{1}}^T\bm{\mathrm{Q}}\bm{\mathrm{1}},
\end{aligned}
\end{equation}
where $\bm{\mathrm{P}}$ and $\bm{\mathrm{Q}}$ are the permutation matrices for the user and group level, respectively, $\lambda$ and $\mu$ are the sparsity constraint parameters for $\bm{\mathrm{P}}$ and $\bm{\mathrm{Q}}$ respectively, $Tr(\cdot)$ denotes the trace of a matrix in the parentheses, and $\bm{\mathrm{1}}$ is the vector of all ones. Then, they used an alternating projected gradient descent approach to optimize the correspondence matrices iteratively. 
Similar works can be found in Refs.~\cite{bayati2009algorithms,bayati2013message,zhang2015multiple,zhang2016pct,zhang2018attributed,heimann2018regal,zhang2019origin,zhang2019multilevel,konar2019iterative,zhang2021RelGCN,tang2023cross,zeng2023parrot,tang2023robust}.

To sum up, local structure consistency-based methods are better suited for aligning social networks, while those based on global structure consistency are more appropriate for biological networks. This distinction arises from the characteristics of the two types of networks. On the one hand, social networks easily obtain observed corresponding node pairs since many users may publicly share their accounts across different platforms or use the same username, profile description, and photos. In contrast, PPI networks often struggle to yield observed corresponding node pairs, especially among less-studied species. On the other hand, the time complexity of global structure consistency-based methods is significantly higher than that of local structure consistency-based methods. PPI networks often have fewer than 1,000 nodes, making it feasible to use global structure consistency-based methods for slight accuracy improvements.
In contrast, social networks can have millions or even billions of nodes, where efficiency is essential, making local structure consistency methods more suitable. However, neither of them is well-suited for KG alignment because KGs typically consist of diverse types of nodes and edges, resulting in significant structural differences between KGs. Structure consistency-based methods rely on topological similarity, but the heterogeneity of KGs makes direct comparisons of their network structures complex and unreliable. Moreover, the semantic information of nodes in KGs plays a crucial role in alignment. Structure consistency-based methods often fail to effectively leverage this semantic information, leading to less accurate alignment results.

\subsection{Machine learning-based methods}
Machine learning-based methods typically involve training a classification model using features extracted from observed corresponding node pairs, which are then used to predict whether unmatched nodes from different networks are corresponding nodes. The features used in training may include node attributes, structural properties, or a combination of them. Once trained, the model takes the unmatched nodes from different networks as input and outputs the likelihood or classification of whether a pair of them is a corresponding node pair or not. Machine learning-based methods provide several advantages over structure consistency-based methods by effectively capturing complex relationships and patterns within networks from different fields, allowing for more accurate alignment and accommodating diverse node attributes and structural variations that may not be easily identified through direct comparison alone.

Machine learning-based methods can be categorized into traditional and deep learning methods. Traditional machine learning-based methods typically extract features from pairs of nodes across different networks, such as attribute similarity, degree similarity, or neighbourhood similarity. Deep learning-based methods leverage artificial neural network techniques to reduce the impact of feature extraction quality on alignment. Simply preprocessing the original network structure and node attribute data can be used as inputs to train the model. It can be further divided into network embedding-based methods and GNN-based methods. Although both embedding-based and GNN-based methods utilize neural networks, they differ in their focus. Network embedding-based methods aim to map nodes into Euclidean or other spaces to obtain usable node representations that capture both the topological structure and attribute information of the nodes. The representations are then used to train classification models. Typically, the training of the classification models and the node embedding process are two independent modules, and the representations obtained through network embedding are not explicitly optimised for network alignment tasks. These representations can be used as inputs for network alignment and other tasks such as link prediction and node classification. In contrast, GNN-based methods enable an end-to-end training strategy by using node attribute features and network structure information as inputs. They simultaneously optimize node representations and alignment tasks, allowing more targeted adjustments to the model to fit the specific requirements of network alignment better. 
Based on recent research trends, we introduce these methods in detail in the following order, starting with the more popular ones:  network embedding-based methods, GNN-based methods, and feature extraction-based methods.

\subsubsection{Network embedding-based methods}
\label{SubSec: Network embedding-based methods}
Network embedding is a technique designed to generate node representations by mapping a network into a hidden space. The result of this process is that each node is represented as a representation vector in the hidden space. These representation vectors have specific properties: nodes that are adjacent or share similar topological structures are mapped to vectors that are positioned closely together, with their similarity measurable by some commonly used distance metrics, such as Euclidean distance or cosine similarity. This approach offers a new perspective on node representation, allowing fundamental features of real networks, such as the scale-free degree distributions, small-world effect, clustering, and community structure, to be interpreted geometrically~\cite{garcia2018multiscale}. Distances among nodes in a hidden space encode a balance between their similarity and popularity, thus determining their likelihood of being linked~\cite{serrano2008self, allard2017geometric}. It leads to a paradigmatic shift in understanding network structures. Many researchers in network alignment are interested in exploring the potential of using this technique for network alignment.

Networks can be embedded into various hidden spaces, including Euclidean \cite{zhang2021systematic} and hyperbolic \cite{kovacs2023model} spaces, etc. We provide a comprehensive overview of studies related to network alignment using node representations in these spaces. Each part will begin with an introduction to the methods used to embed networks into a specific space, followed by an explanation of how node representations are leveraged for network alignment.

\paragraph{A. Euclidean space-based embedding} \mbox{}\\

\begin{figure} 
	\centering
	\includegraphics[scale=1.5]{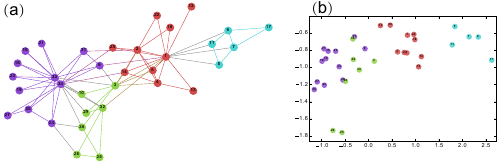}
	\caption{An illustration of embedding a network into an Euclidean space. (a) Karate network. This network consists of 33 nodes, categorized into four groups, each marked with a different color. (b) Embedding the Karate network into a two-dimensional Euclidean space. In this space, the Euclidean distances between the representation vectors of nodes within the same group are smaller.
		\\Source: Reproduced from Ref.~\cite{PerozziBryan2014}. }
	\label{Fig_DeepWalk}
\end{figure}

As shown in Figure~\ref{Fig_DeepWalk}, through network embedding, nodes of the Karate network~\cite{zachary1977information} can be represented as two-dimensional vectors in an Euclidean space. The distance between nodes in the space can reflect their structure proximity or similarity in the original network~\cite{zhang2018network}. Namely, nodes close in the space will likely have similar properties or connections. After the representation vectors are obtained, many network analysis tasks, such as link prediction, node classification, and node clustering, can be quickly and efficiently supported. For example, Nodes of different colours in Figure~\ref{Fig_DeepWalk} (a), which were initially challenging to classify programmatically, become accessible to classify because their representation vectors are close together in the hidden Eclidean space as shown in Figure~\ref{Fig_DeepWalk} (b).

The main problems of leveraging network embedding techniques for network alignment include two key issues: one is how to effectively embed a network into a Euclidean space so that the representation vectors of the nodes that have similar properties or connections are close to each other; the other is how to achieve a well network alignment results based on the representation vectors.

For the problem of how to effectively embed networks into Euclidean spaces, the most representative works include DeepWalk~\cite{PerozziBryan2014}, node2vec~\cite{grover2016node2vec}, and LINE~\cite{TangJian2015}, etc. DeepWalk generates node representations in Euclidean space by employing a technique inspired by the process of learning word vectors~\cite{mikolov2013distributed} from large text corpora. It uses random walks to generate a set of node sequences, analogous to sentences composed of word sequences in text data, and then applies the skip-gram model~\cite{bengio2000neural}, commonly used in natural language processing, to these sequences to learn vector representations for nodes. Given a network $G=(V,E)$, a node sequence is obtained by selecting any node $v_i$ as the initial, randomly choosing a node from the neighbour set $\Gamma(v_i)$ as the next-hop node, and then taking the same step until a sequence of nodes with length $n$ is formed. The skip-gram model is used after many node sequences are obtained. Given a random walk node sequence $\{v_1, v_2, \cdots, v_i, \cdots, v_n\}$, DeepWalk learns representation of node $v_i$ by minimizing the objective function
\begin{equation}
Obj=-log\ p( \{ v_{i-t}^{\alpha},\cdots,v_{i+t}^{\alpha}\}\mid v_i^{\alpha}),
\label{eq_minimizing_deepwalk}
\end{equation}
where $\{ v_{i-t}, \cdots, v_{i+t}\}$ is the sequence composed of the contexts of node $v_i$. $p( \{ v_{i-t},\cdots,v_{i+t}\}\mid v_i)$ denotes the probability that the context is $\{ v_{i-t}^{\alpha},\cdots,v_{i+t}^{\alpha}\}$ when the current node is $v_i^{\alpha}$. By using conditional independence assumption, it is approximately equal to
\begin{equation}
p(\{ v_{i-t}, \cdots, v_{i+t}\}\mid v_i)\approx \prod\limits_{j=-t,j\ne 0}^t{p( v_{i+j}\mid v_{i})}.
\end{equation}
The probability $p(v_{i+j}\mid v_{i})$ in the hidden space can be calculated by 
\begin{equation}
p( v_{i+j}\mid v_{i})=\frac{\exp(( \bm{\mathrm{v}}_{i+j} )^T\cdot ( \bm{\mathrm{v}}_{i})')}{\sum_{k=1}^{|V|}{( \bm{\mathrm{v}}_{k} ) ^T\cdot ( \bm{\mathrm{v}}_{i}) '}},
\end{equation}
where $\bm{\mathrm{v}}_{i}$ and $( \bm{\mathrm{v}}_{i})'$ denote the input and output vectors of node $v_i$ respectively. The representation vectors for all nodes can be achieved by minimizing the Eq.~(\ref{eq_minimizing_deepwalk}), with the help of stochastic gradient descent. 

Grover et al.~\cite{grover2016node2vec} observed that DeepWalk fails to capture sufficient global structural information, such as community structures. They proposed node2vec, introducing a more flexible and controlled random walk strategy. Node2vec allows for biased walks with two parameters. One parameter adjusts the likelihood of returning to the previous node, resembling a depth-first search to traverse all nodes. At the same time, the other encourages visiting new nodes, similar to a breadth-first search. This approach enables node2vec to capture better diverse connectivity patterns, such as the network roles or community structure, thereby producing more informative and discriminative representation vectors for nodes.

Since many large real-world networks contain millions of nodes and billions of edges, obtaining their representation vectors is challenging. Tang et al.~\cite{TangJian2015} proposed LINE, which uses an edge-sampling strategy to approximate the network's overall structure efficiently. For link $e_{ij}=(v_i,v_j)$, the joint probability between node $v_i$ and $v_j$ is
\begin{equation}
p_1(v_i,v_j)=\frac{1}{1+\mathrm{exp}(-{(\bm{\mathrm{v}}_i)}^{\mathrm{T}} \cdot \bm{\mathrm{v}}_j)}.
\end{equation}
The empirical counterpart of $p_1(\cdot , \cdot)$ can be defined as $\widehat{p_1}(\cdot , \cdot)=w_{ij}/ W$, where $w_{ij}$ is the weight of link $e_{ij}$, and $W$ is the summation of the weights for all links. By minimizing the KL-divergence~\cite{manning1999foundations}, which is a method of measuring the similarity of two distributions, of $p_1(\cdot , \cdot)$ and $\widehat{p_1}(\cdot , \cdot)$ over all the links in the network, the representation vectors can be obtained. Namely, the purpose of LINE is to minimize the objective function
\begin{equation}
Obj=\sum_{\forall(v_i,v_j)\in E} \mathrm{KL}(\widehat{p_1}(v_i,v_j),p_1(v_i,v_j)).
\label{eq:embedding_obj1}
\end{equation}
Omitting some constants, the objective function can be rewritten as 
\begin{equation}
Obj=-\sum_{\forall(v_i,v_j)\in E} w_{ij}\log{p_1(v_i,v_j)}.
\label{eq:embedding_obj2}
\end{equation}
By minimizing Eq.~(\ref{eq:embedding_obj2}) over all the links independently, each network node can be represented as a vector in the hidden Euclidean space. The above steps preserve the first-order proximity, which holds that the connected nodes have similar representation vectors. To capture the global network structures, LINE also preserves the second-order proximity by assuming that nodes sharing many links to other nodes are similar. 

In addition to the works mentioned above, numerous other studies are on effectively embedding the network into an Euclidean space. For example, Wang et al.~\cite{wang2017community} not only retained the first-order and second-order proximity of nodes but also preserved the community structures. Furthermore, many scholars have explored dynamic network embedding~\cite{zhiyuli2018modeling} and scale-free network embedding~\cite{FengRui2018}, among others. Various network embedding methods can be found in Refs.~\cite{zhang2018network,cui2019survey,cai2018comprehensive}. 

The second problem to address, as mentioned above, is achieving effective network alignment based on the representation vectors. Intuitively, if a pair of corresponding nodes from different networks have identical or similar representation vectors, accurate alignment could be achieved simply by comparing the distances or similarities between their vectors. However, this poses a significant challenge because the hidden space of each network is unknown to the others~\cite{ZhouFan2018}. In typical network embedding methods, nodes of a single network are represented as vectors within the same Euclidean space because they are optimized together. When dealing with different networks, the representation vectors of nodes from each network reside in separate Euclidean spaces unless special processing is applied, as they are optimized independently. To address this issue, various methods have been developed, which can be broadly categorized into three types based on the core idea of ensuring that corresponding nodes from different networks have identical or similar representation vectors: unifying the space by training a mapping function, joint network embedding and alignment, and embedding networks directly into a unified space.

\begin{figure} 
	\centering
	\includegraphics[scale=0.24]{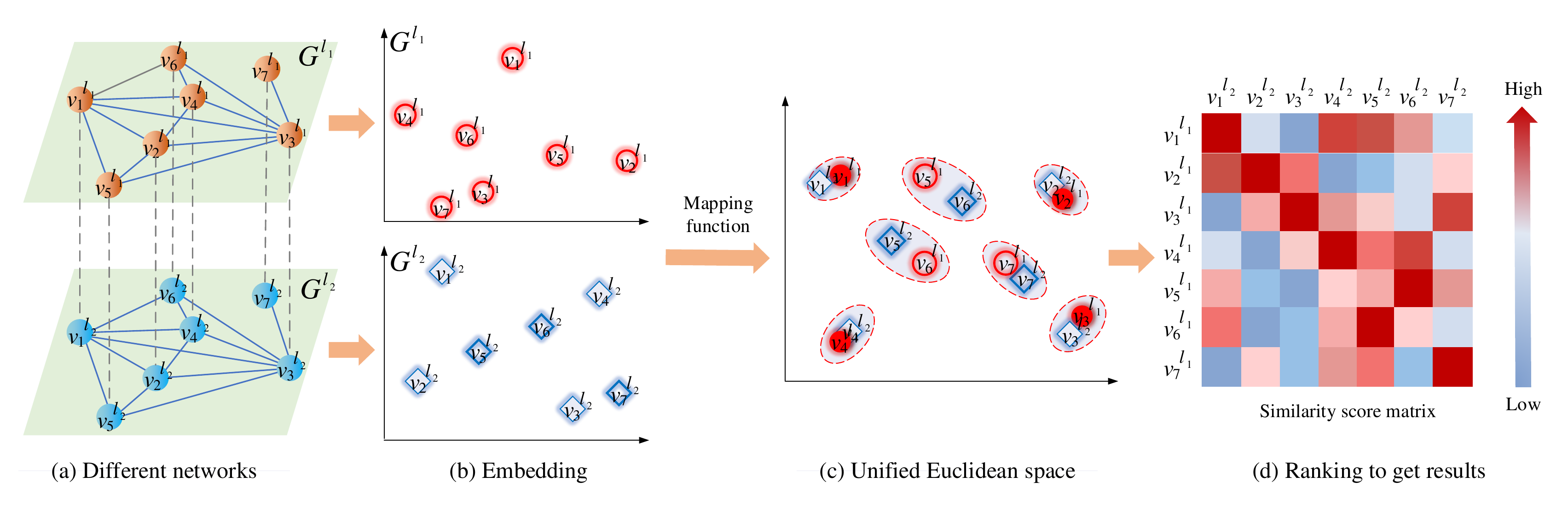}
	\caption{Illustration of leveraging network embedding techniques for network alignment. (a) Different networks. There are two networks, each has seven nodes. (b) Embedding. Using network embedding techniques, each network is embedded into a two-dimensional Euclidean space, where nodes are represented as representation vectors. (c) Unified Enclidean space. A mapping function can be trained using observed corresponding node pairs, projecting the nodes from different networks into an unified Euclidean space. (d) Ranking to get results. The alignment results are obtained by comparing the Euclidean distances between cross-network node representation vectors and sorting them in ascending order.
	\label{Fig_Euclideanspaceforalignment}}
\end{figure}

\textbf{(i) Unifying the space by training a mapping function. }
A simple approach is embedding each network into separate Euclidean spaces and transforming them into a unified one. Finally, the distances between node vectors in the unified space determine which nodes correspond. This approach relies on an assumption: an ideal mapping function can transform the networks into a unified space where the representation vectors of corresponding nodes are identical. Suppose a score ranging from 0 to 1 can denote the similarity of the representation vectors between the nodes from different networks in the unified space. In that case, the score between the vectors of the corresponding nodes should ideally be 1. However, discovering such an ideal mapping function is often challenging~\cite{ZhouFan2018,tang2023interlayer}, since the hidden space of each network is unknown to the others and the sampled contexts of a node may differ during embedding. Until now, no ideal function has been found to make the vectors of all corresponding nodes identical in a unified space. Typically, we can compare the similarity scores. The higher the similarity score of two nodes from different networks, the more likely they are corresponding nodes, as shown in Figure~\ref{Fig_Euclideanspaceforalignment}.

Man et al.~\cite{ManTong2016-IJCAI} proposed PALE, which trains the mapping function by minimizing the loss function
\begin{equation}
    \ell(\phi)=\sum_{(v_i^{l_1},v_j^{l_2}) \in \Psi^o}{||\phi(\bm{\mathrm{v}}_i^{l_1};\theta)-\bm{\mathrm{v}}_j^{l_2}||_F},
    \label{lossfunc_pale}
\end{equation}
where $\bm{\mathrm{v}}_i^{l_1}$ and $\bm{\mathrm{v}}_j^{l_2}$ are the representation vectors for the observed corresponding nodes in $\Psi^o$. $\bm{\mathrm{v}}_i^{l_1}$ resides in the Enclidean space of $Z^{l_1}$, while $\bm{\mathrm{v}}_j^{l_2}$ is in the Enclidean space of $Z^{l_2}$. The embedding method can be any approach that embeds individual networks. $\theta$ represents the collection of all parameters in the mapping function. The mapping function can be linear or non-linear. For the linear mapping function, $\theta$ is a $k \times k$ matrix, so that 
\begin{equation}
    \phi(\bm{\mathrm{v}}_i^{l_1};\theta)=\theta \times \bm{\mathrm{v}}_i^{l_1},
\end{equation}
where $k$ is the dimension of the vectors. For the nonlinear mapping function, $\theta$ is a collection of all parameters for a Multi-Layer neural network~\cite {suter1990multilayer}. If the layer of the neural network is 3, the mapping function is as follows
\begin{equation}
    \phi(\bm{\mathrm{v}}_i^{l_1};\theta)=((\bm{\mathrm{v}}_i^{l_1} \times \theta_1 + b_1)\times \theta_2 + b_2)\times \theta_3+b_3,
\end{equation}
where $\theta_i \in \theta$ and $b_i \in \theta$. After obtaining the mapping function, all vectors in the hidden space $Z^{l_1}$ can be projected into $Z^{l_2}$, thus achieving a unified hidden space. Finally, for an unmatched node $u_i^{l_1}$, its corresponding node can be identified by searching the closest node $u_j^{l_2} \in V^{l_2}$
\begin{equation}
\min \limits_{j}||\phi(\bm{\mathrm{u}}_i^{l_1};\theta)^T-\bm{\mathrm{u}}_j^{l_2}||_F.
\label{eq_sim_vector_distance}
\end{equation}

Equation~(\ref{eq_sim_vector_distance}) reflects the Euclidean distance between nodes in the unified space. The closer two nodes across different networks in the space, the more likely the two nodes will correspond. Other methods can also be used for measurement. For example, Ye et al.~\cite{ye2023vqne} use inner product, which is 
\begin{equation}
\min \limits_{j} \phi(\bm{\mathrm{u}}_i^{l_1};\theta)^T \cdot \bm{\mathrm{u}}_j^{l_2}.
\end{equation}
Du et al.~\cite{du2020cross} use cosine similarity, which is
\begin{equation}
\min \limits_{j} \frac{\phi(\bm{\mathrm{u}}_i^{l_1};\theta)^T \cdot \bm{\mathrm{u}}_j^{l_2}}{||\phi(\bm{\mathrm{u}}_i^{l_1};\theta)||_2 \cdot ||\bm{\mathrm{u}}_j^{l_2}||_2}.
\end{equation}

PALE requires a sufficient number of observed corresponding node pairs to train the mapping function. Under this condition, it is well-suited for aligning social networks and PPI networks, but not for KG alignment. This is primarily because each edge in a KG is typically represented as a triplet in the form of (head node, relation, tail node), i.e., $(v_i,e_{ij},v_j)$. Given any KG, many KG embedding methods embed both nodes and edges into the same hidden space. They often assume that $\bm{\mathrm{v}}_i+\bm{\mathrm{e}}_{ij}=\bm{\mathrm{v}}_j$~\cite{bordes2013translating}, where $\bm{\mathrm{e}}_{ij}$ indicates the representation vector for edge $e_{ij}$. By minimizing the triplet relationships that exist between edges and maximizing those that do not exist, the representation vectors for nodes and edges can be derived, such as minimizing the objective function 
\begin{equation}
Obj=\sum\limits_{(v_i,e_{ij},v_j)\in G} \sum \limits_{(v_i',e_{ij}',v_j') \in G'} (\alpha + (\bm{\mathrm{v}}_i+\bm{\mathrm{e}}_{ij}-\bm{\mathrm{v}}_j)^2 -(\bm{\mathrm{v}}_i'+\bm{\mathrm{e}}_{ij}'-\bm{\mathrm{v}}_j')^2),
\end{equation}
where $\alpha>0$ is a hyperparameter, $(v_i',e_{ij}',v_j')$ represents the relationship not exists~\cite{chen2017multilingual}. For two KG $G^{l_1}$ and $G^{l_2}$, after obtaining the representation vectors for all nodes and edges, the hidden spaces of $G^{l_1}$ and $^{l_2}$ can be unified by training a linear mapping function to map the hidden space from one KG to the other~\cite{chen2017multilingual, zhu2017iterative}. The loss function is
\begin{equation}
\ell(\phi)=\sum \limits_{\substack{(v_i^{l_1},v_j^{l_2}) \in \Psi^o, \\(v_a^{l_1},v_b^{l_2}) \in \Psi^o}}{||\phi(\bm{\mathrm{v}}_i^{l_1};\theta)-\bm{\mathrm{v}}_j^{l_2}||+||\phi(\bm{\mathrm{e}}_{ia}^{l_1};\theta)-\bm{\mathrm{e}}_{jb}^{l_2}||+||\phi(\bm{\mathrm{v}}_a^{l_1};\theta)-\bm{\mathrm{v}}_b^{l_2}||},
\label{lossfunc_MTransE}
\end{equation}
where $v_i^{l_1}$ and $v_j^{l_2}$ are head nodes, $v_a^{l_1}$ and $v_b^{l_2}$ are tail nodes. 

After PALE, Zhou et al.~\cite{ZhouFan2018} proposed DeepLink, which adopts a dual-learning~\cite{he2016dual} method to get a better mapping function to improve alignment accuracy. The mapping function from $Z^{l_1}$ to $Z^{l_2}$ is defined as $\phi$ and from $Z^{l_2}$ to $Z^{l_1}$ is defined as $\phi^{-1}$. All the unmatched nodes can be used to pre-train the mapping function by minimizing the distance between $\bm{\mathrm{u}}_i^{l_1}$ and $\bm{\mathrm{u^{''}}}_i^{l_1}$, where $\bm{\mathrm{u^{'}}}_i^{l_1}=\phi(\bm{\mathrm{u}}_i^{l_1};\theta)$ and $\bm{\mathrm{u^{''}}}_i^{l_1}=\phi^{-1}(\bm{\mathrm{u^{'}}}_i^{l_1};\theta)$. Tang et al.~\cite{tang2022interlayer} argued that the alignment clues are not only the similarity between the vectors in the unified space. The positional relationships of representation vectors between unmatched nodes and their CINs can also provide valuable alignment clues. To capture these clues comprehensively, they proposed a method that considers multiple types of consistency between embedding vectors (MulCev). This approach employs PALE to assess the similarity between representation vectors of unmatched nodes and introduces a distance consistency index to capture the alignment clues in the positional relationships. Similarly, Yang et al.~\cite{yang2022anchor} use FRUI to calculate the similarity between unmatched nodes to replace the distance consistency in MulCev. Yan et al.~\cite{yan2021bright} analyzed the fundamental limitation of network embedding-based methods theoretically and found that the error that occurred from training the network embedding and mapping function can be accumulated, which leads to the issue of space disparity. The solution is a combination of structure consistency-based methods and network embedding-based methods. Huynh et al.~\cite{huynh2021network} constructed multiple representation vectors, such as the vectors obtained by DeepWalk and other network embedding approaches for each node and designed a mechanism to fuse the different types of representation vectors for a node. Yan et al.~\cite{yan2021towards} injected some pseudo-node pairs to enforce the network embedding to be more widely apart among the nodes in the hidden space. It can reduce excessive clustering around real corresponding node pairs and better capture local structural differences, thus allowing for more accurate alignment. Unlike training a mapping function to project the representation vectors of one network into the hidden space of another, Wang et al.~\cite{WangYongqing2019-www} mapped embedding from each network into a common Hamming space and identified corresponding node pairs in the space. Zhang et al.~\cite{zhang2024collaborative} designed a constraint mechanism to ensure that the distance between corresponding nodes is shorter than between each node and its neighbours in the unified hidden space.

\textbf{(ii) Joint network embedding and alignment.} Liu et al.~\cite{LiuLi2016} proposed IONE to embed two networks and simultaneously align them. For a node $v_i$ in any network $G=(V,E)$, its representation vector is obtained via three parts: a node vector $\bm{\mathrm{v}}_i$, an input context vector$ \bm{\mathrm{v}}_{i}^{'}$, and an output context vector $\bm{\mathrm{v}}_{i}^{''}$. Given an edge $e_{ij}\in E$, the probability that $v_i$ contributes to $v_j$ as its input node when compared with how $v_i$ contributes to other nodes is 
\begin{equation}
p_1(v_j|v_i)=\frac{\exp{({\bm{\mathrm{v}}_{j}^{'}}^T \cdot \bm{\mathrm{v}}_i)}}{\sum_{k=1}^{|V|}\exp{ ({\bm{\mathrm{v}}_k^{'}}^T \cdot \bm{\mathrm{v}}_i ) }}
\end{equation}
while the probability that $v_j$ contributes to $v_i$ as its output node when compared with how $v_j$ contributes to other nodes is
\begin{equation}
p_2(v_i|v_j)=\frac{\exp{({\bm{\mathrm{v}}_{i}^{''}}^T \cdot \bm{\mathrm{v}}_j)}}{\sum_{k=1}^{|V|}\exp{ ({\bm{\mathrm{v}}_k^{''}}^T \cdot \bm{\mathrm{v}}_j ) }}.
\end{equation}
Similar to LINE, the empirical counterpart of $p_1(v_j|v_i)$ are defined as $\widehat{p_1}(\cdot , \cdot)=w_{ij}/ d^{out}_i$ and $p_2(v_i|v_j)$ as $\widehat{p_2}(\cdot , \cdot)=w_{ij}/ d^{in}_j$, where $d^{out}_i=\sum_{v_k\in \Gamma_{out} (v_i)} w_{ik}$ and $d^{in}_j=\sum_{v_k\in \Gamma_{in} (v_j)} w_{kj}$. The goal of embedding the two networks is to minimize the objective function
\begin{equation}
\begin{aligned}
    Obj_1=-\sum\limits_{\forall(v_i^{l_1},v_j^{l_1})\in E^{l_1}} {w_{ij} \log p_1(v_j^{l_1}|v_i^{l_1})} \\ - \sum\limits_{\forall(v_i^{l_1},v_j^{l_1})\in E^{l_1}}{w_{ij} \log p_2(v_i^{l_1}|v_j^{l_1})}\\
    -\sum\limits_{\forall(v_i^{l_2},v_j^{l_2})\in E^{l_2}} {w_{ij} \log p_1(v_j^{l_2}|v_i^{l_2})} \\ -\sum\limits_{\forall(v_i^{l_2},v_j^{l_2})\in E^{l_2}}{w_{ij} \log p_2(v_i^{l_2}|v_j^{l_2})}
\end{aligned}.
\end{equation}
Denoting $p_a(v_i^{l_1}|v_k^{l_2})$ as the probability that $v_i^{l_1}$ is the corresponding node of $v_k^{l_2}$ predicted by a pre-trained model, it can act as a bridge between $v_i^{l_1}$ and $v_k^{l_2}$. This means $v_k^{l_2}$ in network $G^{l_2}$ can contribute to the neighbour of node $v_i^{l_1}$ in network $G^{l_1}$ as the input or output node with probability $p_a(v_i^{l_1}|v_k^{l_2})$. To ensure that the representation vectors of two corresponding nodes from different networks are similar in the hidden space, another objective function can be set as
\begin{equation}
\begin{aligned}
    Obj_2=-\sum\limits_{\forall(v_i^{l_1},v_j^{l_1})\in E^{l_1}} {w^{l_1}_{ij} p_a(v_i^{l_1}|v_k^{l_2}) \log p_1(v_j^{l_1}|v_k^{l_2})}\\       
    - \sum\limits_{\forall(v_i^{l_1},v_j^{l_1})\in E^{l_1}}{ w^{l_1}_{ij} p_a(v_j^{l_1}|v_k^{l_2}) \log p_2(v_i^{l_1}|v_k^{l_2})}\\    
    -\sum\limits_{\forall(v_i^{l_2},v_j^{l_2})\in E^{l_2}} {w^{l_2}_{ij} p_a(v_i^{l_2}|v_k^{l_1}) 
 \log p_1(v_j^{l_2}|v_k^{l_1})} \\     
    - \sum\limits_{\forall(v_i^{l_2},v_j^{l_2})\in E^{l_2}}{w^{l_2}_{ij} p_a(v_j^{l_2}|v_k^{l_1}) \log p_2(v_i^{l_2}|v_k^{l_1})}.
\end{aligned}
\end{equation}
By minimizing the combined objective function $ Obj=Obj_1+Obj_2$, networks $G^{l_1}$ and $G^{l_2}$ can be aligned. The objective function $Obj_1$ ensures that the vectors of nodes sharing more links are closer, while $Obj_2$ ensures that corresponding nodes are as similar as possible. 

In Ref.~\cite{liu2019structural}, the community structure is considered to improve the accuracy of IONE. Unmatched nodes are more likely to be identified as a corresponding pair if their shared observed corresponding nodes come from different communities. Wang et al.~\cite{wang2019user} treat each type of common information within node attributes as a separate node injected into the network. If two nodes share a common attribute, they each establish a link with the node representing that specific attribute. The IONE framework is then used to align the different networks. In Ref.~\cite{liu2023wl}, a label propagation strategy based on observed corresponding node pairs is employed to identify potential corresponding node pairs with high probability. These observed pairs, along with the high-probability candidate pairs, are used to guide the training of the joint model interactively, thereby enhancing alignment accuracy. Wang et al.~\cite{wang2019user} incorporated the obtained information to construct user-topic bipartite networks and then extended the IONE method, which initially modelled and solved user-user networks only, to jointly model and solve both the user-user networks and the user-topic bipartite networks. Li et al.~\cite{li2018non} extended the IONE by considering the issue of negative samples in network embedding. By integrating negative samples into the embedding process, the model could learn more discriminative representations for the alignment of KGs.
\begin{figure} 
	\centering
	\includegraphics[scale=1.2]{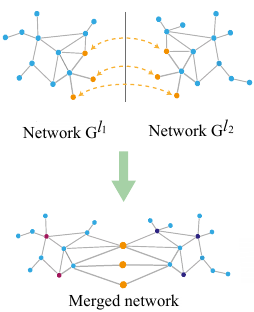}
	\caption{An example of merging two networks to embed them into a unified hidden space. There are two networks $G^{l_1}$ and $G^{l_2}$, which share three observed corresponding node pairs. Using these three pairs, the two networks can be directly merged into a single network, as shown in the lower part. The merged network can then be used for network embedding.
		\\Source: Reproduced from Ref.~\cite{fu2020deep}. }
	\label{Fig_mergenetworks}
\end{figure}

\begin{figure*} [!th]
\centering 
\subfigure[Zachary’s karate club network]{
\includegraphics[width=0.4\linewidth]{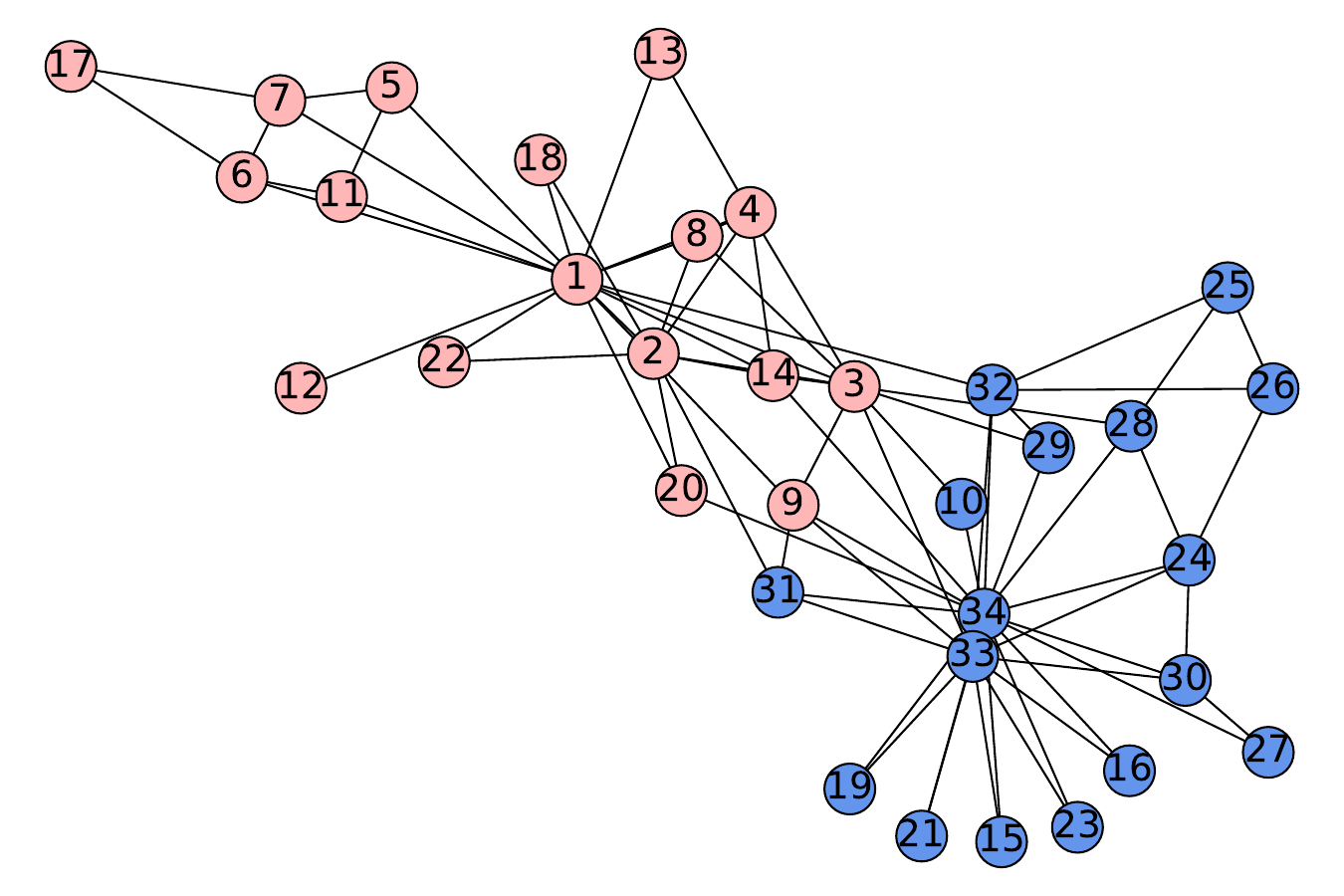}}
\vfill
\subfigure[Hyperbolic space]{
\includegraphics[width=0.3\linewidth]{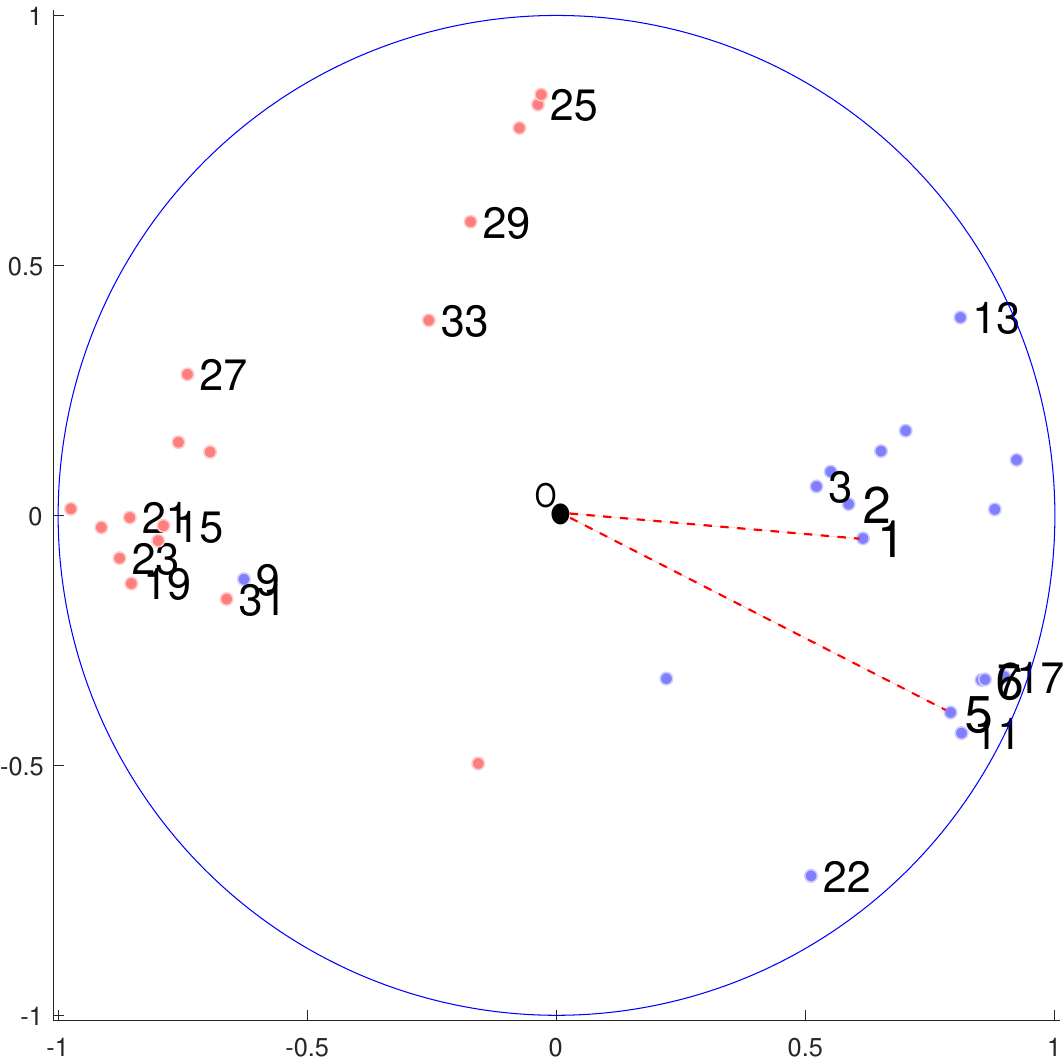}}
\hspace{0.05\linewidth}
\subfigure[Euclidean space]{
\includegraphics[width=0.3\linewidth]{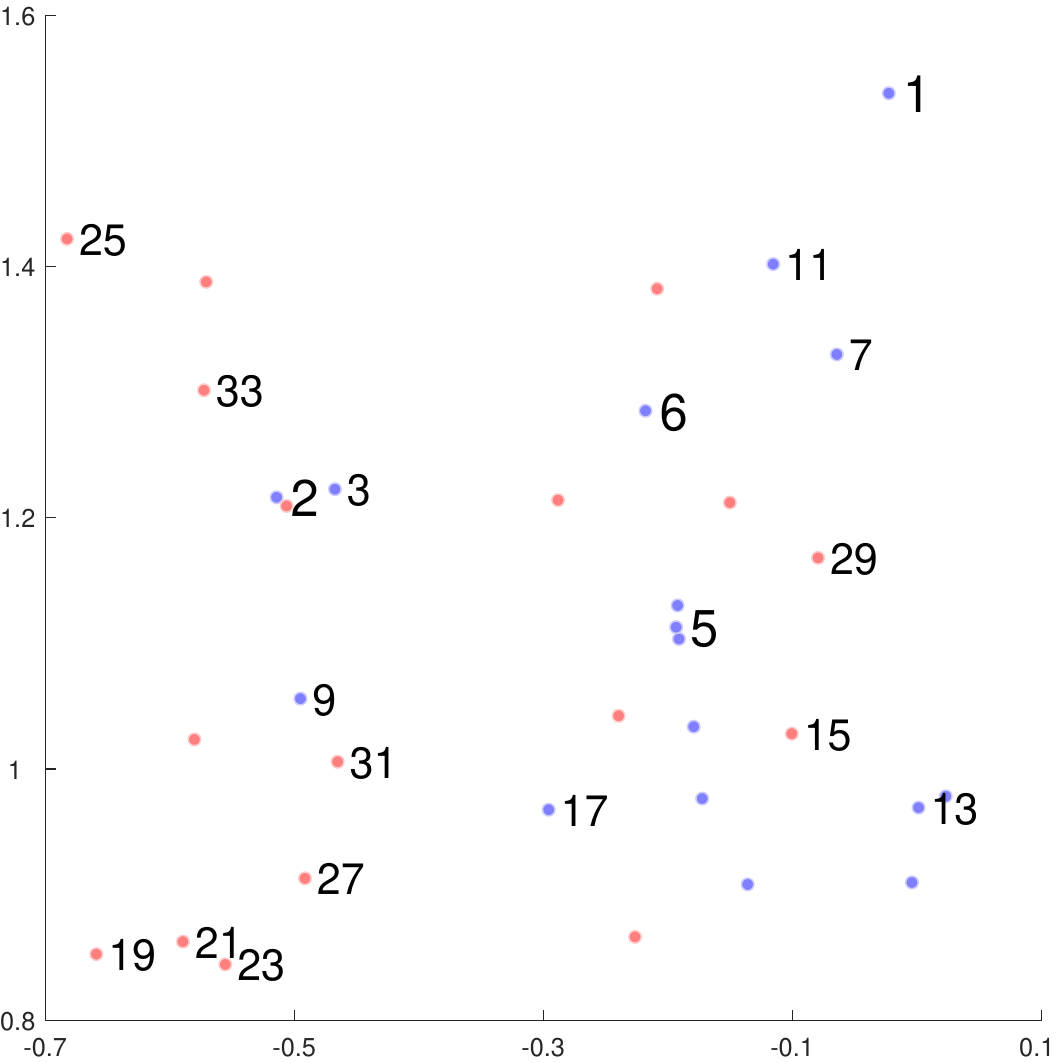}}
\caption{Illustration of embedding a network into different hidden spaces. (a) is a real-world network. It is embedded into a hyperbolic space, as shown in (b) and an Euclidean space, as shown in (c), respectively. In the hyperbolic space, nodes with higher degrees, such as nodes 1 and 2, have a smaller radial distance from the origin compared to nodes with smaller degrees, such as nodes 5 and 6. Additionally, the radial node density grows exponentially with the distance from the origin, while the average degree of nodes decreases exponentially. The similarity between nodes is captured by their angular distance, for example, $\angle 1O5$ in (b). It is evident that popularity is not represented in Euclidean space, and nodes belonging to the same communities are more separable in hyperbolic space.
\\Source: Reproduced from Ref.~\cite{sun2020perfect}.}
\label{fig_example_hyperbolic_space}
\end{figure*}

\textbf{(iii) Embedding networks into a unified space directly.}
Some researchers attempted to merge two networks into one and embed the merged network into a unified hidden space. After obtaining the representation vectors of all nodes, they directly compare the similarity between the representation vectors of unmatched nodes. For example, Fu et al.~\cite{fu2020deep} merged two networks into one using observed corresponding node pairs, as shown in Figure~\ref{Fig_mergenetworks}, and performed random walks on the merged network for the network embedding. Ye et al.~\cite{ye2023vqne} took the same merged operation and performed quantum walks~\cite{yan2022towards} to capture information from both networks. The quantum states from these walks are used as inputs to obtain representation vectors for all nodes in the two networks. Trisedya et al.~\cite{trisedya2019entity} merged two KGs based on the similarity of node names since some nodes across different KGs may have similar names. When the similarity score between two nodes exceeds a certain threshold, they are treated as corresponding nodes, leading to the merging of the two KGs. The network embedding is taken from the merged KG. In Refs.~\cite{du2020cross,tang2023cross}, authors extended DeepWalk by allowing the random walk to jump from one network to another with specific probabilities so that two networks waiting for alignment can be seen as embedded into a unified hidden space.  
\paragraph{B. Hyperbolic space-based embedding} \mbox{}\\
Many physicists have observed that real-world networks such as the Internet~\cite{boguna2010sustaining}, biochemical pathways in cells~\cite{serrano2012uncovering}, and international trade networks~\cite{garcia2016hidden} emerge as discrete samples from hyperbolic space~\cite{allard2017geometric}. In hyperbolic space, both the popularity and similarity of nodes can be effectively represented. Popularity is related to the degrees of the nodes, while similarity represents an aggregate of other attributes influencing the likelihood of interactions~\cite{garcia2018multiscale}. Specifically, in hyperbolic space, a node's popularity is indicated by its radial distance from the origin, and the similarity between nodes is captured by their angular distance~\cite{muscoloni2017machine}, as illustrated in Figure~\ref{fig_example_hyperbolic_space}. 

\begin{figure*} 
\centering 
\subfigure[Embedding DBLP network]{
\includegraphics[width=0.25\linewidth]{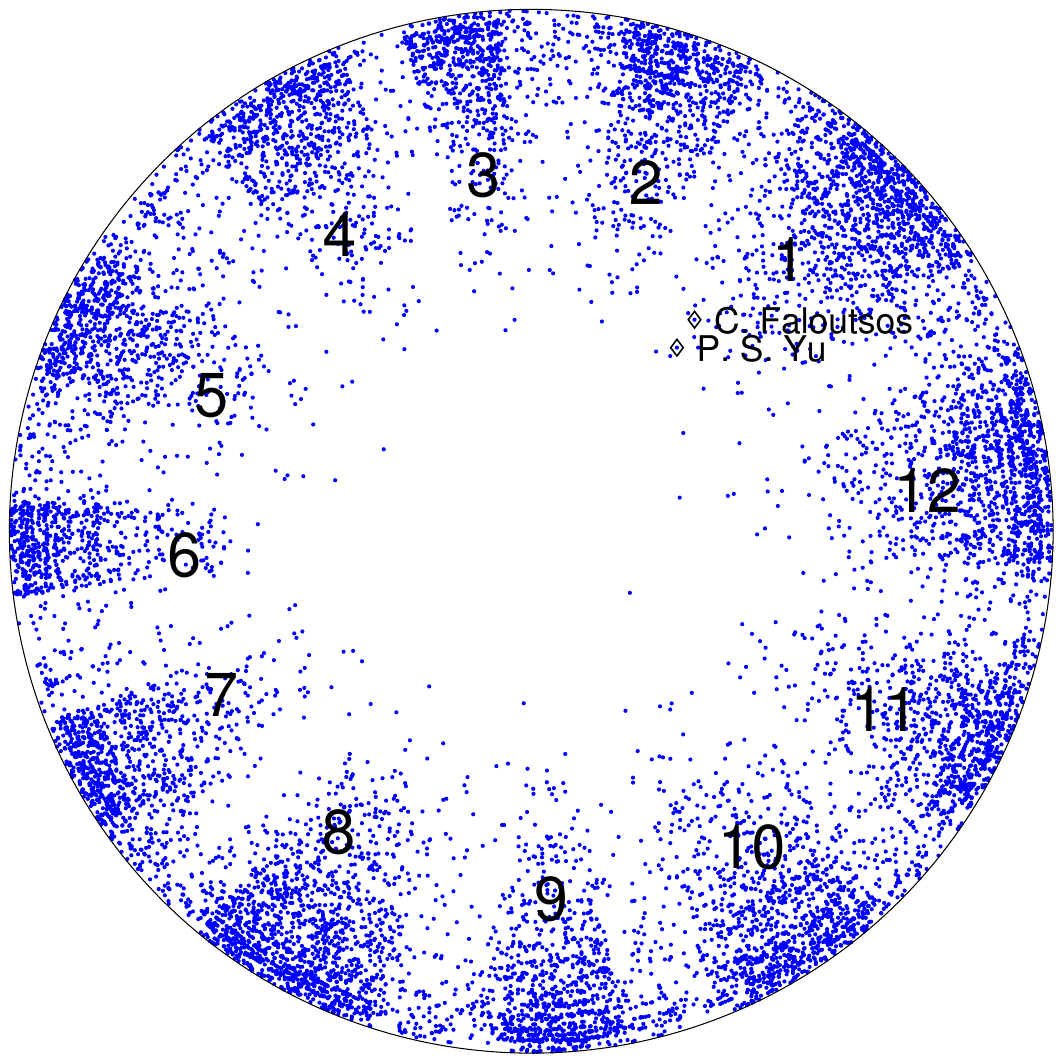}}
\hspace{0.025\linewidth}
\subfigure[The unified hyperbolic space]{
\includegraphics[width=0.25\linewidth]{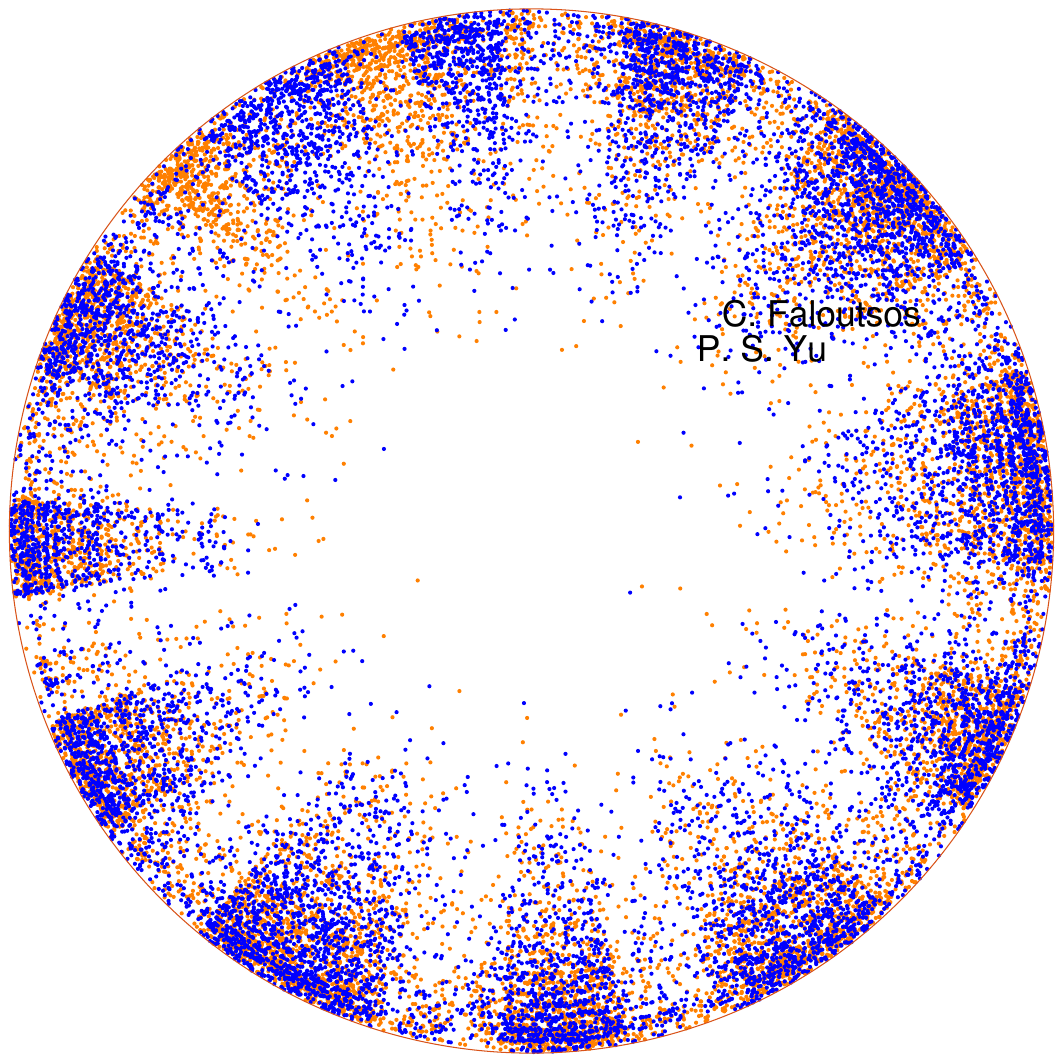}}
\hspace{0.025\linewidth}
\subfigure[Embedding AMiner network]{
\includegraphics[width=0.25\linewidth]{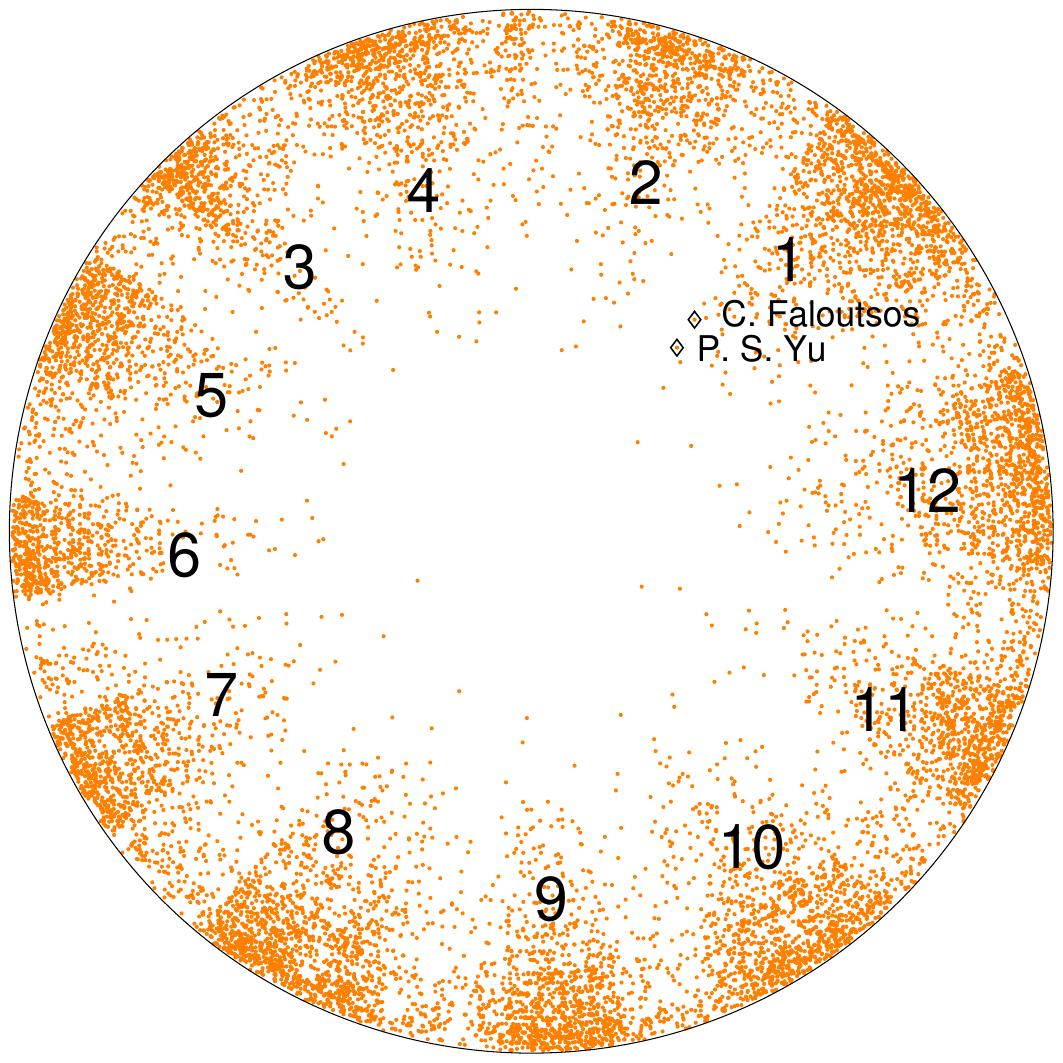}}
\hspace{0.025\linewidth}
\caption{Network alignment by embedding networks into hyperbolic spaces. Two academic collaboration networks were extracted from the DBLP and Aminer datasets. These networks are first embedded into two separate hyperbolic spaces, represented as (a) and (c). Then, utilizing the given observed corresponding node pairs, all nodes from both networks are mapped into a unified hyperbolic space, denoted as (b). The alignment results are obtained by comparing the hyperbolic distances between unmatched nodes across the two networks in (b). Nodes with smaller hyperbolic distances are more likely to be aligned, facilitating network alignment.
\\Source: Reproduced from Ref.~\cite{sun2020perfect}.}
\label{fig_example_align_hyperbolic}
\end{figure*}

After embedding different networks into hyperbolic spaces~\cite{yang2021discrete,yang2022hyperbolic}, the step of using node representation for the network alignment is similar to that of Euclidean space. For example, Li Sun et al.~\cite{sun2020perfect} used the observed corresponding node pairs to map all nodes into a unified hyperbolic space, as shown in Figure~\ref{fig_example_align_hyperbolic}. Zequn Sun et al.~\cite{sun2020knowledge} training a linear mapping function to unify all nodes into a hyperbolic space. 

\subsubsection{GNN-based methods}
Although network embedding techniques have enabled advanced deep learning methods for network alignments, these methods have several limitations when applied to network alignment. 
In network embedding-based methods, network embedding and network alignment are often treated as two separate steps, which leads to node representations that are not specifically optimized for network alignment tasks. Since the generated embeddings are designed to be general-purpose, they can be applied to a wide range of network inference tasks, such as network alignment, link prediction and node classification. However, this generality makes it difficult to fine-tune the embeddings for a specific task like network alignment. Additionally, the separation of the embedding and alignment stages makes the process more cumbersome, as any adjustments to the embeddings require retraining or fine-tuning. This not only introduces inefficiencies but also hampers performance, especially in dynamic or heterogeneous networks, where frequent updates are necessary.
 
\begin{figure*} [ht!]
    \centering
    \includegraphics[width=0.98\textwidth]{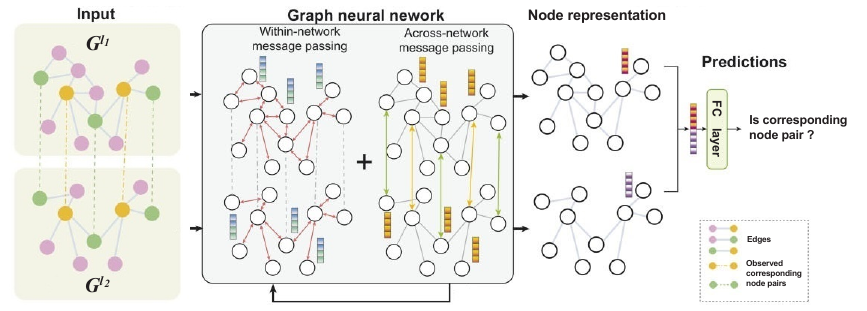}
    \caption{The framework of the GNN-based network alignment methods. The input to the GNN is the adjacency matrices of networks $G^{l_1}$ and $G^{l_2}$. The observed corresponding node pairs are used to compare with the model's predicted correspondence results to construct the loss function. The entire framework provides an end-to-end training strategy that optimizes node representation learning and network alignment tasks simultaneously.\\Source: Reproduced from Ref.~\cite{ding2023supervised}.}
    \label{pic_gnn}
\end{figure*}

In contrast, GNN-based methods utilize an end-to-end training strategy that optimizes node representation learning and network alignment tasks simultaneously. As shown in Figure~\ref{pic_gnn}, for GNN-based methods, the input to the entire model is the adjacency matrices of the networks to be aligned. The observed corresponding node pairs are used alongside the model's predicted correspondence results to construct a loss function, which measures the discrepancy between predicted and true corresponding node pairs. This loss function is then used to optimize the model through backpropagation. In this way, the node representations are specifically optimized for the network alignment task, enabling more targeted and efficient optimization.

The core idea of GNN-based methods is to learn the representation of each node by aggregating the information from its neighbourhood, and GNNs are employed to encode the structural and attributive information into node representations for network alignment~\cite{wang2023gtcalign}. In general, the information aggregation mechanisms of GNN-based methods typically involve different approaches. For instance, graph convolutional networks (GCNs) ~\cite{defferrard2016convolutional,kipf2016semi} use the adjacency matrix to aggregate information from neighbouring nodes; Graph Attention Networks (GATs)~\cite{velickovic2018graph,wang2020knowledge} apply attention mechanisms~\cite{vaswani2017attention,mo2024multi,mo2024temporalhan} to weigh the features of neighbouring nodes. In the following sections, we provide a brief introduction to each type of GNN and offer a comprehensive overview of their applications in network alignment.

\paragraph{A. GCN-based methods}\mbox{}\\

GCNs are neural networks designed to directly operate on network structures, which extend the convolution operation in neural networks, commonly used in image and text domains, to network-structured data. 
The convolution operation on network data, which is the core of GCN, involves a message passing mechanism, where each node updates its feature representation by aggregating the features from its neighboring nodes and applying a linear transformation to these aggregated features. 
In deep learning, the graph convolution operation is formalized as follows:
\begin{equation}
    \mathbf{X}^{(k)}=\sigma\left(\mathbf{\hat{D}}^{-\frac{1}{2}} \mathbf{\hat{A}} \mathbf{\hat{D}}^{-\frac{1}{2}} \mathbf{X}^{(k-1)} \mathbf{W}^{(k-1)}\right),
    \label{eq:gcn}
\end{equation}
where $\mathbf{X}^{k}$ is the node representation matrix at layer $k$, $\mathbf{\hat{A}} = \mathbf{A} + \mathbf{I}$ is the identity matrix with added self-loops, $\mathbf{\hat{D}}$ is the degree matrix of $\mathbf{\hat{A}}$. $\mathbf{{W}^{(k)}}$ is the weight matrix of the $k$-th layer. $\mathbf{W}$ is the weight matrix and $\sigma$ is the activation function. Generally, the inputs of GCNs for an end-to-end network alignment task are two node feature matrices $(\mathbf{X}^{l_1(0)})$ and $(\mathbf{X}^{l_2(0)})$, the adjacency matrices $(\mathbf{A}^{l_1})$ and $(\mathbf{A}^{l_2})$, and observed corresponding node pairs for the prediction task $( \{(v_i^{l_1}, v_i^{l_2}, y_i)\}_{i=1}^|\Psi^o| )$, where $y_i = 1 $ denotes $v_i^{l_1}$ and $v_i^{l_2}$ belong to a corresponding node pair, and $y_i = 0 $ otherwise. Note that the indexing of both $v_i^{l_1}$ and $v_i^{l_2}$ with $i$ does not conflict with the notation $(v_i^{l_1},v_j^{l_2})$ used in other place in this review to represent a corresponding node pair, as the two nodes are treated as a pair. The GCN learns node representations by applying the above convolution operation in Eq.\ref{eq:gcn}. Then, the model use the final learned node representation metrics $\mathbf{X}^{(K)}$ to get a predicted label $\hat{y}_i$ via a prediction function $f$, which is calculated by
\begin{equation}
\hat{y}_i = f(\mathbf{x}_{v_i^{l_1}}, \mathbf{x}_{v_i^{l_2}}).
\end{equation}

The model’s loss is then computed as the cross-entropy loss between predicted labels $\hat{y}_i$ and actual labels $y_i$ by
\begin{equation}
\ell = -\sum_{i=1}^M \left( y_i \log(\hat{y}_i) + (1 - y_i) \log(1 - \hat{y}_i) \right).
\end{equation}

Finally, the GCN's parameters $\{\mathbf{W}^{(l)}\}_{l=1}^K$ and the prediction function $f$'s parameters are optimized by minimizing $\ell$ through backpropagation, thereby enabling the node representations can be specifically optimized for the network alignment task.

In each layer of GCNs, the hidden features of a node are iteratively aggregated with the features of its neighbouring nodes to generate the node’s hidden representation for the subsequent layer. 
By stacking multiple layers, GCNs effectively propagate information across the graph, making them suitable for various tasks such as node classification, link prediction, and graph classification ~\cite{kipf2016semi,wu2020comprehensive}. 
Some GCN-based methods use the embeddings from the final layer as the network representation, while others leverage weighted features from all layers to determine the alignment results.

In recent years, many researchers have excitedly explored the usage of GCNs for the end-to-end network alignment ~\cite{trung2020adaptive,wang2018cross,ye2019vectorized,zhou2019disentangled,zhang2019origin, yang2020re}. Some work \cite{wang2018cross,trung2020adaptive,ye2019vectorized,zhou2019disentangled,park2023power} employed GCNs to embed nodes from different networks into a unified vector space. In contrast, others employed various types of GCNs to achieve representation for a single network collaboratively and aggregate representation information across networks \cite{zhang2019origin, yang2020re}.

Among these advancements, Trung et al. \cite{trung2020adaptive} proposed GAlign, a fully unsupervised network alignment framework based on the GCN model. In this method, nodes from two networks are embedded into higher-order feature vectors through a GCN, and network alignment is then performed based on the similarity of these node embeddings. Unlike approaches that separate the embedding and alignment processes, GAlign enables joint learning of embeddings for different networks, eliminating the need for reconciliation between their embedding spaces.
The core idea of the GAlign training algorithm is a weight-sharing mechanism, where the GCNs of the source network, target network, and augmented networks use the same weight matrix at each layer, ensuring that the embeddings learned from different networks reside in the same embedding space. During the end-to-end training process, the model dynamically learns to generate an alignment matrix by minimizing two losses: the consistency loss ${\ell}_c$ and the noise-adaptive loss ${\ell}_a$, as follows:
\begin{equation}
   \ell_c(G)=\sum_{k=1}^{K}\left\|\hat{\mathbf{D}}^{-\frac{1}{2}} \hat{\mathbf{A}} \hat{\mathbf{D}}^{-\frac{1}{2}}-\mathbf{X}^{(k)} \mathbf{X}^{(k)^T}\right\|_F ,
\end{equation}
\begin{equation}
\ell_a\left(G, G^*\right)=\sum_{v \in G, v^* \in G^*} \sum_{k=1}^K\sigma_{<}\left(\left\|\mathbf{X}^{(k)}(v)-\mathbf{X}^{(k)}\left(v^*\right)\right\|_F\right),  
\end{equation}
where $G^*$ is the augmented network.
This alignment matrix is used for final node matching, aligning the nodes in the source network to their corresponding nodes in the target network.
Similarly, earlier work by Wang et al.\cite{wang2018cross} proposed a network alignment method based on GCN as well. This approach utilizes pre-aligned entities to train GCNs, embedding entities from different knowledge networks into a unified vector space, thereby directly modelling the equivalence relationships among entities. The embeddings incorporate both structural and attribute information of the entities, while node alignment is achieved based on the distances between entities in the embedding space. 
Considering the necessity of clearly separating each node's embedding vector from its neighbouring vectors when learning alignment functions between heterogeneous networks, dNAME \cite{zhou2019disentangled} employs GCNs to learn the latent space of a single network and combines it with a kernel-based regularizer to distinguish the representation of a corresponding node from its neighbours. 
To address the limitation of traditional GCNs in handling multi-relational network alignment tasks, Ye et al. \cite{ye2019vectorized} proposed a vectorized relational graph convolutional network (VR-GCN), which simultaneously generates both entity and relation embeddings to integrate GCN with multi-relational networks. In the convolution process, VR-GCN leverages the role discrimination and translation properties inherent in knowledge networks. Based on this, an alignment framework AVR-GCN was developed for multi-relational network alignment tasks. AVR-GCN employs a weight-sharing mechanism to map the embeddings of different networks into a unified space, aiming to minimize the distance between corresponding nodes from different networks as the objective function for end-to-end supervised training.

In contrast to the above works, ORIGIN~\cite{zhang2019origin} employs two distinct types of GCNs to accomplish the alignment task. Firstly, two independent intra-GCNs aggregate neighbourhood representations within a single network. Then, an inter-GCN is employed to aggregate neighbourhood information across different networks, facilitating the learning of representations. ORIGIN addresses the challenges of non-rigid network alignment by simultaneously learning node representations across multiple networks. Similarly, the study by Yang et al.~\cite{yang2020re} also presents a two-stage entity alignment method based on a relation-enhanced GCN, which first learns feature representations from different networks and subsequently aggregates them.

\paragraph{B. GAT-based methods}\mbox{}\\

Although GCNs have demonstrated great potential in various network alignment tasks, they exhibit certain limitations. Firstly, they cannot effectively handle induction tasks involving dynamic networks because the update of their node representations depends on the fixed network structure. Additionally, GCNs face challenges in handling directed networks, as they do not permit the assignment different weights to different nodes among neighbors. In contrast, GATs dynamically adjust the weights between nodes through an attention mechanism, which depend only on the node features,independent of the network structure.
These characteristics make them more flexible when the network structure changes (e.g. the addition of new nodes) and theoretically better suited for inductive tasks, positioning them as a compelling alternative for advancing end-to-end network alignment methods.

The primary distinction between GCNs and GATs lies in their approaches to feature aggregation and weight assignment strategies: GCNs aggregate neighbour features equally based on convolution operations, while GATs aggregate neighbour features in a weighted manner through a self-attention mechanism.
The input of GATs is a set of node features \(\mathbf{X} = \{\mathbf{x}_1, \mathbf{x}_2, \ldots, \mathbf{x}_N\}\), where \(\mathbf{x}_i \in \mathbb{R}^F\), \(N\) is the number of nodes, and \(F\) denotes the number of features per node. During feature aggregation, GATs first multiply the weight matrix  \(\mathbf{W}\) by the feature embeddings of the two nodes. The resulting vectors are then concatenated and subjected to a dot product with a learnable weight vector \(\vec{a}\) to obtain the attention coefficient \(e_{ij}\), which indicates the importance of node \(i\) to node \(j\):
\begin{equation}
    e_{ij} = a\left(\mathbf{W} \mathbf{x}_i, \mathbf{W} \mathbf{x}_j\right) = \vec{a}^T \left[ \mathbf{W} \mathbf{x}_i \| \mathbf{W} \mathbf{x}_j \right].
\end{equation}
To mitigate the potential issue of extreme variances in attention coefficients among different node pairs, the attention coefficient \(e_{ij}\) is subsequently normalized using the softmax function, resulting in \(\alpha_{ij}\):
\begin{equation}
  \alpha_{ij} = \frac{\exp\left(\text{LeakyReLU}\left(e_{ij}\right)\right)}{\sum_{k \in \Gamma_1(i)} \exp\left(\text{LeakyReLU}\left(e_{ik}\right)\right)},  
\end{equation}
where \(\Gamma_1(i)\) represents the set of neighbors of node \(i\). Finally, the normalized attention coefficients are used to linearly combine the corresponding features, producing the final output feature for each node.
\begin{equation}
   \mathbf{x}'_i = \sigma\left(\frac{1}{K} \sum_{k=1}^K \sum_{j \in \Gamma_1(v_i)} \alpha_{i j}^k \mathbf{W}^k \mathbf{x}_j\right)
\end{equation}
where $K$ is the number of the attention mechanisms used.

In network alignment tasks, the input for GATs typically consists of  two pre-aligned node pairs. First, GATs learn feature representations of the network nodes in a high-dimensional space. then, the optimization of the entire GAT model is directed towards minimizing the distance between the pre-aligned node pairs within this feature space, thereby achieving the goal of aligning the two networks \cite{wang2020knowledge, mao2020mraea}.

\paragraph{C. others}\mbox{}\\
To address the diverse requirements of network alignment tasks, a variety of customized GNN networks have been developed to improve different aspects of network alignment tasks. For instance, the Relational Graph Convolutional Network (R-GCN)~\cite{schlichtkrull2018modeling} was introduced for modeling multi-relational graphs, and the Relation-Aware Dual-Graph Convolutional Network (RDGCN)~\cite{wu2019relation} was proposed to tackle the challenge of capturing and integrating multi-relational information between two graphs. Other models include GraphSAGE~\cite{hamilton2017inductive}, GIN~\cite{xu2018how},Grad-Align~\cite{ park2023power}, HNN~\cite{feng2019hypergraph}, HGCN~\cite{wu2019jointly}, RelGCN~\cite{zhang2021RelGCN}, CG-MuAlign~\cite{zhu2020collective}, and so on, all of which contribute to network alignment tasks. The primary distinction among these GNN-based approaches commonly lies in their methods for generating different network feature representations, which is key to their performance in network alignment.

Zhang et al.~\cite{zhang2023mining} employed GraphSAGE to encode networks at varying granularities, replacing node features with the output embeddings from GraphSAGE. This updated representation of node features integrated both structural and attribute information. Furthermore, the study constructed comparative views within and across networks, utilizing the distances between positive views as contrastive loss to enhance the model's feature extraction capabilities and alignment robustness.
Grad-Align~\cite{park2023power} utilized dual GNNs to compute the similarity of multi-layer embeddings from two networks. The method incorporated weight-matrix-sharing techniques to generate hidden representations for each network consistently and employed a newly designed hierarchical reconstruction loss to accurately capture multi-hop neighborhood structures during training.Grad-Align further analyzed the impact of weight-matrix-sharing mechanisms on node embeddings generated by three types of GNNs, GCN, GraphSAGE, and GIN, in network alignment tasks. The results indicate that GIN consistently outperforms the other models. GIN was introduced to extend the Weisfeiler-Lehman network isomorphism test, maximizing discriminative power among GNNs. Higher expressiveness of node representations increases the likelihood of accurately identifying node correspondences without ambiguity. Thus, the GIN model achieves optimal performance through enhanced representational capability.

HyperAlign~\cite{do2024unsupervised} explored the unsupervised hypergraph alignment problem by using two hypergraphs as input. It pretrained the HNN parameters through a node contrastive learning task, functioning as a ``pseudo-supervised alignment task". Subsequently, it employed a generative adversarial network to align the node embedding spaces of the two distinct networks and predicted node alignment based on the learned similarity of the node embeddings. To encode consistency in network alignment, NeXtAlign~\cite{zhang2021RelGCN} proposed a specialized relational graph convolutional network (RelGCN). RelGCN integrated the relative positional information of nodes into the GCN's message aggregation mechanism, interpreting message passing as the determination of the relative positions of all nodes outside the anchor node. To achieve alignment of various types of entities through a single model, Zhu et al.~\cite{zhu2020collective} introduced a collective GNN for multi-type entity alignment, known as CG-MuAlign. This approach included a novel collective aggregation function tailored for network alignment. CG-MuAlign utilized both attribute information and neighborhood information during the alignment process, employing node-level cross-graph attention and edge-level relation-aware self-attention within the collective aggregation function to address the incompleteness of knowledge network data.
\subsubsection{Feature extraction-based methods}
Feature extraction-based methods, which are the traditional machine learning-based methods, typically involve three main steps. First, extracting features: for each pair of corresponding nodes across different networks, a set of features is extracted. Second, classification model training. Once the features are extracted, they are used to train a classification model (e.g., Support Vector Machines, Decision Trees, Random Forests, etc.). Third, classification model application: the same set of features is extracted from unmatched nodes across pairs across different networks, and these features are inputted to the trained classification model. The classification model then predicts whether the pair of nodes is a corresponding node pair or not. 

Kong et al.~\cite{KongXiangnan2013} extracted several network structure features and attribute features between nodes across different networks to train a SVM classification model. The network structure features include the number of CINs (calculated using Eq.~(\ref{eq_localcons_cn})), the extended Jaccard's coefficient index about the CINs (calculated using Eq.~(\ref{eq_localcons_ejc})), and the extended Adamic/Adar index about the CINs (calculated using Eq.~(\ref{eq_EAA})). The attribute features include the spatial distribution features, temporal features, and text content features. Peled et al.~\cite{peled2016matching} used the number of CINs and the number of second-order CINs identified by identical names as the network structure features and used name similarities, location distances, current employer similarities, text similarities as the attribute features. Similar methods can be found in Refs.~\cite{you2011socialsearch, bartunov2012joint}.

\section{Alignment under different conditions}\label{Sec: Alignment under different conditions}
Networks awaiting alignment often exhibit varying characteristics, such as the presence or absence of node or edge attributes, homogeneity or heterogeneity, directionality (undirected or directed), and whether they are static or dynamic. Each of these conditions introduces unique challenges to the network alignment process. Early research in this field commonly assumed networks to be undirected, homogeneous, and static, with no attributes assigned to nodes or edges. These simplifying assumptions were made to reduce the complexity of the problem, making it more manageable for analysis and enabling researchers to derive conclusions more easily. However, as the field has advanced, the alignment of networks under these simplified conditions has been well-explored. The following sections will delve into the specific challenges and approaches for aligning networks with different characteristics, emphasizing how these features impact the alignment process and shape the development of more sophisticated methods.

\subsection{Alignment for attributed networks} 
\label{attributed}
Besides the purely structural information of networks, many real-world networks contain rich attribute information on both nodes and edges. For example, in OSNs, users fill out profile information such as name, location, occupation, and educational background, and also post content~\cite{ZhouXiaoping2016}; in PPI networks, protein nodes are often associated with their amino acid sequences~\cite{vingron1994sequence}; and in KGs, each node has attribute information composed of detailed descriptions~\cite{pei2019semi}. This attribute information can significantly enhance the accuracy of network alignment if used effectively. However, there are many challenges in utilizing these attributes to facilitate network alignment rather than hinder it. For instance, in social networks, many users may have almost identical profile information, making it difficult to distinguish them; among a user’s numerous posts, only one may be relevant to promoting network alignment, while the rest could introduce noise and confusion. In Sec.~\ref{Sec: Network alignment methods}, we have already introduced different methods for network alignment. Now, we will introduce how each type of method leverages node or edge attribute information.

\subsubsection{Attributes in structure consistency-based methods}

A straightforward approach is to compute the attribute similarity score between two unmatched nodes and add it to their topological similarity score. For example, in IsoRank, the similarity score between unmatched nodes $u^{l_1}_i$ and $u^{l_2}_j$ is given by $\alpha r_{ij} + (1 - \alpha) h_{ij}$, where $h_{ij}$ represents the attribute similarity score between $u^{l_1}_i$ and $u^{l_2}_j$. The control parameter $\alpha$ is typically determined through empirical analysis based on extensive experimentation. In social networks, attribute similarity is often calculated by evaluating the proportion of the same characters in string attributes or the difference in numerical attributes. SimMetrics provides 29 string similarity metric methods for such calculations~\cite{mests2018distributed}\footnote{https://github.com/Simmetrics/simmetrics}. In PPI networks, the basic local alignment search tool (BLAST)~\cite{kelley2004pathblast, altschul1990basic} is commonly used to calculate sequence similarity between unmatched nodes.

\begin{figure} 
	\centering
	\includegraphics[scale=1.2]{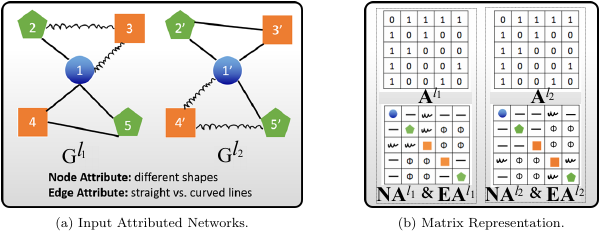}
	\caption{An example of representing the attribute information for the structure consistency-based methods. (a) The input attributed networks. There are two attributed networks $G^{l_1}$ and $G^{l_2}$, each has five nodes. (b) The matrix representation for attributed networks. The upper matrices are the adjacency matrices, which only indicate whether connections exist between nodes but do not capture node or edge attributes. The lower matrices capture node attributes (at the diagonal entries) and edge attributes (at the off-diagonal entries). 
		\\ Source: Reproduced from Ref.~\cite{ZhangSi2016-KDD}. }
	\label{Fig_final_sizhang}
\end{figure}

Some methods explore in greater depth how to effectively incorporate attribute information into structure consistency-based methods. For example, Zhang et al.~\cite{ZhangSi2016-KDD} adhere to the principle that the alignments between two node pairs across different networks should be consistent if the pairs of nodes are similar. This consistency includes their network structure, node attributes, and edge attributes. Denoting $\bm{\mathrm{NA}}$ as the node attribute matrix and $\bm{\mathrm{EA}}$ as the edge attribute matrix, as shown in~\ref{Fig_final_sizhang}, network alignment with attribute information aims to minimize in terms of the alignment matrix $\bm{\mathrm{R}}$:
\begin{equation}
\begin{aligned}
Obj(\bm{\mathrm{R}})=\sum\limits_{v_i^{l_1},v_a^{l_1},v_j^{l_2},v_b^{l_2}} [\frac{\bm{\mathrm{R}}(i,a)}{\sqrt{f(v_i^{l_1},v_a^{l_2})}}-\frac{\bm{\mathrm{R}}(j,b)}{\sqrt{f(v_j^{l_1},v_b^{l_2})}}]^2\\ 
\times \bm{\mathrm{A}}^{l_1}(i,j)\bm{\mathrm{A}}^{l_2}(a,b)\\
\times \mathds{1}(\bm{\mathrm{NA}}^{l_1}(i,i)=\bm{\mathrm{NA}}^{l_2}(a,a))\mathds{1}(\bm{\mathrm{NA}}^{l_1}(j,j)=\bm{\mathrm{NA}}^{l_2}(b,b))\\
\times \mathds{1}(\bm{\mathrm{EA}}^{l_1}(i,j)=\bm{\mathrm{EA}}^{l_2}(a,b)).
\end{aligned}
\label{eq_final_sizhang}
\end{equation}
In Eq.~(\ref{eq_final_sizhang}), $\mathds{1}(\cdot)$ represents the indicator function, $f(v_i^{l_1},v_a^{l_2})$ represents a normalization function that measures the number of neighbouring node pairs, such as $(v_j^{l_1},v_b^{l_2})$, that $v_i^{l_1}$ and $v_a^{l_2}$ share, where these pairs have the same node attribute value and are connected to $v_i^{l_1}$ and $v_a^{l_2}$ via the same link attribute value, $\bm{\mathrm{A}}^{l_1}(i,j)\bm{\mathrm{A}}^{l_2}(a,b)$ ensures the structural consistency, $\mathds{1}(\bm{\mathrm{NA}}^{l_1}(i,i)=\bm{\mathrm{NA}}^{l_2}(a,a))\mathds{1}(\bm{\mathrm{NA}}^{l_1}(j,j)=\bm{\mathrm{NA}}^{l_2}(b,b))$ ensures the node attribute consistency, and $\mathds{1}(\bm{\mathrm{EA}}^{l_1}(i,j)=\bm{\mathrm{EA}}^{l_2}(a,b))$ ensures the link attribute consistency.

\subsubsection{Attributes in network embedding-based methods}
\begin{figure*} [!th]
	\centering
	\includegraphics[scale=1.2]{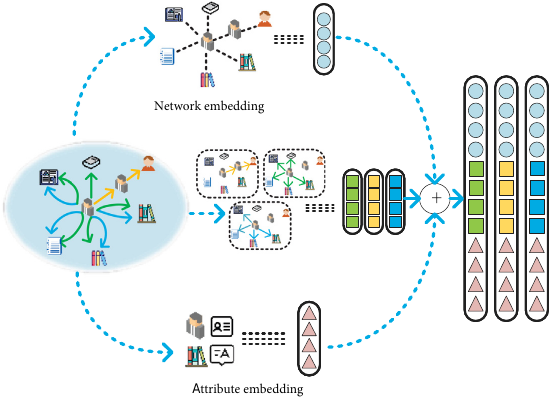}
	\caption{An example of concatenating the different types of representation vectors. For any node $v_i$ in a network, suppose its structure representation vector $\bm{\mathrm{v}}_i^s$ is a four-dimensional vector $[\bm{\mathrm{v}}_{i1}^s,\bm{\mathrm{v}}_{i2}^s,\bm{\mathrm{v}}_{i3}^s,\bm{\mathrm{v}}_{i4}^s]$, its attribute representation vector $\bm{\mathrm{v}}_i^a$ is  $[\bm{\mathrm{v}}_{i1}^a,\bm{\mathrm{v}}_{i2}^a,\bm{\mathrm{v}}_{i3}^a,\bm{\mathrm{v}}_{i4}^a]$, and its any other type representation vector $\bm{\mathrm{v}}_i^o$ is $[\bm{\mathrm{v}}_{i1}^o,\bm{\mathrm{v}}_{i2}^o,\bm{\mathrm{v}}_{i3}^o,\bm{\mathrm{v}}_{i4}^o]$, concatenating these vectors, the final representation vector of node $v_i$ becomes a twelve-dimensional vector: $\bm{\mathrm{v}}_i=[\bm{\mathrm{v}}_i^s,\bm{\mathrm{v}}_i^o,\bm{\mathrm{v}}_i^a]$.
		\\Source: Reproduced from Ref.~\cite{duan2024commerce}. }
	\label{Fig_emb_attandnet}
\end{figure*}

As mentioned in Refs.~\cite{levy2014linguistic,lin2022structured,chen2022efficient}, words, sentences, or even entire documents can be represented as representation vectors while preserving their semantic information. For instance, word vectors learned from large text corpora allow the similarity between words to be measured by the distance between their vectors. The similarity of vectors between two similar words tends to be high, and the relationships between similar word pairs tend to remain consistent. For example, the distance between the vectors of ``princess" and ``queen" is similar to that between ``prince" and ``king". In this way, textual attributes of nodes can be easily represented as attribute representation vectors. These attribute representation vectors can be associated with the structure representation vectors for the network alignment. The integration methods involve concatenating the attribute and structure representation vectors, as well as designing a specific objective function for the model optimization.

For concatenating the attribute and structure representation vectors, as shown in Figure~\ref{Fig_emb_attandnet}, researchers can first embed the network and node attributes separately to obtain distinct representation vectors. These vectors for each node can then be concatenated to form an integrated vector. A method of training mapping function by Eq.~(\ref{lossfunc_pale}) can still be derived to unify the representation vectors from different networks into a unified space, thereby achieving alignment. For example, Li et al.~\cite{li2019partially,li2019adversarial} represent the text attributes as tf-idf vectors and then embed these attributes into the hidden space with the network structure by a node attribute preserving network embedding model proposed in Ref.~\cite{yang2015network}. Zheng et al.~\cite{zheng2021camu} found that the distribution discrepancy of node representations from different networks would positively contribute to the accuracy of network alignment. They integrated node attribute information into representation vectors and proposed a cycle-consistent adversarial mapping model (CAMU) that uses mapping functions and adversarial training for space unified and distribution alignment. Chen et al.~\cite{chen2021adversarial} separately obtained a node's textual representation vector, visual representation vector, check-in representation vector, and structure representation vector and then integrated them. Adversarial learning paradigms are leveraged to make the representation vectors of corresponding nodes from different networks as similar as possible. A similar strategy is then adopted in Ref.~\cite{zhang2019local}.

For the method of designing a specific objective function for the model optimization, Sun et al.~\cite{sun2017cross} computed the similarity scores of attribute vectors between nodes from different KGs. These similarity scores are then combined with the nodes' structure representation vectors to construct a new objective function, which is optimized to improve the alignment. 
When embedding each KG. The objective function is
\begin{equation}
Obj_1=\sum\limits_{(v_i,e_{ij},v_j)\in G} \sum \limits_{(v_i',e_{ij}',v_j') \in G'} ( (\bm{\mathrm{v}}_i+\bm{\mathrm{e}}_{ij}-\bm{\mathrm{v}}_j)^2 - \alpha (\bm{\mathrm{v}}_i'+\bm{\mathrm{e}}_{ij}'-\bm{\mathrm{v}}_j')^2).
\end{equation}
Denoting $\bm{\mathrm{C}}$ as the matrix of attribute representation vectors in a network, $\bm{\mathrm{R}}^{(l_1,l_2)}$ as the similarity matrix of attribute representation vectors between nodes across networks, $\bm{\mathrm{R}}^{(l_1)}$ as the similarity matrix of attribute representation vectors between nodes in $G^{l_1}$, $\bm{\mathrm{R}}^{(l_2)}$ as the similarity matrix of attribute representation vectors between nodes in $G^{l_2}$, these similarity matrices can be calculated by
\begin{equation}
\bm{\mathrm{R}}^{(l_1,l_2)}=\bm{\mathrm{C}}^{l_1}{\bm{\mathrm{C}}^{l_2}}^T, \quad \bm{\mathrm{R}}^{(l_1)}=\bm{\mathrm{C}}^{l_1}{\bm{\mathrm{C}}^{l_1}}^T, \quad \bm{\mathrm{R}}^{(l_2)}=\bm{\mathrm{C}}^{l_2}{\bm{\mathrm{C}}^{l_2}}^T.
\end{equation}
Denoting $\bm{\mathrm{V}}$ as the matrix of structure representation vectors in a network, the objective function for the alignment process is 
\begin{equation}
Obj_2=||\bm{\mathrm{V}}^{l_1}-\bm{\mathrm{R}}^{(l_1,l_2)}\bm{\mathrm{V}}^{l_2}||^2_F+ \beta(||\bm{\mathrm{V}}^{l_1}-\bm{\mathrm{R}}^{l_1}\bm{\mathrm{V}}^{l_1}||^2_F+||\bm{\mathrm{V}}^{l_2}-\bm{\mathrm{R}}^{l_2}\bm{\mathrm{V}}^{l_1}||^2_F),
\end{equation}
where $\beta$ is a parameter to balance the different types of similarity matrices. The final objective function is $Obj=Obj_1+\delta Obj_2$. In Ref.~\cite{trisedya2019entity}, the objective function to learning the attribute representations are also considered into the designed specific objective function. 

\subsubsection{Attributes in GNN-based methods}
When using GNN-based  network alignment methods to handle attributed networks, node attributes are first represented as high-dimensional feature vectors using various techniques, such as word embeddings or other pre-trained model embeddings. The attribute feature vectors of different nodes are then concatenated to form a feature matrix 
\bm{\mathrm{X}}, which serves as the input to the GNN. During model training, different GNN models employ various strategies to aggregate neighbourhood information, such as mean aggregation, max aggregation, or attention mechanisms. Through iterative updates, the feature matrix \bm{\mathrm{X}} is progressively optimized, improving the accuracy of node embeddings and enhancing the effectiveness of network alignment.

\subsubsection{Attributes in feature extraction-based methods}
In feature extraction-based methods, there are typically two approaches: either attribute and network structure information are both used to extract features for training a classification model, or network structure information is used as a filtering condition. In the first approach, attribute information often plays a primary role, while features derived from network structure typically serve a supplementary function. For example, in Ref.~\cite{zafarani2013connecting}, 414 distinct features were extracted solely from OSN usernames, while in many studies, features extracted from network structure between unmatched nodes are often limited to degree similarity or the number of commonly identified neighbours (CINs). Liu et al.~\cite{LiuSiyuan2014,liu2015structured} proposed a framework called HYDRA, which models attributes such as name, gender, age, user topics, and language style separately from network structure information. The objective functions for attribute modelling and network structure modelling are combined to address the alignment problem as a multi-objective optimization problem. Zhong et al.~\cite{zhong2018colink} proposed CoLink, which uses attribute-based and relationship-based models separately and defines a co-training algorithm to reinforce these two models iteratively. The candidate matched node pairs with high confidence obtained from these two models are merged into the matched corresponding node pair set and retrained these two models until they converge.  

In the second approach, network structure is often used as a filtering criterion to reduce the number of node pairs that need to be combined for feature extraction.
For example, Jain et al.~\cite{JainParidhi2013} first search in $G^{l_2}$ based on the attributes of the node $u_i^{l_1}$ in $G^{l_1}$, such as username, name, age, posts, and timestamp, along with the friendship information of its neighbours, to identify a set of candidate nodes. Then, they perform a string, numeric, and character type attributes comparison and image comparison between $u_i^{l_1}$ and each node in the candidate set, ranking the candidates to obtain the alignment results.
Nie et al.~\cite{NieYuanping2016} obtain the similarity between unmatched nodes across two networks by comparing the frequency of a node mentioning its neighbours, the topic similarity of posts made by their neighbouring nodes, and how these factors change over time. The alignment results are achieved by ranking the similarity scores.
Qin et al.~\cite{qin2020two} start by comparing the username similarity between the neighbours of seed nodes to identify potentially matched node pairs. They then evaluate the candidate-matched node pairs based on several similarity metrics, including username similarity, location similarity, educational background similarity, and the similarity of their one-hop and two-hop neighbours. These similarity measures are used as features to train classification models such as Naive Bayes and Support Vector Machine. Finally, the trained classification models determine whether two network nodes are matching counterparts.

\subsection{Alignment for heterogeneous networks} 
\label{homogeneous}

\begin{figure*} [ht!]
    \centering
    \includegraphics[width=0.95\textwidth]{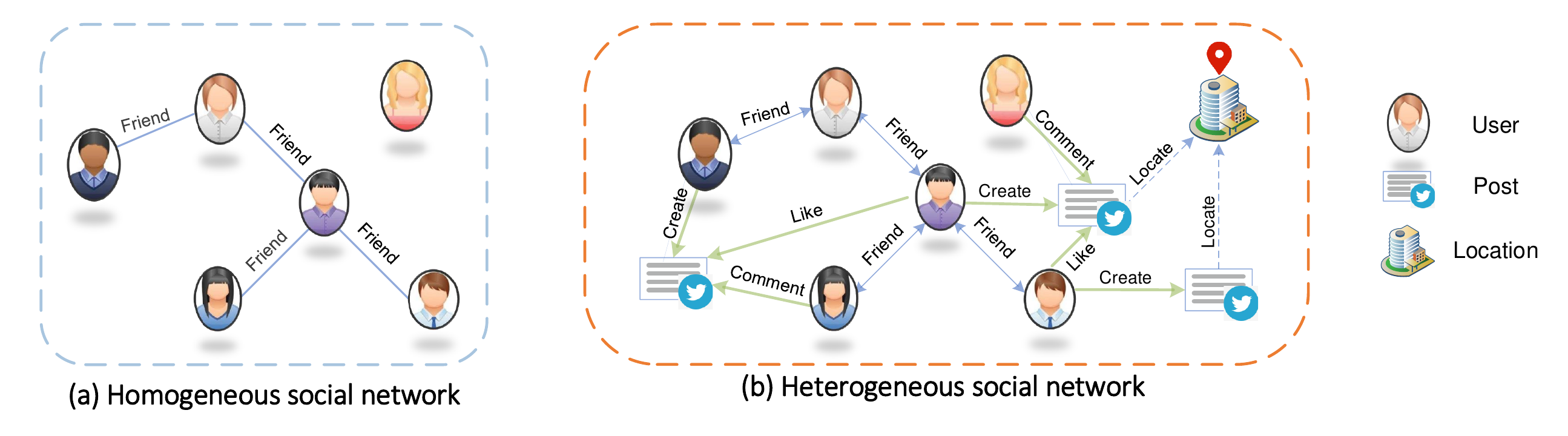}
    \caption{Comparison between homogeneous OSNs and heterogeneous OSNs. (a) Homogeneous social network: A network where the only node type is user, and the only edge type is friendship. (b) Heterogeneous social network: A network with multiple node types and edge types. The node type include user, post, and location. The edge types include friendship, user interactions with posts (such as create, like, or comment), and relationship between a post and location.
    }
    \label{pic_hetero_OSN}
\end{figure*}

Network alignment primarily focuses on homogeneous networks, where the types of nodes or edges are assumed to be unique. However, real-world networks are often heterogeneous and consist of multiple types of nodes and edges~\cite{zafarani2013connecting}. For example, in OSNs, node types contain users, posts, locations, and so forth; edge types include user friendships, user interactions with posts such as likes, shares, or comments, and relationships between posts and locations, as shown in Figure~\ref{pic_hetero_OSN}. Similarly, in PPI networks, protein domains interact with each other. The networks involve nodes of protein and domain, with relationships of protein-protein, domain-protein, and domain-domain. Noting that the type of nodes and edges differs from the attributes of nodes and edges. In academic cooperation networks, node types include authors, papers, and research topics, while edge types include co-authorship, author papers, and citation relationships between papers. 
 
The main approach for handling heterogeneous networks is to extract meta-paths, sequences of node types and edge types that connect two nodes in a heterogeneous network, capturing the relationships between different types of entities~\cite{dong2017metapath2vec}. These meta-paths can be summarized into meta-diagrams. Meta-diagrams extend the concept of meta-paths by representing multiple meta-paths in a single structure, capturing the complex interactions between nodes and edges of different types. Based on the meta-paths and meta diagrams, the similarity between unmatched nodes can be calculated, or the embedding of heterogeneous networks can be performed. Therefore, the methods for aligning heterogeneous networks alignment can be simply categorized into meta-path-based methods and other methods. 

\begin{figure*} [ht!]
    \centering
    \fbox{\includegraphics[width=0.9\textwidth]{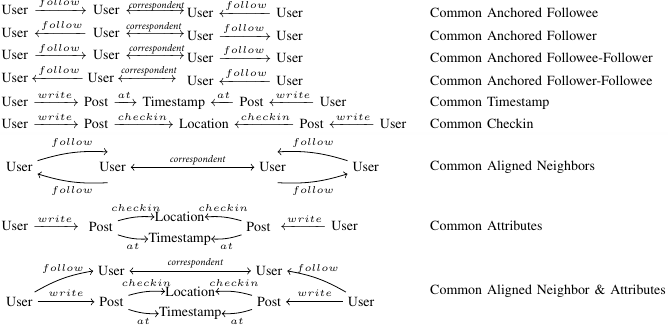}}
    \caption{Illustration of meta diagrams. There are nine types of meta diagrams that capture the complex interactions between different types of nodes and edges. \\ Source: Reproduced from Ref.~\cite{ren2019meta}}
    \label{pic_metadiagram}
\end{figure*}

For meta path-based methods, Ren et al.~\cite{ren2019meta} defined a set of meta diagrams for heterogeneous network alignment. As shown in Figure~\ref{pic_metadiagram}, nine types of representative meta diagrams belong to the meta diagram set $\mathcal{MD}$. Given users $u^{l_1}_i$ and $u^{l_2}_j$ across different heterogeneous social networks, based on any meta diagram $\mathcal{MD}_k \in \mathcal{MD}$, it is possible to extract several instances that satisfy any meta diagram. The set of any type of meta diagram instances connecting $u^{l_1}_i$ and $u^{l_2}_j$ can be represented as $\mathcal{I}_{\mathcal{MD}_k}(u^{l_1}_i, u^{l_2}_j)$. Each user node can have many meta-diagram instances going into or out of him/her. Denoting all the meta diagram instances going out form $u^{l_1}_i$ (or going into $u^{l_2}_j$), based on the meta diagram $\mathcal{MD}_k$, as $\mathcal{I}_{\mathcal{MD}_k}(u^{l_1}_i, :)$ (or $\mathcal{I}_{\mathcal{MD}_k}(:,u^{l_2}_j)$), the similarity score between $u^{l_1}_i$ and $u^{l_2}_j$ based on meta diagram $\mathcal{MD}_k$ can be calculated as the following meta proximity concept:
\begin{equation}
r_{\mathcal{MD}_k}(u^{l_1}_i, u^{l_2}_j)=\frac{2|\mathcal{I}_{\mathcal{MD}_k}(u^{l_1}_i, u^{l_2}_j)|}{|\mathcal{I}_{\mathcal{MD}_k}(u^{l_1}_i, :)|+|\mathcal{I}_{\mathcal{MD}_k}(:,u^{l_2}_j)|}.
\end{equation}
By summing the similarity scores of different types of meta diagrams, the similarity between $u^{l_1}_i$ and $u^{l_2}_j$ can be obtained. The overall alignment of heterogeneous networks can be achieved by comparing the similarity scores across all user nodes. It is important to note that counting the instances of meta diagrams among user nodes is a complex task, as it involves a graph isomorphism step to match the various types of meta diagrams. Many other researchers are dedicated to achieving more cost-effective, accurate, and time-efficient heterogeneous social network alignment within this framework, such as SHNA~\cite{ren2020scalable}, PNA~\cite{zhang2015pna}, ActiveIter~\cite{ren2019activeiter}, TransLink~\cite{zhou2019translink}, and iDev~\cite{fan2019idev}. 

For other methods, Li et al.~\cite{li2020type} designed a type-aware alignment method that utilizes GAT to construct a complete end-to-end alignment framework for heterogeneous networks. The framework generates a user node's $i$-th type-aware representation vector by integrating the representation vectors of all its neighbours that belong to the $i$-th type. It then fuses the different type-aware representation vectors of the user node into a fusion one. Finally, the type information and fusion information together serve as the input for the alignment part of the framework. 

\subsection{Alignment for directed network} 
\label{directed}

For the alignment of directed networks, a simple approach is to assume that these networks are undirected. Specifically, for a directed network $G$, if $\bm{\mathrm{A}}(i,j)=1$, then set $\bm{\mathrm{A}}(j,i)=1$ and vice versa. Many network alignment studies have been conducted under this assumption to handle directed networks. However, some researchers believe that the direction of edges plays a critical role in network alignment, and they have developed specialized methods to address this condition. This specialized handling is mainly applied in social network alignment and KG alignment, as both types of networks are often directed. In social networks, user relationships may have directionality, such as follower-followee relationships. At the same time, in KGs, directed edges represent specific relationships between entities, such as "is a" or "part of." The direction of edges in these networks carries important semantic information, making it essential to develop methods that preserve and leverage this aspect during the alignment process. We will provide a brief overview of the approaches that specifically handle directed networks in the context of different alignment methods.

In structure consistency-based methods, it is possible to distinguish between the contributions of incoming and outgoing edges to the similarity score between unmatched nodes. Narayanan and Shmatikov~\cite{narayanan2009anonymizing} developed the algorithm for directed networks, which employs unmatched nodes’ in-degree and out-degree, as well as the identified nodes, to calculate the similarity score between unmatched nodes. It is defined as
\begin{equation}
r_{ij}^{NS}=\frac{c^{in}}{\sqrt{d^{in}(u^\beta_j)}}+\frac{c^{out}}{\sqrt{d^{out}(u^\beta_j)}},
\label{eq_NS}
\end{equation}
where $c^{in}$ and $c^{out}$ are the numbers of common incoming and outgoing identified neighbours of nodes $u^{l_1}_i$ and $u^{l-2}_j$ respectively, and $d^{in}(u^{l_2}_j)$ and $d^{out}(u^{l_2}_j)$ are the in-degree and out-degree of node $u^{l_2}_j$ respectively. 

\begin{figure*} [ht!]
    \centering
    \includegraphics[width=0.5\textwidth]{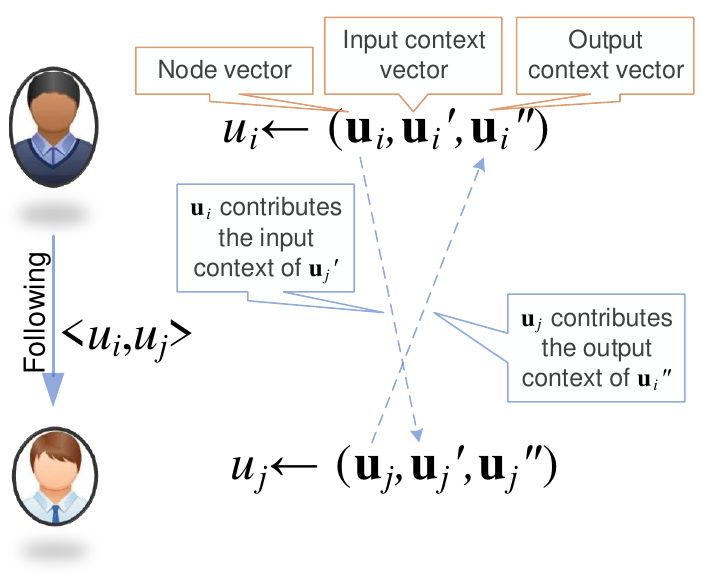}
    \caption{Illustration of leveraging direction of edges for the network embedding. On an OSN platform, user $u_i$ follows user $u_j$, which can be represented as a directed edge$<u_i,u_j>$. The final representation vectors of $u_i$ and $u_j$ are obtained through the mutual contributions of their node vectors, input context vectors, and output context vectors. Specifically, the node vector $\bm{\mathrm{u}}_i$ of $u_i$ contributes to the input context vector ${\bm{\mathrm{u}}_j}^{'}$ of $u_j$, while the output context vector ${\bm{\mathrm{u}}_i}^{''}$ of $u_i$ is influenced by the node vector $\bm{\mathrm{u}}_j$ of $u_j$. The contribution relationships among these three types of vectors reflect the direction of edges.
    \\Source: Reproduced from Ref.~\cite{LiuLi2016}
    }
    \label{pic_IONE}
\end{figure*}

\begin{figure*} [ht!]
    \centering
    \includegraphics[width=0.6\textwidth]{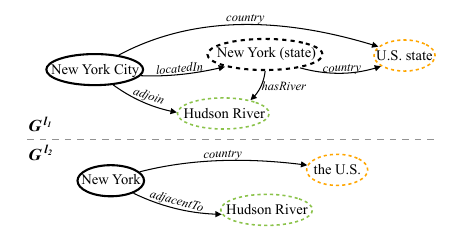}
    \caption{An example of directed edge in KGs. There are two similar KGs, where the nodes represent various proper nouns. The edges between nodes capture meaningful, directed relationships between these nouns. For example, in KG $G^{l_1}$, the node New York (state) has a directed edge pointing to the node Hudson River, with the edge attribute hasRiver,  indicating that the state has a river named Hudson River.
    \\Source: Reproduced from Ref.~\cite{li2019semi}
    }
    \label{pic_kecg}
\end{figure*}

In machine learning-based methods, the direction of edges are utilized to reflect how a node's representation influences other nodes, ensuring that the final representation vectors of all nodes are derived through the directed edges. For example, Liu et al.~\cite{LiuLi2016,liu2019structural,liu2023wl} represented a node by a node vector $\bm{\mathrm{u}}_i$, an input context vector ${\bm{\mathrm{u}}_i}^{'}$, and an output context vector ${\bm{\mathrm{u}}_i}^{''}$. This representation was used to capture the contribution of edge direction to the network alignment of nodes. For a directed edge $<u_i,u_j>$ from $u_i$ to $u_j$, the node vector $\bm{\mathrm{u}}_i$ of $u_i$ contributes to the input context vector ${\bm{\mathrm{u}}_j}^{'}$ of $u_j$, while the output context vector ${\bm{\mathrm{u}}_i}^{''}$ of $u_i$ is influenced by the node vector $\bm{\mathrm{u}}_j$ of $u_j$, as illustrated in Figure~\ref{pic_IONE}. The mutual contributions among the three types of vectors are used to optimize and solve for the final node vectors. 
Bordes et al.~\cite{bordes2013translating} took the idea of letting the representation vector of each head node $\bm{\mathrm{v}}_i$ plus the representation vector of the edge $\bm{\mathrm{e}}_{ij}$ from $v_i$ to $v_j$ equal to the representation vector $\bm{\mathrm{v}}_j$ of the tail node $v_j$ in the triplets. This method captures the directed nature of KGs, where edges are represented as triplets (head node, relation, tail node), as shown in Figure~\ref{pic_kecg}. In this way, the representation vectors of nodes in a KG are derived based on the directed edges, effectively capturing both the topological and direction information inherent to the KG structure. Similar approaches have been employed in various KG alignment studies, such as in Refs.~\cite{chen2017multilingual, zhu2017iterative, zhu2019neighborhood, sun2018bootstrapping, lin2019guiding, pei2019improving, pei2019semi}. 

\subsection{Alignment for dynamic networks} 
\label{dynamic}
Real-world networks often change over time and can be described by dynamic networks. A dynamic network depicts the relationships between nodes and how these relationships emerge, disappear, or evolve. Most existing network alignment studies treat networks as static, but relying solely on static information can result in a loss of accuracy, as the evolving nature of networks may reveal some distinctive patterns for accurate alignment~\cite{sun2019dna}. The core challenge lies in capturing the valuable information provided by the evolving nature of networks for alignment purposes. We will introduce how structure consistency and machine learning-based methods capture this information effectively. 

\begin{figure*} [ht!]
    \centering
    \includegraphics[width=0.8\textwidth]{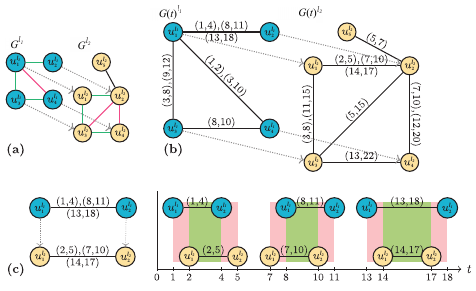}
    \caption{Example of utilizing dynamic information for the network alignment in the structure consistency-based methods. (a) Two static networks $G^{l_1}$ and $G^{l_2}$. Solid lines represent edges between nodes, while dashed lines indicate corresponding node pairs. The solid lines are categorized into two types: green lines represent conserved edges, meaning the relationships between the same nodes exist in both networks; red lines represent non-conserved edges, indicating that the edges exist in only one network but not in the other. (b) Two dynamic network $G(t)^{l_1}$ and $G(t)^{l_2}$. A solid line between two nodes indicates that they interact at least once during the network’s lifetime. A single solid line can represent multiple events. Each event is described as $(t_s, t_e)$, where $t_s$ is the start time, and $t_e$ is its end time. For example, the event between $u_3^{l_1}$ and $u_4^{l_2}$ is active from $t_s^{l_1}=8$ to $t_e^{l_2}=10$. Network alignment for dynamic networks aims to align the two networks by leveraging the valuable information provided by their evolving nature. (c) The concepts of CET and NCET. The green shaded area represents CET, indicating the event between two nodes are active in both networks. The red shaded area represents NCET, indicating the event between two nodes are active in only one network. 
    \\Source: Reproduced from Ref.~\cite{vijayan2017alignment}
    }
    \label{pic_DynaMAGNA}
\end{figure*}

For structure consistency-based methods, the duration of interactions can be used to weigh the contribution of edges to the similarity score. Longer-lasting interactions may contribute more heavily to the similarity score between unmatched nodes, reflecting the stability and strength of relationships over time, while shorter interactions may have a lesser impact. For example, Vijayan et al.~\cite{vijayan2017alignment} proposed DynaMAGNA++, which maximizes the duration of conserved edges between different networks for dynamic network alignment. They assumed that the dynamic network $G(t)(V, T)$ consisting of a node-set $V$ and an event set $T$, where an event is a temporal edge, consisting of a 4-tuple $(v_i,v_j,t^{l_{1}}_{s},t^{l_{1}}_{e})$. If there is an edge connecting two nodes in the $t^{th}$ snapshot, then there is an event between the two nodes that is active from time $t$ to $t+1$. Conserved event time (CET) and non-conserved event time (NCET) are proposed to measure the duration of conserved edges. For events $(u^{l_{1}}_{i},u^{l_{1}}_{a},t^{l_{1}}_{s},t^{l_{1}}_{e})$ and $(u^{l_{2}}_{j},u^{l_{2}}_{b},t^{l_{2}}_{s},t^{l_{2}}_{e})$ in $G(t)^{l_1}$ and $G(t)^{l_2}$ respectively, $CET((u^{l_{1}}_{i},u^{l_{1}}_{a}),(u^{l_{2}}_{j},u^{l_{2}}_{b}))$ is the amount of time that $(u^{l_{1}}_{i},u^{l_{1}}_{a},t^{l_{1}}_{s},t^{l_{1}}_{e})$ and $(u^{l_{2}}_{j},u^{l_{2}}_{b},t^{l_{2}}_{s},t^{l_{2}}_{e})$ are active at the same time if $(u^{l_{1}}_{i},u^{l_{2}}_{j})$ and $(u^{l_{1}}_{a},u^{l_{2}}_{b})$ are corresponding node pairs. $NCET((u^{l_{1}}_{i},u^{l_{1}}_{a}),(u^{l_{2}}_{j},u^{l_{2}}_{b}))$ is the amount of time that $(u^{l_{1}}_{i},u^{l_{1}}_{a},t^{l_{1}}_{s},t^{l_{1}}_{e})$ and $(u^{l_{2}}_{j},u^{l_{2}}_{b},t^{l_{2}}_{s},t^{l_{2}}_{e})$ are active at the different time. The final alignment results are obtained by maximizing CET and minimizing NCET for all node pairs, as illustrated in Figure~\ref{pic_DynaMAGNA}. Similar approaches are adopted in Refs.~\cite{vijayan2018aligning,aparicio2019temporal}.

For machine learning-based methods, techniques that model temporal information, such as long short-term memory networks (LSTM)~\cite{graves2012long,manessi2020dynamic}, are often employed to perform dynamic network alignment. These methods leverage LSTM's ability to capture sequential patterns of the dynamic networks. For exmaple, Sun et al.~\cite{sun2019dna} propose a dynamic social network
alignment (DNA) framework. The framework begins by applying a random walk with a restart to calculate proximity scores between nodes and others. These scores are used to construct ego networks for each node, capturing their closest relationships. The temporal evolution of these ego networks is then analyzed by tracking changes in proximity scores across different time snapshots. These evolving proximity sequences are input into an LSTM model to generate representation vectors that capture dynamics and structure information. Finally, the alignment results are obtained by unifying the hidden space and calculating the similarity of the representation vectors. Balakrishnan et al.~\cite{balakrishnan2023network} also employ LSTM to model the dynamics of networks. Wang et al.~\cite{peng2024accurate} and Yan et al.~\cite{yan2021dynamic} utilized GNN to capture structural information and a self-attention-based encoder to model dependencies across different snapshots over time.

\subsection{Alignment without seeds} 
\label{seeds}
Seeds are the observed corresponding node pairs providing to network alignment algorithms or models. For structure consistency-based methods, seeds offer critical information for guiding the alignment process. In local structure consistency-based methods, seeds form the basis for calculating similarity scores between unmatched nodes; without them, reliable similarity measures are nearly impossible. In global structure consistency-based methods, the similarity score between seeds can be preset to 1, allowing their neighboring nodes to gain more accurate similarity estimates, which then propagate outward to refine the similarity scores across the entire network. For machine learning-based methods, having a sufficient number of high-quality seeds as labeled data is crucial for training an effective alignment model. Generally, the more seeds available, the better the model’s potential performance, as a larger training set provides more information for learning accurate alignments~\cite{bishop2006pattern}.

However, obtaining seeds is often challenging. In social network alignment, privacy concerns hinder the identification of corresponding node pairs across OSN platforms. As users and platforms become more more conscious of cybersecurity, it is increasingly rare for users to share identifiable information, like telephone numbers or email addresses, across multiple OSNs, and they often use different usernames on different platforms. In biological network alignment, the scarcity of experimentally validated protein or gene correspondences complicates the acquisition of seeds. Studies such as those by Singh et al.~\cite{singh2008global} and Clark et al.~\cite{clark2014comparison} highlighted the challenges in aligning biological networks due to the limited availability of validated data. For KG alignment, the process of labeling corresponding node pairs is highly labor-intensive due to the vast number of nodes and edges. As a result, many researchers are exploring how to perform network alignment without relying on seeds. 

According to the different sources of discriminative information the seedless network alignment methods rely on, they can be categorized into three types: using hand-crafted rules to identify potential seeds, leveraging sufficient attribute information, and capturing valuable structural information from the network structure more carefully.

\textbf{(i) Using hand-crafted rules to obtain potential seeds.} In OSNs, a fraction of users like explicitly linking their accounts across multiple OSNs in their profile pages~\cite{korula2014efficient}. For example, some people post Facebook or Foursquare URLs on Google+. Additionally, some individuals may use the same username, photos, email, etc., across different OSNs. These provide seeds for social network alignment. Therefore, some researchers investigate the unsupervised network alignment methods by using these common attributes~\cite{li2018distribution}. For example, to obtain seed nodes, Zhong et al.~\cite{zhong2018colink} define some hand-crafted rules, including user name uniqueness, attribute value mapping, and relationship propagation. Then, they use the identified seeds as input for other alignment methods requiring seeds.

\textbf{(ii) Leveraging attribute information sufficiently.} Heimann et al.~\cite{heimann2018regal} propose a cross-network matrix factorization method for the node representation, which can preserve both the structural and attribute similarities of the nodes. To capture the structural similarities, they define a $k$-hop neighbour degree distribution vector $\bm{\mathrm{d}}_i^k$ for each node $u_i$ in any network $G$. The $b$-th element of $\bm{\mathrm{d}}_i^k$ represents the number of $k$-hop neighbours of node $u_i$ with a degree of $b$. By taking the weighted average of degree distribution vectors across different hop neighbours, the neighbour degree distribution vector for node $u_i$ can be obtained to measure structural similarity between cross-network nodes. Meanwhile, they define an attribute vector $\bm{\mathrm{a}}_i$ for node $u_i$ where the $b$-th element represents a normalized value of a specific node attribute. Then, the cross-network node similarity of nodes $u_i^{l_1}$ and $u_j^{l_2}$ can be calculated by
\begin{equation}
\bm{\mathrm{R}}(i,j)=\exp( -\delta_1 ||\bm{\mathrm{d}}_i^{l_1}-\bm{\mathrm{d}}_j^{l_2}||^2_2-\delta_2 \mathrm{dist}(\bm{\mathrm{a}}_i^{l_1},\bm{\mathrm{a}}_j^{l_2})),
\end{equation}
where $\mathrm{dist}(\cdot,\cdot)$ is an attribute-based distance. The goal for network representation is then set as finding a $(|V^{l_1}|+|V^{l_2}|)\times k$ matrices $\bm{\mathrm{Y}}$ and $\bm{\mathrm{Z}}$ such that $\bm{\mathrm{R}}\approx \bm{\mathrm{Y}}\bm{\mathrm{Z}}^T$ where $\bm{\mathrm{Y}}$ is the node representation matrix, $\bm{\mathrm{Z}}$ is not needed, and $k$ is the dimension of the representation vectors. To get $\bm{\mathrm{Y}}$ efficiently, the authors select $k \ll (|V^{l_1}|+|V^{l_2}|)$ `` landmark'' nodes randomly from networks $G^{l_1}$ and $G^{l_2}$ so that obtaining a approximation matrix $\tilde{\bm{\mathrm{R}}}$. Referencing Ref.~\cite{drineas2005nystrom}, $\tilde{\bm{\mathrm{R}}}$ is given by $\tilde{\bm{\mathrm{R}}}=\bm{\mathrm{C}}\bm{\mathrm{W}}^*\bm{\mathrm{C}}^T$, where $\bm{\mathrm{C}}$ is a $(|V^{l_1}|+|V^{l_2}|) \times k$ matrix by calculating the similarities between nodes in the two networks and the landmark nodes, $\bm{\mathrm{W}}^*$ is a $k \times k$ matrix by calculating the pseudoinverse of the similarities matrix between landmark nodes to landmark nodes. Taking singular value decomposition for matrix $\bm{\mathrm{W}}^*$, $\tilde{\bm{\mathrm{R}}}$ will equal to $\bm{\mathrm{C}}(\bm{\mathrm{U}}\bm{\mathrm{\Sigma}}\bm{\mathrm{V}}^T)\bm{\mathrm{C}}^T=(\bm{\mathrm{C}}\bm{\mathrm{U}}\bm{\mathrm{\Sigma}}^{1/2})(\bm{\mathrm{\Sigma}}^{1/2}\bm{\mathrm{V}}^T\bm{\mathrm{C}}^T)=\tilde{\bm{\mathrm{Y}}}\tilde{\bm{\mathrm{Z}}}^T$. The node representation matrix $\bm{\mathrm{Y}}$ is approximated by $\tilde{\bm{\mathrm{Y}}}=\bm{\mathrm{C}}\bm{\mathrm{U}}\bm{\mathrm{\Sigma}}^{1/2}$. The representation vectors for nodes $u_i^{l_1}$ and $u_j^{l_2}$ are the $i$-th row and $(j+|V^{l_1}|)$-th row of matrix $\tilde{\bm{\mathrm{Y}}}$, respectively. This way, the representation vectors of nodes across different networks are comparable. 

Xie et al.~\cite{xie2018unsupervised} hold the opinion that every attribute associated with a node can reflect the node's uniqueness in a specific aspect to some extent and thus help to distinguish it from others. They constructed a unified network by leveraging the attributes of different nodes to align two social networks. Each user account from both networks is represented as a new user node, and every data object, such as a username, is represented as a data node in the unified network. User-object connections, derived from additional data, and user-user relationships from the original networks are represented as links in this unified network. Network embedding is then performed on the unified network, and the similarity of object nodes is used to bring potential corresponding nodes closer in the hidden space. As a result, even without seeds of user nodes, network alignment can be achieved with a certain level of accuracy.

\textbf{(iii) Capturing valuable structural information from the network structure more carefully.} Pedarsani et al.~\cite{pedarsani2013bayesian} construct network structural statistics such as node degree, clustering coefficient, and the density of some neighbours, which they refer to as fingerprints. These fingerprints are used first to identify the most likely matched pairs. These pairs then provide additional features, such as fingerprints, to help align other nodes, eliminating the need for seeds. It is worth noting that the initial matched pairs are based on the similarity of structural statistics, making this method applicable only to the alignment of similar networks.

Fu et al.~\cite{fu2015effective} proposed a seed-free algorithm based on the principle that nodes are as similar as their neighbours. The algorithm iteratively updates the similarity scores for all possible pairs among the $|V^{l_1}|\times||V^{l_2}|$ potential node pairs until convergence. The alignment results are then determined based on the final similarity scores. For any pair of cross-network nodes $v_i^{l_1}$ and $v_i^{l_2}$, the initial similarity score $r^{(0)}(v_i^{l_1},v_i^{l_2})$, denoted as $\bm{\mathrm{R}}^{(0)}(i,j)$, is set to $1$. In the $k$-th iteration, the algorithm constructs a weighted complete bipartite graph $\bm{\mathrm{B}}_{ij}^{(k+1)}$ to match all neighbours of $v_i^{l_1}$ and $v_i^{l_2}$. In the bipartite graph, any link $(i',j')$ is weighted by $\bm{\mathrm{R}}^{(k)}(i',j')$. The algorithm then identifies the maximum weighted match of $\bm{\mathrm{B}}_{ij}^{(k+1)}$ and the set of matched links are denoted as $M_{ij}^{(k+1)}$. The similarity score $\bm{\mathrm{R}}^{(k+1)(i,j)}$ then can be updated by 
\begin{equation}
\bm{\mathrm{R}}^{(k+1)}(i,j)=\sum\limits_{(i',j')\in M_{ij}^{(k+1)}} \bm{\mathrm{R}}^{(k)}(i',j').
\end{equation}
The above process is performed for all possible $|V^{l_1}|\times||V^{l_2}|$ pairs between nodes from the two networks in each iteration until the similarity score matrix $\bm{\mathrm{R}}$ converges. 

Zhou et al.~\cite{ZhouXiaoping2018-IEEE} proposed a friend relationship user identification without prior knowledge (FRUI-P) method, representing nodes as vectors by extracting features from their neighbours. The similarities between nodes across networks are then calculated by comparing these representation vectors. The final alignment results are obtained by iteratively repeating the node representations and recalculating the similarities multiple times. Li et al.~\cite{li2018distribution} consider all nodes and perform network alignment from the node distribution level. A network $G$ is represented as a distribution by considering the nodes and their features as distribution elements. Each node is treated as a weighted point in a hidden space, where the node's position is defined by its representation vector, and the weight represents the node's importance, such as its degree. In this manner, networks $G^{l_1}$ and $G^{l_2}$ can be depicted as two discrete probability distributions over these weighted points. A projection function $\phi$ is then learned to minimize the Earth Mover's Distance (EMD)~\cite{rubner2000earth} between the projected distribution of $G^{l_1}$ and the distribution of $G^{l_2}$. The ground distance between the projected point in the EMD is subsequently used as the indicator for network alignment. This approach of projecting two networks into different distributions and then measuring their Earth Mover's Distance (EMD) has been widely adopted by researchers to address the problem of having few or no seeds, such as in Refs.~\cite{ren2020banana,zhang2022behavior}.

\section{Discussions and outlooks}
\label{Discussions and outlooks}

In this review, we introduced the network alignment problem and its relevant background, analyzing network alignment issues in different fields such as social network analysis and bioinformatics. We provided a detailed overview of the current advancements in different methods for achieving network alignment, including global and local structure consistency-based methods, machine learning-based methods such as network embedding-based methods, GNN-based methods. We also examined how network alignment can be effectively implemented in the context of attributed networks, heterogeneous networks, directed networks, and dynamic networks. Although network theory provides valuable tools for modelling complex systems, treating each system as an isolated unit for independent analysis may be insufficient when specific components exist across multiple systems simultaneously. Through network alignment, multiple systems can be integrated into a single structure for network dynamics analysis, allowing the comparison of node and edge evolution patterns across different systems. This provides solutions and insights for addressing problems in various fields.

The advent of the big data era and continuous advancements in hardware and software technologies, coupled with the development of advanced computational methods, have opened new pathways for building more accurate, efficient, scalable, and generalizable network alignment methods. In recent years, new technologies such as reinforcement learning~\cite{li2021rlink, zhou2021unsupervised}, adversarial learning~\cite{ren2019dual,derr2021deep,wang2024anchor,zhou2020unsupervised}, meta-learning~\cite{zhou2020fast,xu2023sinkhorn}, active learning~\cite{malmi2017active,cheng2019deep}, transfer learning~\cite{li2014matching,wu2024rethinking}, and contrastive learning~\cite{xiong2021contrastive,liu2021self,zhang2023mining} have all been successfully applied to network alignment, supporting the development of more powerful alignment algorithms and tools. Looking ahead, the powerful capabilities of emerging technologies and methods, such as diffusion models~\cite{kong2023autoregressive} and large language models~\cite{deng2024pentestgpt}, may further enhance the capabilities of network alignment, providing yet another leap forward in the field.

Another critical aspect of network alignment is the inherent interplay of multiple stakeholders' interests, especially in social and biological networks. In social network analysis, there is a clear divergence between the needs of commercial organizations, government agencies, and research institutions to analyze network data and the privacy concerns of OSN platforms and users. While businesses and governments seek to leverage network data for insights to enhance marketing and governance, OSN platforms and users prioritize data anonymity and security to prevent privacy breaches. A similar dynamic exists in biological network alignment, mainly when data is sourced from multiple research teams or institutions. Institutions with unique biological data face a dilemma: on the one hand, they are required to share data for scientific advancement due to academic collaborations; on the other, they fear that data sharing could undermine their competitive advantage. As a result, strategies like network structure perturbation are adopted to protect proprietary data while still impairing other teams' alignment algorithms.
This situation presents research and technical challenges for network alignment: how to maximize disruption to existing alignment algorithms with minimal perturbation~\cite{zhang2020adversarial,TANG2022109095,tang2024degrading,shao2023toak} and develop more robust and resistant alignment tools to handle such perturbations and incomplete data. These challenges represent classic game theory problems, with data holders and alignment developers constantly optimizing strategies to find a dynamic balance. Such research not only helps stakeholders achieve their goals but also deepens our understanding of network alignment.

Additionally, alignment for higher-order networks is equally important. In many cases, a "link" does not merely connect two nodes but can connect more than just two. For example, a group of users in social networks~\cite{alvarez2021evolutionary,tian2024higher} or a group of coauthors of an academic research article~\cite{battiston2020networks}, where each node in the group has pairwise interactions with all other nodes in the group. Higher-order interactions can describe this phenomenon where a hyperlink can involve groups of three or more nodes~\cite{wang2024epidemic, qian2024modeling,li2024social,peng2024message,zhao2023robustness,lai2023robustness}. Existing research has shown that higher-order interactions have heavily impacts on the dynamics of network systems, including cascade~\cite{zhao2024robustness}, synchronization~\cite{lucas2020multiorder,zhang2021unified,gallo2022synchronization}, propagation~\cite{de2020social}, epidemic spreading~\cite{gu2024epidemic,li2023coevolution}, and game~\cite{alvarez2021evolutionary,guo2021evolutionary}. Although a few studies~\cite{peng2024network,huang2023egomuil,zhao2018learning,tan2014mapping} have explored the alignment of higher-order networks, the effects of higher-order interactions on network alignment, as well as the outcomes under different statistical and topological characteristics, still require further in-depth investigation. 

\section*{Acknowledgements}
We want to acknowledge the partial support from the National Natural Science Foundation of China (No. 62202320), Natural Science Foundation of Sichuan Province (2024NSFSC1449), Fundamental Research Funds for the Central Universities (No. 2023SCU12126), Project of Humanities and Social Sciences Research of Chongqing Municipal Education Commission in 2024 (No. 24SKGH048),
Science and Technology Research Program of Chongqing Municipal Education Commission (No. KJQN202200429),
Program for Youth Innovation in Future Medicine, Chongqing Medical University (No. W0150), Science and Technology Project of Sichuan Provincial Administration of Traditional Chinese Medicine(No.2023MS047), National Natural Science Foundation of China (No. 12022113), and Fundamental Research Program of Shanxi Province under Grant No.
202303021223009.

\section*{CRediT authorship contribution statement}
Rui Tang: Writing-review \& editing, Writing-original draft, Methodology, Supervision, Conceptualization. 
Ziyun Yong: Writing-review \& editing, Writing-original draft, Methodology, Investigation.
Shuyu Jiang: Writing-original draft, Methodology, Investigation. 
Xingshu Chen: Writing-original draft, Supervision. 
Yaofang Liu: Writing-review \& editing, Writing-original draft, Supervision, Conceptualization. 
Yi-Cheng Zhang: Writing-review \& editing, Supervision. 
Gui-Quan Sun: Writing-review \& editing, Writing-original draft, Supervision, Conceptualization. 
Wei Wang: Writing-review \& editing, Writing-original draft, Methodology, Supervision, Conceptualization. Yaofang Liu(sclzlyf001@swmu.edu.cn), Gui-Quan Sun(gquansun
@126.com), and Wei Wang(wwzqbx@hotmail.com) are the co-corresponding authors. 

\begin{spacing}{0}
\bibliography{main}
\bibliographystyle{elsarticle-num}
\end{spacing}\textit{}

\end{sloppypar}
\end{document}